# Do voters choose rationally, irrationally or at random? Data and theory for proportional elections.

V. Hösel[1], J. Müller[1,2], A. Tellier[3],

Affiliations:
1 Centre for Mathematical Sciences, Technische Universität München, Boltzmannstr. 3, D-85747 Garching/Munich, Germany

2 Institute of Computational Biology, Helmholtz Zentrum München - German Research Center for Environmental Health, Ingolstädter Landstr. 1, D-85764 Neuherberg, Germany;

3 Section of Population Genetics, Center of Life and Food Sciences Weihenstephan, Technische Universität München, 85354 Freising, Germany

**Corresponding Author:**
Johannes Müller. e-mail: johannes.mueller@mytum.de

**Keywords:**
Elections, stochastic process, Voter model with party dynamics, infinite allele Moran model, Ewens sampling formula.

**Abstract**

Data of proportional elections show a striking feature: If the parties are ranked according to the number of their voters, the number of votes grows exponentially with the rank of the party. This so-called Zipf's law has been reported before. We first show this correlation in results from recent elections in Germany, France and republican primaries in the USA. However the mechanism generating such feature remains so far unexplained.

We develop a mathematical model of voter grouping that is only based on the word of mouth and not on political contents. The model is close to the infinite allele model and the Ewens sampling formula that are well known in population genetics. Strikingly, the model generates output agreeing very well with the observed election data.

We further identify a cannibalism effect in Germany, whereby parties above the 5% threshold withdraw votes from parties just below the threshold. We demonstrate that the number of parties present in the parliament can be predicted given the number of parties standing for election. Finally, the steady loss of big-tent parties in Germany in the last 20 years relates to the increasing number of parties since 1975. We discuss the interpretation and consequences of these findings for modern democracies and the rationality or lack thereof of voters' choices.

**Significance Statement**

It is known that data from elections exhibit several statistical pattern. Such patterns seem to contradict our idea of elections as the rational decision of mature citizens. Up to now there are only few or no convincing available explanations for these structures. By means of a basic mathematical model that does not take into account any political content we are able to reproduce the pattern in data from recent elections in Germany, France and primaries in the USA. Based on such models it is possible to reveal properties in election data and in the dynamics of opinion formation that is not recognizable otherwise. However, it is central to interpret models and results properly to deepen our understanding of democracies and election results, particularly in view of the recently increasing criticisms of the democratic process.

## Introduction

Elections are at the heart of modern democracies. It is of outmost interest to understand the inherent dynamical processes. Statistical analyses of election data reveal several statistical patterns that repeat themselves over a whole range of democracies and elections, namely as the relative number of votes per candidate within a given party (1–3) with respect to spatial correlations (4–6) or other properties (vote participation, hierarchical model)(7, 8). Perhaps the most striking one is the Zipf's law (9–11) which states that considering the parties/candidates in dependency to the number of voters, a double logarithmic plot appear to be linear. The slope is sometimes not equal to one (12–15) but can often be one as well (16–18). In contrast to these papers, we investigate elections with only few candidates/parties (but at least 10), and find a similar result: if the parties are ranked according to the number of votes, the logarithm of the number of votes depends linearly on the rank. This indicates an exponential growth of the number of voters with the rank of the parties. We find this relationship in proportional elections in Germany since 1949, the first round of presidential elections in France (2005, 2012 and 2017) and the Republican primaries in the USA (2016)(Figure 1a,b and SI Figures). Such a repeated pattern, at several spatial scale (for Germany and France results over large towns, regional unit or whole country) hints to an underlying dynamical process that seems to be incompatible with our idea of election as a rational decision of mature citizens. In particular in view of recent developments in western democracies it is necessary to discuss possible explanations, and to find an appropriate interpretation of these observations.

Much academic work and media analyses are done to predict and better understand the results of elections, that is the way voters decide rationally or irrationally, and how parties position themselves to maximize their success (19–21). Most models in this field focus on the weight of political opinion of voters and the targets (demographic, socio-economic groups) of parties. The present paper is based on an alternative argument, namely a neutral dynamics, for which the content of the opinion (e.g. left- or right wing, liberal or conservative, …) does not play a role. There exist few examples for this kind of dynamics such as the voter model (22) in which voters copy the opinion of randomly chosen neighboring voters. The noisy voter model (5, 23) or Sznajd models (15, 24) are also popular versions of (nonlinear) voter models. In a similar spirit to the present paper, Fortunato and Castellano (3) are able to explain a unimodal distribution appearing in some election data (Italy, Poland, and Finland) by a linear branching. However, the key is to propose an underlying mechanism generating such pattern. We are only aware of the approach based on Sznajd models on graphs, see e.g. Gonzales et al. (15), as an attempt to reproduce the occurrence of the Zipf's law. However, the assumption of locality as the driving force in opinion formation seems rather unrealistic to explain common results in different countries (see more detail in discussion).

In order to give a more persuading explanation of these empirical findings, we develop a model based on a stochastic process to capture the outcome of voters' decision. The model resembles the famous voter model (22, 23) but is based on two well established models of population genetics: the infinite allele model (*e.g.* 25) and the Ewens sampling formula (25–27). In particular, the foundation and liquidation of groups (or parties) are an integral part of the model, which we therefore call "voter model with party dynamics". Despite the models relatively simplicity, the analysis of data in a quantitative manner was possible, in particular regarding predictions on data structure. We indeed recover the log-linear structure observed in the order statistics of election results. Based on this model we are able to identify and describe the cannibalism effect as a consequence of the threshold introduced in the proportional election system. Moreover, we suggest to infer the number of parties present in the parliament based on the the number of parties that stand for election. We defer all mathematical work to the supplementary information (SI), and focus below on the model approach and results in order to insist on the consequences and conclusions of our work for modern democracies.

### Motivation for the study: the elections in the FRG

We mainly consider data from the Federal Republic of Germany (FRG), and provide therefore a short sketch of the most important facts concerning that election system. Every voter has two votes, the 'first' and 'second' vote. With the first vote, a direct candidate is elected. In the present work, we exclusively address the more important second votes, by which an elector votes for a party. The parties that receive more than 5% of these votes send representatives into the parliament and the number of mandates of a party is proportional to the number of second votes.

Data for these elections are shown in Figure 1a, b, table 1, and SI section 4 . The parties are ranked according to the number of votes, and the logarithm of votes depends linearly on the rank of a party (the fit to a linear correlation is measured by the $R^2$, Figure 1c and SI Figure 5). Strikingly, the $R^2$ values observed in elections in Germany, the USA republican primaries , and France are almost always above 0.9. This log-linear structure is conserved over several years and over magnitudes of organizational units (cities, states, country). In Germany, even the demographic change and societal changes due to the German reunification did not apparently affect this data structure. The only case for which this linear relation is only weakly observed is the election in the Federal Republic of Germany in 1949. This election was the first one after the Second World War, and the elective system was different as each citizen has had only one vote, thereby possibly explaining the weak correlation.

This structure in the data, namely the strong log-linear correlation between votes and the rank, is rather unexpected and is so far unexplained in the literature. When inspecting the log-linear relation over organizational units of different size (cities, states, country), we find the slope to be relatively similar and only the intercept depends on the size of the considered unit (table 1). Obviously, the mechanism creating this structure scales with the size of the population. This scale invariance in the population size is a strong requirement for the model we introduce next.

### Model overview and analytical results

We consider a population of voters. Each voter supports a party, meaning he/she plans to vote for this party (see also SI, Figure 1). We aim to model the dynamics of the affiliation of voters to party and the time evolution of the number of parties. The dynamics is not based on political aims and competition of ideas, but driven by the word of mouth according to the following rules: a voter either just copies the opinion of another voter, or creates a new party. At each time step, a randomly chosen person rethinks his/her opinion. This person may decide to keep his/her opinion with probability $v$, and if not there are two possibilities: 1) with a given small probability $(1-v)u$, he/she can found a new party, or 2) he/she selects randomly one individual in the population, and supports the party of that individual (with probability $(1-v)(1-u)$). When the last supporter of a party leaves, that party is eliminated. In this way we obtain a voter model that allows for the formation/destruction (or birth and death) of parties. We repeat the process until an equilibrium in the party structure is reached, that is when the number of parties reaches a stable value. This model is thus a birth and death process analogous to the well-known infinite allele Moran model in population genetics. The distribution of party size and number is described by the Ewens

Sampling formula (25, 26).

In our model, the equilibrium solution, i.e., the invariant measure of this stochastic process, is not directly the result obtained at the elections. We find that many parties with only very few supporters appear, and this situation is not realistic. Small groups are less likely to bear the effort required to be a party in country-wide elections, so that parties with a subcritical number of supporters would not stand for election. The relative critical size $z$ (minimal size possible/population size) is introduced as the last parameter of the model. As a consequence, the model behavior scales with the population size.

As the infinite allele Moran model and the Ewens sampling formula are well understood, our analysis (see SI section 1 for an overview, and SI section 2 for the demonstrations) can be based on those results. In particular, we can condition on the number of parties that stand for election K (SI, sections 2.1 and 2.2). It turns out that the invariant measure is in this case determined by three parameters: the number of voters $n$, the running number of parties $K$, and the relative critical party size $z$. Moreover, it is straight forward to characterize the invariant measure of the model. Determine K identical and independent realizations $X_1,...X_K$ of random variables with

$$P(X = j) = \frac{c}{j} \text{ for } zn \leq j \leq n \text{ and } P(X = j) = 0 \text{ else, where } c^{-1} = \sum_{j=zn}^{n} \frac{1}{j}$$

In order to address the log-linear structure, we order the realizations, $X_{(1)} \leq X_{(2)} \leq ... \leq X_{(K)}$. We demonstrate that the model satisfies a variant of Zipf's law, namely (SI, section 2.3)

$$\lim_{n \to \infty} E(X_{(i+1)} / X_{(i)}) = \int_z^1 \left(\frac{1}{z} - \frac{1}{x}\right)\left(\frac{\ln(1/x)}{\ln(1/z)}\right)^{\ln(z)\Theta - 1} dx$$

where $\Theta = K / \ln(1/z)$. Note that this expression parallels the classical Watterson estimator for the rescaled mutation rate $\Theta$ (28, Section 2.2]. The right hand side of the equation is independent of rank $i$. That is, the quotients of subsequent group sizes are constant in expectation. This observation is strongly related to the Zipf's law; as without stochasticity, the logarithm depends linearly on the rank if and only if the quotient of subsequent groups is constant. If we consider the logarithm instead, we find no exact linear relation:

$$\lim_{n \to \infty} E(\ln(X_{(i)})) - \ln(n) = -\sum_{j=0}^{i-1} \int_z^1 \frac{1}{x}\left(1 - \frac{\ln(1/x)}{\ln(1/z)}\right)^{j-1}\left(\frac{\ln(1/x)}{\ln(1/z)}\right)^{K-j}\left(K - j - K\frac{\ln(1/x)}{\ln(1/z)}\right)dx.$$

However, simulations indicate that the dependency of $E(\ln(X_{(i)}))$ is approximately linear with the rank $i$.

In order to apply the model to data, we refine our assumptions. In elections the number of voters is known. We include this assumption in our model by conditioning the realizations $X_1,..., X_K$ on the known total number of voters, $\sum_{j=1}^{K} X_j = n$. As it is shown in SI, section 2.4, neither is the expected quotient of subsequent groups constant, nor does the expected logarithm of the group sizes depends linearly on the rank. Nevertheless, simulations indicate that the logarithmic data depend almost linearly on the rank.

## Results of the model fit to data

We compare data of several elections with our model and find surprisingly a very good agreement (Figure 1a, b, and SI section 5 with data of elections and boxplots of 100 realizations of our model). Two parameters ($K, n$) can be readily read off the data. The third parameter, namely the relative critical party size, does only weakly depend on the number of parties/the number of voters (SI, section 3). For the investigation of the slope, we fix this parameter to 0.3 per mille of the total number of voters. Particularly, it becomes of importance below that the slope of the log-linear relation is well met by the model over all elections in the FRG that we consider (Figure 1d). Since the voter number only has a weak effect on the slope, and the relative minimal party size $z$ is fixed, the slope is well predicted by the number of parties present in an election. We use model-based data analysis to infer more detailed information about voting behavior and the electoral system.

## Results and direct implications: the cannibalism effect

We inspect the residuals of the linear model (Figure 2c), by plotting the data of different elections against the linear regression, using the 5% threshold as a reference point. We find that the number of voters for parties above the 5% threshold is systematically underestimated, while votes directly below the threshold are overestimated (Figure 2c). The voter numbers for small parties are basically met, even perhaps slightly underestimated (for very small parties).

We suggest that this deviation from the correlation is due to strategic thinking by the voters. In fact, the voters who originally would want to vote for parties just below the threshold may expect that their vote would eventually not count (as the party would not be present in the parliament). Therefore, these voters change their mind and vote for a party above the threshold. The bigger parties thus cannibalize the slightly smaller ones. On the end of the distribution, the voters of very small parties may be driven by their political conviction only, and are not affected by the hypothetical presence of their party in the parliament. Though this systematic bias is doubtless present, and interesting in itself, it is only a second order effect that we do not address further here.

## Results and direct implications: number of parties in the parliament

We can predict the slope $s$ of the log-linear relation from the number of parties that stand for election $K$, and we know the total number of voters $n$. Consequently, we can compute the intercept $a$ of the log-linear relation and are able to predict the expected outcome of the election. Approximately, the size of the i'th group $x_i$ satisfies

$$x_i = 10^{a+si}, \quad \sum_{i=1}^{K} x_i = n \quad \Rightarrow \quad x_i = n \frac{10^s - 1}{10^{s(K+1)} - 1} 10^{si}.$$

Namely, we can compute the number of parties that are present in the parliament as a result of the elective process (that is parties reaching the 5% threshold, $\#\{i \mid x_i / n > 0.05\}$). Recall that the slope of the log-linear regression mainly depends on the number of parties that stand for election, so that we can draw the expected number of parties in the parliament (Figure 2b).

Our predictions can be compared with the election data (Figure 2a). We find a good agreement, with a tendency towards overestimation. This overestimation can be corrected by taking the cannibalism effect into account. The cannibalism effect increases the difficulty of a party to cross the 5% threshold, and consequently reduces the number of parties able to enter the parliament. The number of parties present in the parliament cannot grow to arbitrary high numbers, but we expect that at the present time we already reached this maximum in Germany.

**Discussion**

We start this study by highlighting the statistical structure of electoral data in Germany but also recent elections in the USA (Republican primaries) and presidential elections in France when more than 10 parties or candidates are present. This motivates us to build a stochastic, mechanistic model that yields the observed structure of the log-linear regression between number of voters and rank of the party. The model has some intuitive ingredients, as the importance of the word-of-mouth in opinion dynamics.

In effect, each individual chooses initially a group, and the dynamics starts, i.e. voters can change their mind and move to another group or found a new one; groups vanish if the last member leaves. Groups can be parties or candidates (for the US and French elections). The groups do not exhibit in our model any specific characteristic (ideas, socio-economic policies, sociological identity,…), and therefore the moving from one group to another occurs at random (with a given probability). Our model is thus a "neutral" model of voter choice analogous to population genetics model of neutral alleles which frequency varies as a function of genetic drift in a population. The predictions of our model are in line with the statistical features of the data, and we identify the cannibalism effect that promotes large parties on the costs of smaller ones. Finally we provide a relation between the expected number of parties entering the parliament and the running number of parties.

In recent years in the FRG, the large big-tent parties steadily loose votes, and more parties enter the parliament. In view of our analysis, we identify the reason for this effect as the steady increase of running and founded parties since approximately 1975 (Figure 2d). We suggest that the marginalization of big tent parties is the consequence of a long-term process, and not only a recent trend. The increase in the total number of parties is due to a fragmentation of the political spectrum into specialized parties, focusing on specific issues (e.g. environmental problems, women equality, elderly people place in the society, development of internet and information technology, promoting animal welfare,…). As these parties persist (often on a low level of votes, though), this yields the decrease in voters of large parties. Thus, the marginalization of big-tent parties is not necessarily based on a societal shift towards political extremes.

We now interpret further the results of this study. On the first glance, our findings seem to contradict our understanding of elections, whereby rational and responsible citizens think about actual and future tasks and challenges of the society. Our model presents a voter as an individual primarily driven by social interactions and influenced mainly by the opinions of his/her neighborhood, and not by rational considerations about the direction the society should take in the next years. Models of this type (voter model, Sznajd model) have been proposed before. In these papers (see e.g. Gonzalez et al. (15)) local dynamics on a graph, where individuals interact with neighbors only were considered, also yielding the Zipf law. It is questionable whether in the current age of mass media and social networks these assumptions are realistic. Moreover, results from spatially structured data analysis indicate a simple logarithmic dependence of the correlation on the distance are in agreement with two-dimensional diffusion processes (6) which again hints that it is not the microscopic scale that drives the dynamics. A new aspect in the present study is the insight that the party (birth and death) dynamics and the truncation of small groups play a central role. This view doubtless capture one but not all important aspects in the dynamics of opinions. With the advent of social media, we predict that the word of mouth becomes important and generalized, and the opportunity for voters to change their mind (the probability *v*) becomes larger. The results of future elections and voter distributions can be studied using our model to test this hypothesis. Particularly, the birth/death process of parties is expected to accelerate.

In present times, democracy is criticized, and alternatives are proposed from an adaptation of the electoral process, e.g. by incorporation of random elements and chance (29), up to a fundamental change of democracy to epistocracy (30). Proponents of these ideas could be tempted, in order to support their point of view, to use our observation that elections have an inherited structure which is not in line with a rational decision process. However, we propose three other interpretations which could be more appropriate and fruitful.

First, the ultimate decision process, and the shaping of governments and ruling majorities, takes place in the competition of large parties for voters. Most part of the data discussed here is concerned with smaller parties that only represent a relatively small part of the population (note that we take the logarithm of the voter numbers). This effect is best observed in the data of the presidential elections in France 2017 (see SI, section 4). Basically four candidates, representing drastically different political ideas, did compete and attracted fairly similar amount of votes. Nevertheless, when including all candidates' results, the data show the log-linear structure discussed in the present paper. In the French presidential election, there is also a major cannibalism effect, because only the first two candidates qualify for the second round, which distorts the vote distribution as seen around the 5% threshold in the FRG results.

Second, the model opens the door for problem driven decision processes. A central, new aspect in our model is the dynamics of party creation/destruction. This process is formulated in a neutral way, as parties are formed at a certain probability, and destroyed, if – by chance – the last supporter leaves. However, when inspecting the political aims of new parties, it appears often that their focus is driven by actual political issues. In other words, elections and the parliamentary system are crucially influenced by the question of whether new topics are handled adequately by big-tent parties. If not, the citizens may feel that new parties and movements are necessary to handle them. Our results indicate that new, even small, parties are able to influence the complete political landscape and are important for the distribution of votes.

Third, our model cannot disentangle between two components of the switch probability *v*: 1) the decision of voters *per se*, or 2) the adequacy between voters' need and party programs. In other words, our results can indicate some elements of randomness in the way that voters may switch their opinion and/or that parties change policies between elections. The feedback loop between voters need and priorities and parties' policies being based upon a complex multi-factorial set of ideas filtered by rationale and irrational decisions (at the individual and at the party level), which may generate the neutral distribution observed in our results. This point of view is in agreement with *e.g.* (18).

The aim of considerations as the present one is not a complete description of the democratic process, but the insight into some basic mechanisms acting in the system, such as the rationale or randomness of voters. We believe that our results reveal fundamental trends such as the steady increase of the number of political parties, and the cannibalism effect that influences the political geometry. The democratic system, which in essence should be based on equal opportunities and individuals' participation to the society rather than only rely on elections, is here not questioned.

**Acknowledgment**: We thank Faidra Stavropoulou for intensive discussions that did substantially improve the paper.

**Table**

Table 1: Unit of organization (city, state, or country), number of parties, number of active voters, and parameter of the fit of the linear model for the FRG election in the year 2013.

| unit | #parties | #voters | intercept | slope | $R^2_{adj}$ |
|---|---|---|---|---|---|
| Stuttgart | 20 | 284541 | 1.651 | 0.156 | 0.9438 |
| Munich | 20 | 650216 | 1.933 | 0.165 | 0.9889 |
| Baden Wuerttemberg | 20 | 5642019 | 2.953 | 0.156 | 0.9759 |
| Bavaria | 20 | 6580755 | 3.009 | 0.161 | 0.9826 |
| Fed. Rep. Germany | 30 | 43726856 | 3.146 | 0.121 | 0.9473 |

**Figure Legends**

Figure 1: Occurrence of the Zipf's law in FRG election results and model fitting. (a)-(b) The logarithm of the number of voters over the rank of a party (data as filled circles) for organizational units of different size, together with a linear fit (solid line), indication of the 5% threshold (horizontal, dashed line) and results of 100 runs of the stochastic model (boxplot). Filled circles are for whole of FRG in (a) for the 2013 elections and (b) for 2017. (c) $R^2$ for the linear fit of the data combined for cities (Munich, Stuttgart), regional states (Bavaria, Baden Württemberg), and FRG per election year. (d) The slope of the linear fit for the log-rank data over the number of parties, together with the model prediction (solid line). The data of the election in 1949 are indicated by +, the others by circles.

Figure 2: Test of our model prediction and inference using all elections in FRG. (a), (b) number of parties in the parliament over number of parties in the election (solid lines for model prediction, circles for data) (c) Boxplot of residuals. Data are adapted such that the 5% threshold is always located between rank 28 and 29 (dashed vertical line). (d) Number of parties at the level of the FRG that obtained second votes per election.

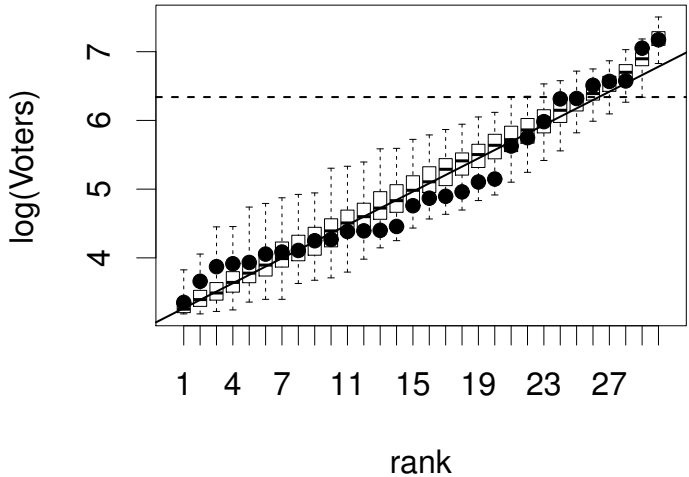 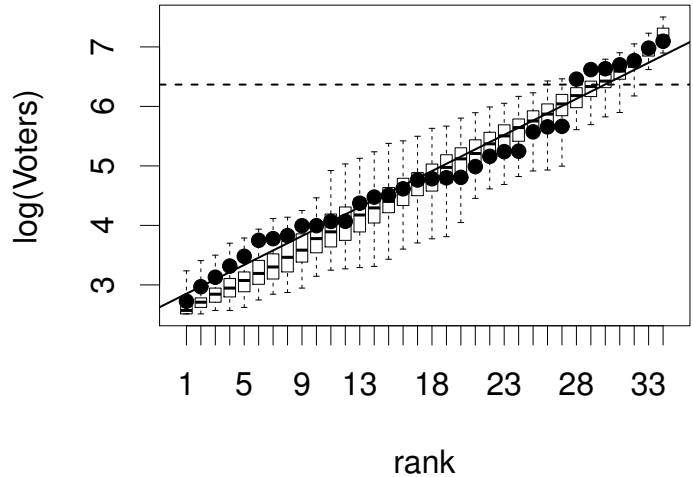
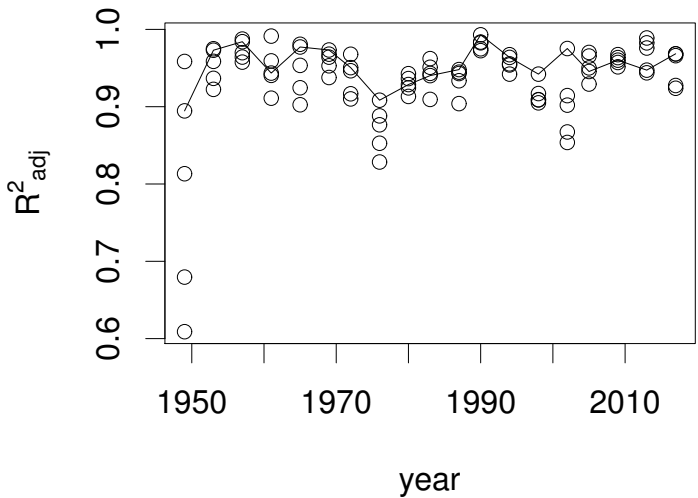 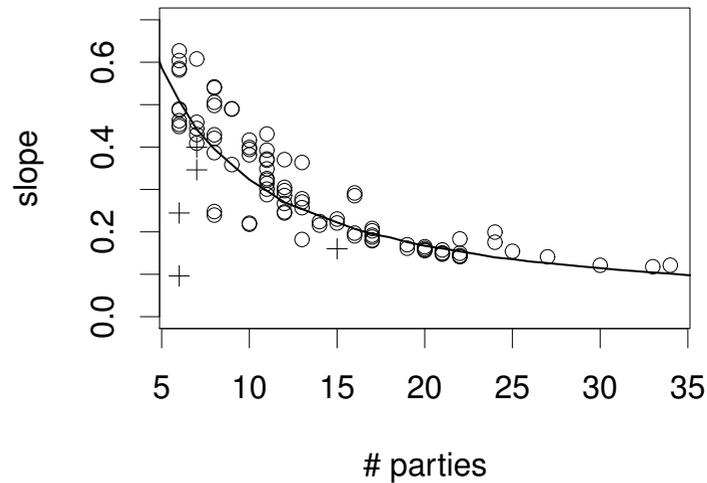

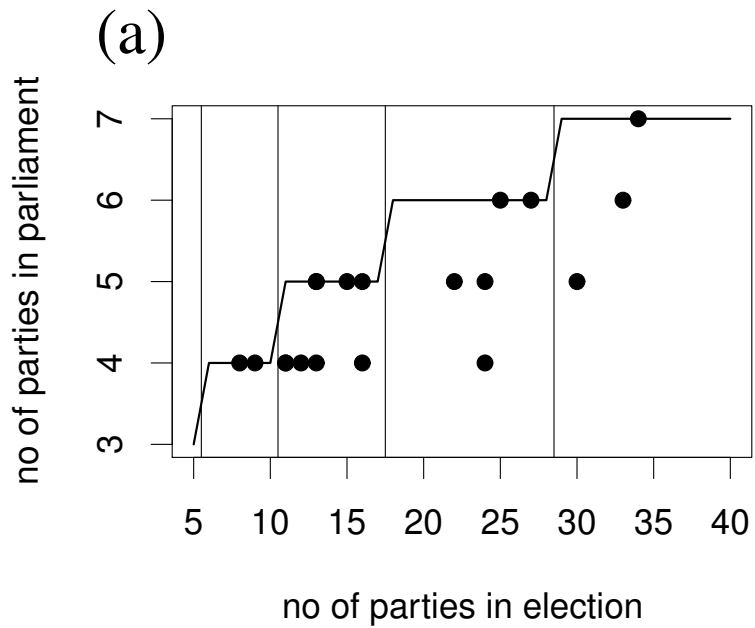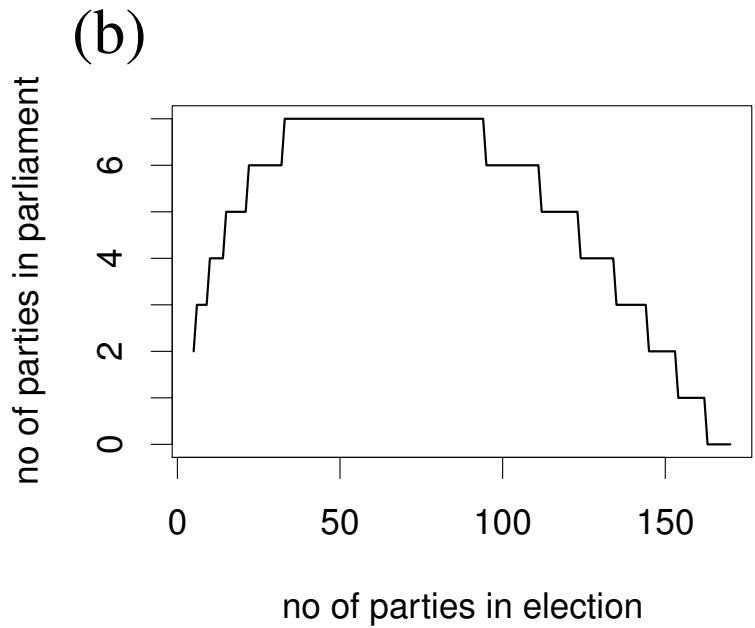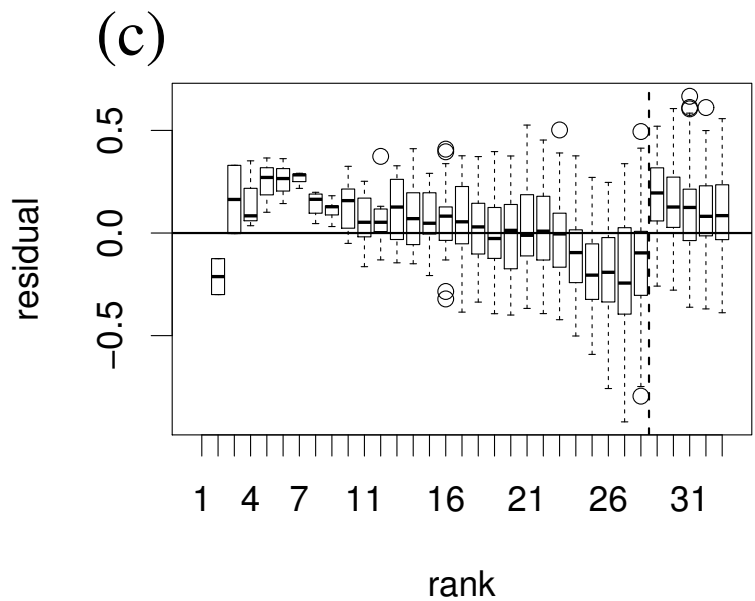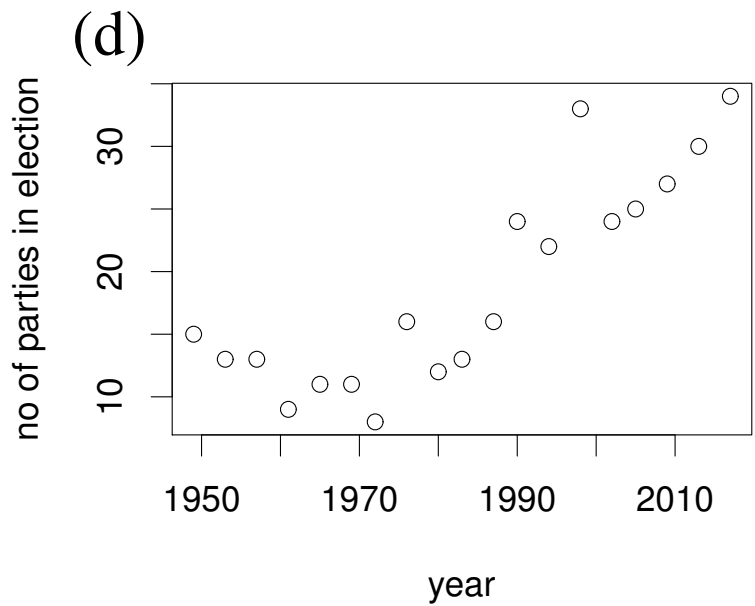

# Supplementary material: Mathematical analysis and data

Hösel, Tellier, Müller

March 17, 2018

## 1 Overview over the mathematical analysis

### 1.1 Data driven modeling

We summarize the considerations made in the main text. The starting point of our considerations is the observation that the number of votes per party satisfies approximately a log-linear dependency on the party's rank. The slope of this log-linear relation seems to be rather independent of the number of voters: If we inspect the parameters of the linear model, we only find the intercept to vary with the size of the group. The slope is fairly constant. That is, we expect that the mechanism is the same on different levels of organization (city, state, country).

The decision process leads to group formation – we identify the voters of a given party with one group. This process inherits, of course, stochasticity. That is, if we consider any partition of the population, the resulting decomposition has a certain probability to meet the "true" decomposition, observed in elections. As the data indicate that the slope is basically independent on the size of the organizational unit, we expect that the fundamental ruling mechanisms are fairly independent of the population size (number of voters). Even more, if we first take a subset of the complete population (the voters in Bavaria, say), and the a subset of the subset (the voters in Munich), yields the same result as if we directly take the small sample (Munich). In statistics, this property is known as sampling consistency condition [7, 6]. The probability measures that exhibit the sampling consistency condition are well characterized: they can be constructed by the so-called paint ball process [7]. One distribution out of those is especially famous: the Ewens sampling distribution [4, 3]. This distribution has many applications, in particular in population genetics [2], and moreover, can be generated by a stochastic process – the infinite allele Moran model with mutations.

We adapt the infinite allele model in the context of opinion formation of voters (see Figure 1). Let us consider a population of $n$ voters – non-voters are neglected. Each voter is a supporter of a proto-party. The difference between a party and a proto-party becomes clear below, by now we may identify the two terms. Voters change their opinion in the following way: a randomly selected voter thinks over his/her opinion. With a certain probability $v$, he/she stays with his/her opinion. With probability $1-v$ the person is prepared to change the proto-party he/she is supporting. If this is the case, this person either constitutes a new proto-party with probability $u$, or selects randomly one person of the population and adopts the opinion of that individual. If the last supporting voter of a given proto-party changes his/her mind such that this proto-party has no supporters any more, this proto-party is dissolved. In particular, no political aims or believes are involved in this model – it is a model neutral.

In the long run, the stochastic process approaches its invariant measure. There is some kind of equilibrium for the number of proto-parties (a distribution), and each individual is the follower of a proto-party. In an election, however, proto-parties that are too small will not stand for election. Only proto-parties with a supercritical number of followers become parties. It turns out later that the relative critical size $z = n_0/n$ is even more handy parameter. Followers of a tiny proto-party give their vote to another party, where a party is selected proportional to its size. We call this model the voter model with party-dynamics.

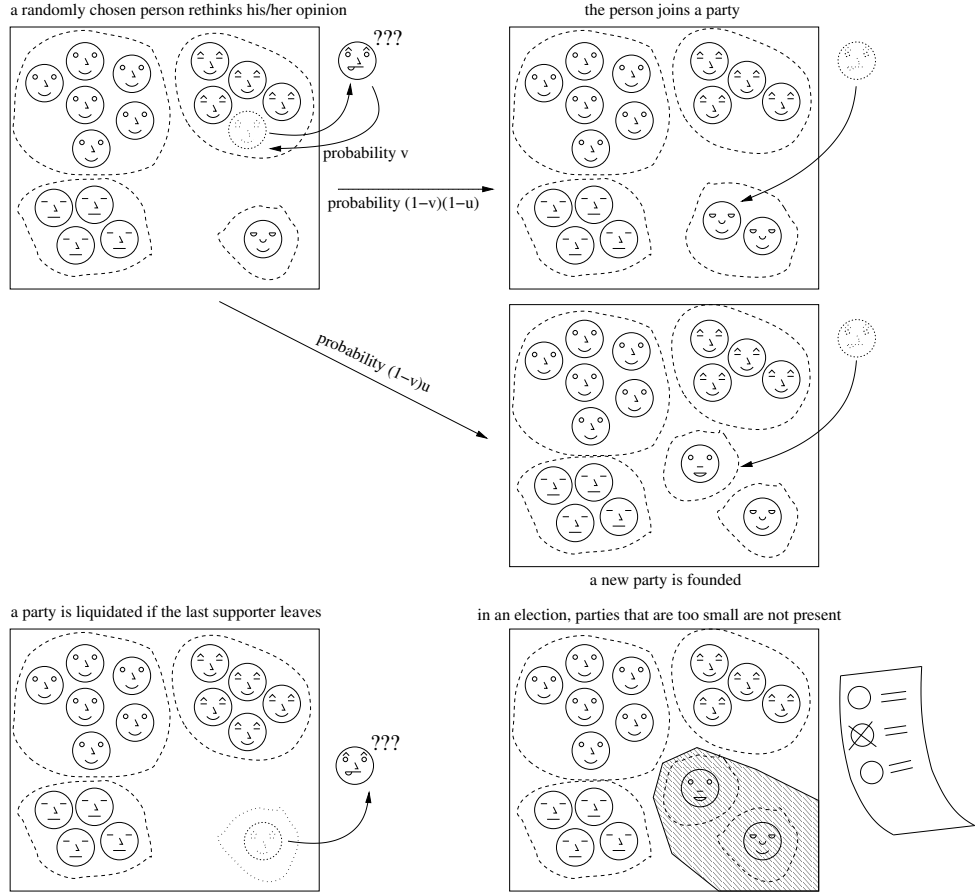

Figure 1: Scheme of the model.

The voter model with party dynamics and $n_0 = 0$ is equivalent with an infinite-allele Moran model with mutation. It is well known that the resulting composition structure follows the Ewens sampling formula, which has only one parameter $\theta = 2\,u\,n/(1-v)$, the rescaled party-constitution rate (or the rescaled mutation rate in the context of population genetics). Our model has one additional parameter: the threshold $z \in (0,1)$. All individuals in groups with a size smaller $n_0 = z\,n$ are distributed to the groups with size larger or equal $z\,n$.

## 2  Invariant measure of the model, Ewens Sampling Formula and Zipf's law

### 2.1  The Ewens Sampling Formula

We recall some basic facts about the Ewens Sampling foprmula. The Ewens Sampling Formula is a probability measure that describes partitions of a set (population) of size $n$. The Ewens Sampling

Formula is appealing in that this probability measure is generated by the invariant measure of the infinite allele model, that is, of a mechanistic and interpretable process. If $z_0 = 0$, this process is identical with our voter model with party dynamics.

Let us first consider the original Ewens Sampling Formula. In a sample of size $n$ (given), we observe $K$ groups ($K$ is a random variable). The sizes are given by $a_1, \ldots, a_K$. Let $c_1$ denote the number of groups of size 1, $c_2$ the number of groups of size 2, etc. The partition can be characterized by the vector of frequencies $C = (c_1, \ldots c_n)$, where $n = \sum_{i=1}^n i\, c_i$ and $K = \sum_{i=1}^n c_i$. The Ewens Sampling Formula states the probability to observe a given partition structure $C$. Mostly, it is parametrized by a scaled mutation probability $\theta$, and can be formulated as (e.g. [3, page 22])

$$C \sim (Y_1, \ldots Y_n | \sum i Y_i = n) \tag{1}$$

where $Y_i \sim \text{Pois}(\theta/i)$ are independent Poisson random variables.

In our application it will be interesting to condition the distribution on the number of groups [3, chapter 3.1], [10], [5, chapter 9.5]. Note that the number of groups $K = K(C) = \sum_i c_i$ is a random variable. Then,

$$P(K = k | \theta) = \frac{\theta^k}{\theta_{(n)}} |S_n^k| \tag{2}$$

where $S_n^k$ are the number of permutations of $\{1, \ldots, n\}$ with exactly $k$ cycles. The Ewens Sampling Formula reads ($k = \sum c_j$ in the given realization)

$$P(C|\theta) = \frac{n!}{\theta_{(n)}} \prod_{j=1}^n \frac{(\theta/j)^{c_j}}{c_j!} = \frac{n!\, \theta^k}{\theta_{(n)}} \prod_{j=1}^n \frac{(1/j)^{c_j}}{c_j!}.$$

Therefore,

$$P(C|\theta, K = k) = \frac{P(c_1, \ldots, c_n \text{ and } K = k|\theta)}{P(K = k|\theta)} = \frac{P(c_1, \ldots, c_n|\theta)}{P(K = k|\theta)} = \frac{n!}{|S_n^k|} \prod_{j=1}^n \frac{(1/j)^{c_j}}{c_j!}.$$

In a slight abuse of notation, we define for $a_i \geq 0$, $\sum a_i > 0$

$$\text{Multinom}(n, (a_1, \ldots, a_m))$$

to denote the multinomial distribution with $n$ trials, and $p_i = a_i / \sum_{j=1}^m a_j$. If we compare the result above with the probability function of a multinomial distribution, we conclude

$$(C|K = k) \sim \text{Multinom}(k, (1, 1/2, \ldots, 1/n)) \ \Big|\ \sum_{i=1}^n i c_i = n. \tag{3}$$

## 2.2 Invariant measure of the model

In our model we know the number of parties before the election takes place. That is, the natural parametrization of our model uses the minimal group size possible $n_0 = \lfloor z\, n \rfloor$, the population size $n$, and the number of observed groups $K_+$ (that is, groups with a supercritical size, $a_i \geq \lfloor z\, n \rfloor$). The aim of the present section is to show that, conditioned on these parameters, the invariant measure of our model can be well approximately by a multinomial distribution.

In our case, we are only interested in groups with size above the critical group size $n_0$. For the random variable $C = (c_1, \ldots, c_n)$ define

$$K_+ = K_+(C) = \sum_{i=n_0}^{n} c_i, \quad N_+ = N_+(C) = \sum_{i=n_0}^{n} i\, c_i, \quad K_- = \sum_{i=1}^{n_0-1} c_i = K - K_+, \quad N_+ = \sum_{i=1}^{n_0-1} i\, c_i = n - N_+.$$

Furthermore, we define the projections

$$\Pi_+(c_1, \ldots, c_n) = (0, \ldots, 0, c_{n_0}, \ldots, c_n), \quad \Pi_-(c_1, \ldots, c_n) = (c_1, \ldots, c_{n_0-1}, 0 \ldots, 0).$$

Our ultimate goal is an approximation of the distribution of $\Pi_+(C)|\theta, K_+$. If we do not only condition on $K$ and $K_+$, but also on $N_+$, then

$$(\Pi_+(C)|K = k, \ K_+ = k_+, \ N_+ = n_+) \sim \mathrm{Multinom}\left(k_+, (0, \ldots, 0, 1/n_0, \ldots, 1/n)\right) \ \bigg|\ \sum_{i=n_0}^{n} i c_i = n_+.$$

Note that the r.h.s. does not depend on $K$ any more. Thus, we may drop the condition on $K$,

$$(\Pi_+(C)|K_+ = k_+, \ N_+ = n_+) \sim \mathrm{Multinom}\left(k_+, (0, \ldots, 0, 1/n_0, \ldots, 1/n)\right) \ \bigg|\ \sum_{i=n_0}^{n} i c_i = n_+.$$

If $K_+ = k_+$ is fixed, we have a hierarchical model: We obtain a conditioned multinomial distribution for $\Pi_+(C)|\theta, K_+ = k_+, \ N_+ = n_+$, where $\theta$ does not appear directly any more; however, the distribution of $N_+$ depends on $\theta$. That is, $n_+$ can be considered as a hyperparameter with a given ($\theta$-dependent) distribution. In that, the situation is somewhat similar to that before: if we observe all groups, the distribution of $C|\theta, K$ only depends on $\theta$ via the distribution of $K$. Also (3) can be interpreted as an hierarchical model with hyperparameter $K$, which has the distribution indicated in (2).

In order to infer the distribution of $N_+$, we note $N_+ = n - N_-$. Next we use (see [3, Theorem 1.19], [1]) that for $n$ large, approximately $c_i \sim Y_i$, where $Y_i \sim \mathrm{Pois}(\theta/i)$ are independent random variables. Note that in this step some heuristics are involved, as this theorem is only true for $i$ fixed and $n \to \infty$. For $n$ large we have approximately

$$N_- \sim \sum_{i=1}^{n_0-1} Y_i \sim \mathrm{Pois}\Big(\sum_{i=1}^{n_0-1} \theta/i\Big).$$

Note that this Poisson random variable may attend arbitrary large numbers, such that in this approximation $N_+ \leq n$ is not always given. However,

$$E(N_-) \approx \sum_{i=1}^{n_0-1} \theta/i \approx \theta \log(n_0).$$

This argument indicates that the hyperparameter $N_+$ can be represented as $n - N_-$, where $N_-$ has a Poisson distribution with expectation $\mathcal{O}(\log(n))$. Thus, for $n$ large, $N_+ \approx n$, and we find in this sense

$$(\Pi_+(C)|\theta, \ K_+ = k_+, \ N_+ = n) \approx \mathrm{Multinom}(k_+, (0, \ldots, 0, 1/n_0, \ldots, 1/n)) \ \bigg|\ \sum_{i=n_0}^{n} i c_i = n_+. \qquad (4)$$

This statistics can be reformulated in a handy way if the underlying urn model for a multinomial distribution is considered. We return to the group sizes $(a_1, \ldots, a_{k_+})$; recall that the sample configuration $C = (c_{n_0}, \ldots, c_n)$ is computed from a realization of $(a_1, \ldots, a_{k_+})$. We define i.i.d. random variables $X_\ell$, $\ell \in \{1, \ldots, k_+\}$, with values in $\{n_0, \ldots, n\}$ where

$$P(X_\ell = i) = c/i, \tag{5}$$

and, as before,

$$c^{-1} = \sum_{i=n_0}^{n} i^{-1} \approx \ln((n_0)/(n+1)) \approx \ln(1/z), \tag{6}$$

then

$$(a_1, \ldots a_{k_+}) | \, N_+ = n \ \sim \ (X_1, \ldots, X_{k_+}) \ \Big| \ \sum_{\ell=1}^{k_+} X_\ell = n, \tag{7}$$

The approximate distribution of configurations $(\Pi_+(C)|\theta, K_+ = k_+, N_+ = n)$ given in (4) is identical with the distributions of configurations generated by $(a_1, \ldots a_{k_+})$ defined in (7). The invariant measure of our dynamic model, the voter model with party dynamics, is well described by the construction (7).

There is a direct relation of this model to Zipf's law: We draw a histogram of many realizations $(X_1, \ldots, X_{k_+})$ while pooling these random variables. For $n$ and $k_+$ large, $n \gg k_+$, this histogram of many realizations follows approximately the distribution of $X_\ell$. The histogram follows a power law (with exponent 1) in the relevant region between $n_0$ and $n$, which is called Zipf's law [9] in the present context.

However, we are not interested in the structure of many realizations, but in that of one realization $(X_1, \ldots, X_{k_+})$. We claim, that also for one realization (with $n \gg k_+$, $n \gg n_0$, and $k_+$ large) we find again a power law (with different exponent) for the rank data. We approach this question in the next sections.

## 2.3 Rank statistics – unconditioned case

In this section, we derive an expression for the expectation of the size ratio of subsequent groups (ordered according to their rank), respectively about the expectation of the logarithm of the group size. Here we do not condition on the total population size. That is, we consider $K$ independent realizations $X_1, \ldots, X_K$ of i.i.d. random variables that assume values in $\{n_0, \ldots, n\}$, where $0 < n_0 < n$, $P(X_i = j) = c/j$ for $j \in \{n_0, \ldots, n\}$ and 0 else, $c^{-1} = \sum_{j=n_0}^{n} j^{-1}$. We order these realizations according to size $X_{(1)} \leq X_{(2)} \ldots \leq X_{(K)}$, and investigate $E(X_{(\ell+1)}/X_{(\ell)})$ respectively $E(\ln(X_{(\ell+1)}))$. The central result of this section is the independence of that expectation w.r.t. $\ell$ for $n$ large:

**Theorem 2.1** *Let $\theta = K/\ln(1/z)$. For $\ell \in \{1, \ldots, K-1\}$ we find*

$$\lim_{n \to \infty} E(X_{(\ell+1)}/X_{(\ell)}) \ = \ G(\theta, z) := 1 + \theta \int_z^1 \left(\frac{1}{z} - \frac{1}{y}\right) \left(\frac{\ln(1/y)}{\ln(1/z)}\right)^{\ln(1/z)\,\theta - 1} dy. \tag{8}$$

That is, the expectation of the quotient of subsequent group sizes is constant in the rank. The next proposition indicates that $E(X_{(\ell+1)}/X_{(\ell)})$ only depends weakly on $z$ (for $z$ small) if $\theta$ is larger one and kept fixed.

**Proposition 2.2** *We find for $\theta > 1$ that*

$$\lim_{z \to 0} G(\theta, z) \ = \ \frac{\theta}{\theta - 1}. \tag{9}$$

For the logarithm, we obtain the following theorem.

**Theorem 2.3** *For $\ell = 1, \cdots, K$, we find*

$$\lim_{n \to \infty} E(\ln(X_{(\ell+1)}) - \log(n))$$
$$= -\sum_{j=0}^{\ell-1} \binom{K}{j} \int_z^1 \frac{1}{x} \left(1 - \frac{\ln(1/x)}{\ln(1/z)}\right)^{j-1} \left(\frac{\ln(1/x)}{\ln(1/z)}\right)^{K-j} \left(K - j - K\frac{\ln(1/x)}{\ln(1/z)}\right) \, dx. \quad (10)$$

This theorem indicates that the logarithm of the population size does not exactly depends in a linear way on the rank $\ell$, but only approximatively.

We prove the theorems resp. proposition in the next sections.

### 2.3.1 Order statistics

As a first step we obtain $P(X_{(\ell)} = i)$, based on the well known formulas for the distribution functions of order statistics.

In the following, we use the convention that a sum extending from $a$ to $b$ with $a > b$ is zero, in particular

$$\sum_{i=n+1}^{n} (\cdots) := 0.$$

**Proposition 2.4**

$$P(X_{(1)} = i_1) = \left(\sum_{j=i_1}^{n} \frac{c}{j}\right)^K - \left(\sum_{j=i_1+1}^{n} \frac{c}{j}\right)^K$$

$$P(X_{(\ell)} = i_1) = \frac{c^K}{i_1} \sum_{j=0}^{\ell-1} \binom{K}{j} \left\{ \left(\sum_{w=n_0}^{i_1-1} \frac{1}{w}\right)^{j-1} \left(\sum_{m=i_1+1}^{n} \frac{1}{m}\right)^{K-j-1} \left[K \left(\sum_{w=n_0}^{i_1-1} \frac{1}{w}\right) - jc^{-1}\right] \right\} + \mathcal{O}(i_1^{-2}).$$

**Proof:** Since $P(X_\ell \geq i_1) = \sum_{i=i_1}^{n} c/i$ for $\ell = 1, \ldots, K$, we have $P(X_{(1)} \geq i_1) = \left(\sum_{j=i_1}^{n} \frac{c}{j}\right)^K$ and

$$P(X_{(1)} = i_1) = P(X_{(1)} \geq i_1) - P(X_{(1)} \geq i_1 + 1) = \left(\sum_{j=i_1}^{n} \frac{c}{j}\right)^K - \left(\sum_{j=i_1+1}^{n} \frac{c}{j}\right)^K$$

Furthermore, we find for the $\ell$'th order statistics $X_{(\ell)}$

$$P(X_{(\ell)} \geq i) = P(\text{at most } \ell - 1 \text{ realizations are smaller } i)$$
$$= \sum_{j=0}^{\ell-1} \binom{K}{j} P(X_1 < i)^j \, P(X_1 \geq i)^{K-j} = c^K \sum_{j=0}^{\ell-1} \binom{K}{j} \left(\sum_{w=n_0}^{i-1} \frac{1}{w}\right)^j \left(\sum_{m=i}^{n} \frac{1}{m}\right)^{K-j}.$$

Hence, by means of Taylor expansion we obtain

$$
\begin{aligned}
P(X_{(\ell)} = i_1) &= P(X_{(\ell)} \geq i_1) - P(X_{(\ell)} \geq i_1 + 1) \\
&= c^K \sum_{j=0}^{\ell-1} \binom{K}{j} \left\{ \left( \sum_{w=n_0}^{i_1-1} \frac{1}{w} \right)^j \left( \sum_{m=i_1}^{n} \frac{1}{m} \right)^{K-j} - \left( \sum_{w=n_0}^{i_1} \frac{1}{w} \right)^j \left( \sum_{m=i_1+1}^{n} \frac{1}{m} \right)^{K-j} \right\} \\
&= c^K \sum_{j=0}^{\ell-1} \binom{K}{j} \left\{ \left( \sum_{w=n_0}^{i_1-1} \frac{1}{w} \right)^j \left( \sum_{m=i_1+1}^{n} \frac{1}{m} + \frac{1}{i_1} \right)^{K-j} - \left( \sum_{w=n_0}^{i_1-1} \frac{1}{w} + \frac{1}{i_1} \right)^j \left( \sum_{m=i_1+1}^{n} \frac{1}{m} \right)^{K-j} \right\} \\
&= c^K \sum_{j=0}^{\ell-1} \binom{K}{j} \Bigg\{ \left( \sum_{w=n_0}^{i_1-1} \frac{1}{w} \right)^j \left[ \left( \sum_{m=i_1+1}^{n} \frac{1}{m} \right)^{K-j} + \frac{K-j}{i_1} \left( \sum_{m=i_1+1}^{n} \frac{1}{m} \right)^{K-j-1} \right] \\
&\quad - \left[ \left( \sum_{w=n_0}^{i_1-1} \frac{1}{w} \right)^j + \frac{j}{i_1} \left( \sum_{w=n_0}^{i_1-1} \frac{1}{w} \right)^{j-1} \right] \left( \sum_{m=i_1+1}^{n} \frac{1}{m} \right)^{K-j} \Bigg\} + \mathcal{O}(i_1^{-2}) \\
&= \frac{c^K}{i_1} \sum_{j=0}^{\ell-1} \binom{K}{j} \Bigg\{ \left( \sum_{w=n_0}^{i_1-1} \frac{1}{w} \right)^{j-1} \left( \sum_{m=i_1+1}^{n} \frac{1}{m} \right)^{K-j-1} \\
&\quad \times \left[ (K-j) \left( \sum_{w=n_0}^{i_1-1} \frac{1}{w} \right) - j \left( \sum_{m=i_1+1}^{n} \frac{1}{m} \right) \right] \Bigg\} + \mathcal{O}(i_1^{-2}) \\
&= \frac{c^K}{i_1} \sum_{j=0}^{\ell-1} \binom{K}{j} \Bigg\{ \left( \sum_{w=n_0}^{i_1-1} \frac{1}{w} \right)^{j-1} \left( \sum_{m=i_1+1}^{n} \frac{1}{m} \right)^{K-j-1} \left[ K \left( \sum_{w=n_0}^{i_1-1} \frac{1}{w} \right) - jc^{-1} \right] \Bigg\} + \mathcal{O}(i_1^{-2})
\end{aligned}
$$

$\square$

Therewith, we find

$$
\begin{aligned}
E(X_{(1)}) &= \sum_{i=n_0}^{n} i \left( \left( \sum_{j=i}^{n} \frac{c}{j} \right)^K - \left( \sum_{j=i+1}^{n} \frac{c}{j} \right)^K \right) \\
&= \left( \sum_{i=n_0}^{n} i \left( \sum_{j=i}^{n} \frac{c}{j} \right)^K \right) - \left( \sum_{i=n_0}^{n} (i+1) \left( \sum_{j=i+1}^{n} \frac{c}{j} \right)^K \right) + \left( \sum_{i=n_0}^{n} \left( \sum_{j=i+1}^{n} \frac{c}{j} \right)^K \right) \\
&= n_0 + \left( \sum_{i=n_0}^{n} \left( \sum_{j=i+1}^{n} \frac{c}{j} \right)^K \right) \quad (11)
\end{aligned}
$$

where we used that $\sum_{j=n_0}^{n} c/j = 1$ and $\sum_{j=n+1}^{n}(\ldots) = 0$.

Let us introduce some more notation. Denote by $X_{(\ell);n_0,n,K}$ the random variable as introduced above; the additional indices characterize all parameters of the random variable.

If we condition on $X_{(1)} = i_1$, then $X_{(2)}, \ldots, X_{(K)}$ is the order statistics of $K-1$ random variables with values in $i_1, \ldots, n$. That is, we obtain realizations $X_{(2)}, \ldots, X_{(K)}$ by determining $K-1$ realizations of random variables $Y_i$ with values in $i_1, \ldots, n$, where $P(Y_i = j) = \tilde{c}_{i_1,n}/j$. Here, as before, $\tilde{c}_{i_1,n}^{-1} = \sum_{j=i_1}^{n} j^{-1}$. Then,

$$(X_{(2)}, \ldots, X_{(K)}) | X_{(1)} = i_1 \sim (Y_{(1)}, \ldots, Y_{(K-1)}).$$

In particular,
$$X_{(2);n_0,n,K}|X_{(1);n_0,n,K} = i_1 \sim X_{(1);i_1,n,K-1}$$
and, similarly, for $\ell = 1, \ldots, K-1$,
$$X_{(\ell+1);n_0,n,K}|X_{(\ell);n_0,n,K} = i_1 \sim X_{(1);i_1,n,K-\ell} \tag{12}$$

### 2.3.2 Size ratio

Before we come to the point where we investigate $\lim_{n \to n} E(X_{(\ell+1)}/X_{(\ell)})$, we first indicate two algebraic relations. All computations below are not deep, but lengthy and involving.

**Proposition 2.5**
$$\sum_{m=0}^{\ell-1}(-1)^m \binom{K}{\ell-1-m}\binom{\ell-(\ell-1-m)-1}{m} K = \frac{1}{(\ell-1)!}\prod_{j=0}^{\ell-1}(K-j).$$

**Proof:** First we note that
$$\frac{1}{(\ell-1)!}\prod_{j=0}^{\ell-1}(K-j) = \frac{K!}{(\ell-1)!(K-\ell)!} = \frac{K!}{(\ell-1)!(K-1-(\ell-1))!} = K\binom{K-1}{\ell-1}$$

Then,
$$\sum_{m=0}^{\ell-1}(-1)^m \binom{K}{\ell-1-m}\binom{\ell-(\ell-1-m)-1}{m} K$$
$$= K\sum_{m=0}^{\ell-1}(-1)^m \binom{K}{\ell-1-m} = K\sum_{m=0}^{\ell-2}(-1)^m\left[\binom{K-1}{\ell-1-m-1} + \binom{K-1}{\ell-1-m}\right] + (-1)^{\ell-1}\binom{K}{0}K$$
$$= K\left[\binom{K-1}{\ell-2} + \binom{K-1}{\ell-1}\right]$$
$$+ K\sum_{m=2}^{\ell-1}(-1)^{m+1}\binom{K-1}{\ell-1-m} + K\sum_{m=1}^{\ell-2}(-1)^m \binom{K-1}{\ell-1-m} + (-1)^{\ell-1} K$$
$$= K\left[\binom{K-1}{\ell-2} + \binom{K-1}{\ell-1}\right] + K(-1)^\ell\binom{K-1}{0} - K\binom{K-1}{\ell-2} + (-1)^{\ell-1} K = K\binom{K-1}{\ell-1}$$
□

**Proposition 2.6** For $n \in \{0, \ldots, \ell-2\}$,
$$\sum_{m=0}^{n}(-1)^m\left\{\binom{K}{n-m}\binom{\ell-(n-m)-1}{m}K - \binom{K}{n+1-m}(n+1-m)\binom{\ell-(n+1-m)-1}{m}\right\} = 0$$

**Proof:** With
$$\binom{K}{n+1-m}(n+1-m) = \binom{K-1}{n+1-m}K$$

we find

$$\sum_{m=0}^{n}(-1)^m \binom{K}{n-m} K \binom{\ell-(n-m)-1}{m}$$
$$-\sum_{m=0}^{n}(-1)^m \binom{K}{n+1-m}(n+1-m)\binom{\ell-(n+1-m)-1}{m}$$
$$= K\sum_{m=0}^{n}(-1)^m \left\{\binom{K}{n-m}\binom{\ell-(n-m)-1}{m} - \binom{K-1}{n-m}\binom{\ell-(n-m)-2}{m}\right\}$$
$$= K\sum_{m=0}^{n}(-1)^{n-m}\left\{\binom{K}{m}\binom{\ell-m-1}{n-m} - \binom{K-1}{m}\binom{\ell-m-2}{n-m}\right\}$$
$$= K(-1)^n \binom{\ell-1}{n} + K\sum_{m=1}^{n}(-1)^{n-m}\left\{\binom{K}{m}\binom{\ell-m-1}{n-m}\right\}$$
$$-K\sum_{m=0}^{n-1}(-1)^{n-m}\left\{\binom{K-1}{m}\binom{\ell-m-2}{n-m}\right\} - K\binom{K-1}{n}$$
$$= K(-1)^n \binom{\ell-1}{n} - K\sum_{m'=0}^{n-1}(-1)^{n-m'}\left\{\binom{K}{m'+1}\binom{\ell-m'-2}{n-m'-1}\right\}$$
$$-K\sum_{m=0}^{n-1}(-1)^{n-m}\left\{\binom{K-1}{m}\binom{\ell-m-2}{n-m}\right\} - K\binom{K-1}{n}$$
$$= K(-1)^n \binom{\ell-1}{n} - K\sum_{m=0}^{n-1}(-1)^{n-m}\left\{\left(\binom{K-1}{m+1}+\binom{K-1}{m}\right)\binom{\ell-m-2}{n-m-1}\right\}$$
$$-K\sum_{m=0}^{n-1}(-1)^{n-m}\left\{\binom{K-1}{m}\binom{\ell-m-2}{n-m}\right\} - K\binom{K-1}{n}$$
$$= K(-1)^n \binom{\ell-1}{n} - K\sum_{m=0}^{n-1}(-1)^{n-m}\left\{\binom{K-1}{m+1}\binom{\ell-m-2}{n-m-1}\right\}$$
$$-K\sum_{m=0}^{n-1}(-1)^{n-m}\left\{\binom{K-1}{m}\left(\binom{\ell-m-2}{n-m}+\binom{\ell-m-2}{n-m-1}\right)\right\} - K\binom{K-1}{n}$$
$$= K(-1)^n \binom{\ell-1}{n} - K\sum_{m=0}^{n-1}(-1)^{n-m}\left\{\binom{K-1}{m+1}\binom{\ell-(m+1)-1}{n-(m+1)}\right\}$$
$$-K\sum_{m=0}^{n-1}(-1)^{n-m}\left\{\binom{K-1}{m}\binom{\ell-m-1}{n-m}\right\} - K\binom{K-1}{n}$$
$$= K(-1)^n \binom{\ell-1}{n} + K\sum_{m=1}^{n}(-1)^{n-m}\left\{\binom{K-1}{m}\binom{\ell-m-1}{n-m}\right\}$$
$$-K\sum_{m=0}^{n-1}(-1)^{n-m}\left\{\binom{K-1}{m}\binom{\ell-m-1}{n-m}\right\} - K\binom{K-1}{n} = 0.$$

$\square$

**Proposition 2.7**

$$\lim_{n\to\infty} E(X_{(\ell+1)}/X_{(\ell)}) - 1$$
$$= \frac{1}{(\ell-1)!} \prod_{j=0}^{\ell-1}(K-j) \int_0^{1/z-1} \left(1 - \frac{\ln(u+1)}{\ln(1/z)}\right)^{K-\ell} \int_1^{1+u} \frac{\ln(v)^{\ell-1} \ln(1/z)^{-\ell}}{v^2} \, dv \, du \quad (13)$$

**Proof:** As $X_{(\ell+1);n_0,n,K}|X_{(\ell);n_0,n,K} = i_1 \sim X_{(1);i_1,n,K-\ell}$, we have $(c_{i,n} := (\sum_{h=i}^n h^{-1})^{-1})$

$$E(X_{(\ell+1)}/X_{(\ell)}) = E(E(X_{(\ell+1)}/X_{(\ell)}|X_{(\ell)})) = \sum_{i_1=n_0}^n \frac{1}{i_1} E(X_{(\ell+1)}|X_{(\ell)} = i_1) P(X_{(\ell)} = i_1)$$

$$= \sum_{i_1=n_0}^n \frac{1}{i_1} E(X_{(1);i_1,n,K-\ell}) \, P(X_{(\ell)} = i_1)$$

$$= \sum_{i_1=n_0}^n \frac{1}{i_1} \left[ i_1 + \left( \sum_{i=i_1}^n \left( \sum_{j=i+1}^n \frac{c_{i_1,n}}{j} \right)^{K-\ell} \right) \right] P(X_{(\ell)} = i_1)$$

$$= 1 + \sum_{i_1=n_0}^n \frac{1}{i_1} \left( \sum_{i=i_1}^n \left( \sum_{j=i+1}^n \frac{c_{i_1,n}}{j} \right)^{K-\ell} \right) P(X_{(\ell)} = i_1).$$

Hence,

$$E(X_{(\ell+1)}/X_{(\ell)}) - 1$$
$$= \sum_{j=0}^{\ell-1} \sum_{i_1=n_0}^n \sum_{i=i_1}^n \frac{c^K}{(i_1/n)^2} \left( \sum_{j'=i+1}^n \frac{c_{i_1,n}}{j'} \right)^{K-\ell} \left[ \binom{K}{j} \left\{ \left(\sum_{w=n_0}^{i_1-1} \frac{1}{w}\right)^{j-1} \left(\sum_{m=i_1+1}^n \frac{1}{m}\right)^{K-j-1} \times \right. \right.$$
$$\left. \left. \times \left( K \sum_{w=n_0}^{i_1-1} \frac{1}{w} - jc^{-1} \right) + \frac{1}{n} \mathcal{O}((i_1/n)^{-3}) \right] \frac{1}{n^2}.$$

Introducing $y = i/n \in [\ln(1/z)^{-1}, 1]$, $w = i_1/n \in [y, 1]$, we find for $n$ large

$$\sum_{j'=i+1}^n \frac{c_{i_1,n}}{j'} = \frac{\ln((n+1)/i)}{\ln((n+1)/i_1)} + \mathcal{O}(i_1^{-1}) = \frac{\ln(1/y)}{\ln(1/w)} + \mathcal{O}(n^{-1})$$

$$\sum_{w=n_0}^{i_1-1} \frac{1}{w} = \ln((i_1-1)/n_0) + \mathcal{O}(i_1^{-1}) = \ln\left(\frac{(i_1-1)/n}{(n_0/n)}\right) + \mathcal{O}(i_1^{-1}) = \ln(w/z) + \mathcal{O}(n^{-1})$$

$$\sum_{m=i_1+1}^n \frac{1}{m} = \ln(n/(i_1+1)) + \mathcal{O}(i_1^{-1}) = \ln(1/w) + \mathcal{O}(n^{-1})$$

As $n_0 = \lfloor z\,n \rfloor$, we note furthermore that $\lim_{n\to\infty} c^{-1} = \lim_{n\to\infty} \ln(n/n_0) = \ln(1/z)$. In the equation $E(X_{(\ell+1)}/X_{(\ell)}) - 1$ we recognize two nested Riemann sums; for $n \to \infty$, we find that these terms converge to nested integrals

$$\lim_{n\to\infty} E(X_{(\ell+1)}/X_{(\ell)}) - 1$$

$$= \sum_{j=0}^{\ell-1} \int_z^1 \int_w^1 \frac{\ln(1/z)^{-K}}{w^2} \left(\frac{\ln(1/y)}{\ln(1/w)}\right)^{K-\ell} \binom{K}{j} \ln(w/z)^{j-1} \ln(1/w)^{K-j-1} \Big(K \ln(w/z) - j \ln(1/z)\Big) dy\, dw$$

$$= \sum_{j=0}^{\ell-1} \binom{K}{j} K \ln(1/z)^{-K} \int_z^1 \ln(1/y)^{K-\ell} \int_z^y \frac{1}{w^2} \frac{\ln(w/z)^j \ln(1/w)^{K-j-1}}{\ln(1/w)^{K-\ell}} dw\, dy$$

$$- \sum_{j=0}^{\ell-1} \binom{K}{j} j \ln(1/z)^{-K+1} \int_z^1 \ln(1/y)^{K-\ell} \int_z^y \frac{1}{w^2} \frac{\ln(w/z)^{j-1} \ln(1/w)^{K-j-1}}{\ln(1/w)^{K-\ell}} dw\, dy$$

$$= \sum_{j=0}^{\ell-1} \binom{K}{j} K \ln(1/z)^{-K} \int_z^1 \ln(1/y)^{K-\ell} \int_z^y \frac{1}{w^2} \ln(w/z)^j \ln(1/w)^{\ell-j-1} dw\, dy$$

$$- \sum_{j=0}^{\ell-1} \binom{K}{j} j \ln(1/z)^{-K+1} \int_z^1 \ln(1/y)^{K-\ell} \int_z^y \frac{1}{w^2} \ln(w/z)^{j-1} \ln(1/w)^{\ell-j-1} dw\, dy$$

Now we transform the integrals: Let $y = z(u+1)$, then

$$E(X_{(\ell+1)}/X_{(\ell)})$$

$$= 1 + \sum_{j=0}^{\ell-1} \binom{K}{j} K \ln(1/z)^{-K} \int_0^{1/z-1} \ln\left(\frac{1}{z(u+1)}\right)^{K-\ell} z \int_z^{z(1+u)} \frac{1}{w^2} \ln(w/z)^j \ln(1/w)^{\ell-j-1} dw\, du$$

$$- \sum_{j=0}^{\ell-1} \binom{K}{j} j \ln(1/z)^{-K+1} \int_0^{1/z-1} \ln\left(\frac{1}{z(u+1)}\right)^{K-\ell} z \int_z^{z(1+u)} \frac{1}{w^2} \ln(w/z)^{j-1} \ln(1/w)^{\ell-j-1} dw\, du.$$

And next, let $v = w/z$,

$$E(X_{(\ell+1)}/X_{(\ell)})$$

$$= 1 + \sum_{j=0}^{\ell-1} \binom{K}{j} K \ln(1/z)^{-K} \int_0^{1/z-1} \ln\left(\frac{1}{z(u+1)}\right)^{K-\ell} \int_1^{1+u} \frac{1}{v^2} \ln(v)^j \ln(1/(z\,v))^{\ell-j-1} dv\, du$$

$$- \sum_{j=0}^{\ell-1} \binom{K}{j} j \ln(1/z)^{-K+1} \int_0^{1/z-1} \ln\left(\frac{1}{z(u+1)}\right)^{K-\ell} \int_1^{1+u} \frac{1}{v^2} \ln(v)^{j-1} \ln(1/(z\,v))^{\ell-j-1} dv\, du$$

$$= 1 + \sum_{j=0}^{\ell-1} \binom{K}{j} K \ln(1/z)^{-\ell} \int_0^{1/z-1} \left(1 - \frac{\ln(u+1)}{\ln(1/z)}\right)^{K-\ell} \int_1^{1+u} \frac{1}{v^2} \ln(v)^j \ln(1/(z\,v))^{\ell-j-1} dv\, du$$

$$- \sum_{j=1}^{\ell-1} \binom{K}{j} j \ln(1/z)^{-\ell+1} \int_0^{1/z-1} \left(1 - \frac{\ln(u+1)}{\ln(1/z)}\right)^{K-\ell} \int_1^{1+u} \frac{1}{v^2} \ln(v)^{j-1} \ln(1/(z\,v))^{\ell-j-1} dv\, du.$$

We expand $\ln(1/(zv))^{\ell-j-1} = [-\ln(v) + \ln(1/z)]^{\ell-j-1}$, and collect terms with equal powers of $\ln(v)$ and $\ln(1/z)$: If we use the abbreviation

$$A = \left(1 - \frac{\ln(u+1)}{\ln(1/z)}\right)$$

we have

$$E(X_{(\ell+1)}/X_{(\ell)}) - 1$$

$$= \sum_{j=0}^{\ell-1} \binom{K}{j} K \ln(1/z)^{-\ell} \int_0^{1/z-1} A^{K-\ell} \int_1^{1+u} \frac{1}{v^2} \ln(v)^j \ln(1/(zv))^{\ell-j-1} \, dv \, du$$

$$- \sum_{j=1}^{\ell-1} \binom{K}{j} j \ln(1/z)^{-\ell+1} \int_0^{1/z-1} A^{K-\ell} \int_1^{1+u} \frac{1}{v^2} \ln(v)^{j-1} \ln(1/(zv))^{\ell-j-1} \, dv \, du$$

$$= \sum_{j=0}^{\ell-1} \binom{K}{j} K \ln(1/z)^{-\ell} \times$$

$$\times \int_0^{1/z-1} A^{K-\ell} \int_1^{1+u} \frac{1}{v^2} \left( \sum_{m=0}^{\ell-j-1} \binom{\ell-j-1}{m} (-1)^m \ln(v)^m \ln(1/z)^{\ell-j-1-m} \right) \ln(v)^j \, dv \, du$$

$$- \sum_{j=1}^{\ell-1} \binom{K}{j} j \ln(1/z)^{-\ell+1} \times$$

$$\times \int_0^{1/z-1} A^{K-\ell} \int_1^{1+u} \frac{1}{v^2} \left( \sum_{m=0}^{\ell-j-1} \binom{\ell-j-1}{m} (-1)^m \ln(v)^m \ln(1/z)^{\ell-j-1-m} \right) \ln(v)^{j-1} \, dv \, du$$

$$= \sum_{j=0}^{\ell-1} \binom{K}{j} K \int_0^{1/z-1} A^{K-\ell} \int_1^{1+u} \frac{1}{v^2} \left( \sum_{m=0}^{\ell-j-1} \binom{\ell-j-1}{m} (-1)^m \ln(v)^{m+j} \ln(1/z)^{-j-1-m} \right) dv \, du$$

$$- \sum_{j=0}^{\ell-1} \binom{K}{j} j \int_0^{1/z-1} A^{K-\ell} \int_1^{1+u} \frac{1}{v^2} \left( \sum_{m=0}^{\ell-j-1} \binom{\ell-j-1}{m} (-1)^m \ln(v)^{m+j-1} \ln(1/z)^{-j-m} \right) dv \, du$$

By $n = m + j$, we re-order the sums, noting that

$$\sum_{j=0}^{\ell-1} \sum_{m=0}^{\ell-j-1} \text{term}(j, m) = \sum_{n=0}^{\ell-1} \sum_{m=0}^{n} \text{term}(n-m, m)$$

Therwith,

$$E(X_{(\ell+1)}/X_{(\ell)}) - 1$$

$$= \sum_{n=0}^{\ell-1} \sum_{m=0}^{n} (-1)^m \binom{K}{n-m} \binom{\ell-(n-m)-1}{m} K \int_0^{1/z-1} A^{K-\ell} \int_1^{1+u} \frac{\ln(v)^n \ln(1/z)^{-n-1}}{v^2} \, dv \, du$$

$$- \sum_{n=0}^{\ell-1} \sum_{m=0}^{n} (-1)^m \binom{K}{n-m} \binom{\ell-(n-m)-1}{m} (n-m) \int_0^{1/z-1} A^{K-\ell} \int_1^{1+u} \frac{\ln(v)^{n-1} \ln(1/z)^{-n}}{v^2} \, dv \, du$$

$$= \sum_{m=0}^{\ell-1} (-1)^m \binom{K}{\ell-1-m} \binom{\ell-(n-m)-1}{m} K \int_0^{1/z-1} A^{K-\ell} \int_1^{1+u} \frac{\ln(v)^{\ell-1} \ln(1/z)^{-\ell}}{v^2} \, dv \, du$$

$$+ \sum_{n=0}^{\ell-2} \sum_{m=0}^{n} (-1)^m \binom{K}{n-m} \binom{\ell-(n-m)-1}{m} K \int_0^{1/z-1} A^{K-\ell} \int_1^{1+u} \frac{\ln(v)^n \ln(1/z)^{-n-1}}{v^2} \, dv \, du$$

$$- \sum_{n=1}^{\ell-1} \sum_{m=0}^{n-1} (-1)^m \binom{K}{n-m} \binom{\ell-(n-m)-1}{m} (n-m) \int_0^{1/z-1} A^{K-\ell} \int_1^{1+u} \frac{\ln(v)^{n-1} \ln(1/z)^{-n}}{v^2} \, dv \, du$$

$$= \sum_{m=0}^{\ell-1}(-1)^m \binom{K}{\ell-1-m}\binom{\ell-(\ell-1-m)-1}{m} K \int_0^{1/z-1} A^{K-\ell} \int_1^{1+u} \frac{\ln(v)^{\ell-1}\ln(1/z)^{-\ell}}{v^2}\,dv\,du$$

$$+ \sum_{n=0}^{\ell-2}\sum_{m=0}^{n}(-1)^m \binom{K}{n-m}\binom{\ell-(n-m)-1}{m} K \int_0^{1/z-1} A^{K-\ell} \int_1^{1+u} \frac{\ln(v)^n \ln(1/z)^{-n-1}}{v^2}\,dv\,du$$

$$- \sum_{n=0}^{\ell-2}\sum_{m=0}^{n}(-1)^m \binom{K}{n+1-m}\binom{\ell-(n+1-m)-1}{m}(n+1-m) \int_0^{1/z-1} A^{K-\ell} \int_1^{1+u} \frac{\ln(v)^n \ln(1/z)^{-n-1}}{v^2}\,dv\,du$$

With propositions 2.5 and 2.6 the result follows.

$\square$

Now we are in the position to prove theorem 2.1.
**Proof:** [of Theorem 2.1] We start off with

$$E(X_{(\ell+1)}/X_{(\ell)}) - 1$$

$$= \frac{1}{(\ell-1)!} \prod_{j=0}^{\ell-1}(K-j) \int_0^{1/z-1}\left(1 - \frac{\ln(u+1)}{\ln(1/z)}\right)^{K-\ell} \int_1^{1+u} \frac{\ln(v)^{\ell-1}\ln(1/z)^{-\ell}}{v^2}\,dv\,du$$

$$= \frac{\ln(1/z)^{-\ell}}{(\ell-1)!} \prod_{j=0}^{\ell-1}(K-j) \int_0^{1/z-1}\left(1 - \frac{\ln(u+1)}{\ln(1/z)}\right)^{K-\ell} \int_1^{1+u} \frac{\ln(v)^{\ell-1}}{v^2}\,dv\,du$$

If we focus on the inner integral, we find ($\nu = \ln(v)$, $d\nu = \frac{1}{v}dv$, $v = e^\nu$)

$$\int_1^{1+u} \frac{\ln(v)^{\ell-1}}{v^2}\,dv = \int_0^{\ln(1+u)} \nu^{\ell-1} e^{-\nu}\,d\nu = \gamma(\ell, \ln(1+u))$$

where $\gamma(n,x)$ denotes the (lower) incomplete $\Gamma$ function. In particular, $\gamma(n,x) = (n-1)\gamma(n-1,x) - x^{n-1} e^{-x}$. Thus, for $\ell \geq 2$,

$$E(X_{(\ell+1)}/X_{(\ell)}) - 1$$

$$= \frac{\ln(1/z)^{-\ell}}{(\ell-1)!} \prod_{j=0}^{\ell-1}(K-j) \int_0^{1/z-1}\left(1 - \frac{\ln(u+1)}{\ln(1/z)}\right)^{K-\ell} \int_1^{1+u} \frac{\ln(v)^{\ell-1}}{v^2}\,dv\,du$$

$$= \frac{\ln(1/z)^{-\ell}}{(\ell-1)!} \prod_{j=0}^{\ell-1}(K-j) \int_0^{1/z-1} \gamma(\ell, \ln(1+u))\left(1 - \frac{\ln(u+1)}{\ln(1/z)}\right)^{K-\ell}du$$

$$= -\frac{\ln(1/z)^{-\ell+1}}{(\ell-1)!} \prod_{j=0}^{\ell-2}(K-j) \int_0^{1/z-1} \gamma(\ell, \ln(1+u))(1+u)\frac{d}{du}\left(1 - \frac{\ln(u+1)}{\ln(1/z)}\right)^{K-(\ell-1)}du$$

$$= \frac{\ln(1/z)^{-\ell+1}}{(\ell-1)!} \prod_{j=0}^{\ell-2}(K-j) \int_0^{1/z-1}\left(\gamma(\ell, \ln(1+u)) + \frac{\ln(1+u)^{\ell-1}(1+u)}{(1+u)^2}\right)\left(1 - \frac{\ln(u+1)}{\ln(1/z)}\right)^{K-(\ell-1)}du$$

$$= \frac{\ln(1/z)^{-\ell+1}}{(\ell-1)!} \prod_{j=0}^{\ell-2}(K-j) \int_0^{1/z-1}\left((\ell-1)\gamma(\ell-1, \ln(1+u))\right)\left(1 - \frac{\ln(u+1)}{\ln(1/z)}\right)^{K-(\ell-1)}du$$

$$= \frac{\ln(1/z)^{-(\ell-1)}}{(\ell-2)!} \prod_{j=0}^{\ell-2}(K-j) \int_0^{1/z-1} \gamma(\ell-1, \ln(1+u))\left(1 - \frac{\ln(u+1)}{\ln(1/z)}\right)^{K-(\ell-1)}du$$

$$= E(X_{(\ell)}/X_{(\ell-1)}) - 1.$$

Per finite induction we find that $E(X_{(\ell)}/X_{(\ell-1)})$ is independent off $\ell$ (for those $\ell$ that are feasible). If we take $\ell = 1$, we have $(y = z(1+u),\ \theta = K/\ln(1/z))$

$$\begin{aligned}
E(X_{(2)}/X_{(1)}) &= 1 + \frac{K}{\ln(1/z)} \int_0^{1/z-1} \left(1 - \frac{\ln(u+1)}{\ln(1/z)}\right)^{K-1} \int_1^{1+u} \frac{1}{v^2}\, dv\, du \\
&= 1 + \frac{K}{\ln(1/z)} \int_0^{1/z-1} \left(\frac{1}{z} - \frac{1}{z(1+u)}\right) \left(\frac{\ln(1/(z(1+u)))}{\ln(1/z)}\right)^{K-1} z\, du \\
&= 1 + \theta \int_z^1 \left(\frac{1}{z} - \frac{1}{y}\right) \left(\frac{\ln(1/y)}{\ln(1/z)}\right)^{\theta \ln(1/z)-1} dy = G(\theta, z).
\end{aligned}$$

$\square$

### 2.3.3 Limit $z \to 0$

We expect this expression mainly to depend on $\theta = K/\ln(1/z)$. That is, our formula reads

$$E(X_{(\ell+1)}/X_{(\ell)}) = G(\theta, z) = 1 + \theta \int_z^1 \left(\frac{1}{z} - \frac{1}{y}\right) \left(\frac{\ln(1/y)}{\ln(1/z)}\right)^{\ln(1/z)\theta - 1} dy$$

In order to discuss the dependencies of $G(\theta, z)$, we keep $\theta$ fixed and take the limit $z \to 0$.

**Proposition 2.8**

$$\lim_{z \to 0} G(\theta, z) = \frac{\theta}{\theta - 1} \tag{14}$$

**Proof:** We use the transformation $y = z(1+u)$, and $w = \ln(1+u)$, and introduce $x = \ln(1/z)$:

$$\begin{aligned}
&\lim_{z \to 0} \int_z^1 \left(\frac{1}{z} - \frac{1}{y}\right) \left(\frac{\ln(1/y)}{\ln(1/z)}\right)^{\ln(1/z)\theta - 1} dy \\
&= \lim_{z \to 0} \int_0^{1/z-1} \left(\frac{1}{z} - \frac{1}{z(1+u)}\right) \left(\frac{\ln(1/(z(1+u)))}{\ln(1/z)}\right)^{\ln(1/z)\theta - 1} z\, du \\
&= \lim_{z \to 0} \int_0^{1/z-1} \frac{u}{1+u} \left(1 - \frac{\ln(1+u)}{\ln(1/z)}\right)^{\ln(1/z)\theta - 1} du \\
&= \lim_{x \to \infty} \int_0^x (e^w - 1) \left(1 - \frac{w}{x}\right)^{x\theta - 1} dw
\end{aligned}$$

In order to compute this limit, we first note that $\zeta(u) = \ln(1-u) + u/(1-u)$ has the derivative $\zeta'(u) = u/(1-u)^2 \geq 0$ for $u \in [0, 1)$; since $\zeta(0) = 0$, we have $\zeta(u) \geq 0$ for $u \in [0, 1)$. Since $\frac{d}{dx}(1 - w/x)^x = (1 - w/x)^x \zeta(w/x) \geq 0$, we have $0 \leq (1 - w/x)^x \leq e^{-w}$, and $(1 - w/x)^x$ tends in a monotonously increasing way to $e^{-w}$.
Next we note (recall that $\theta > 1$)

$$\limsup_{x \to \infty} \int_0^x (e^w - 1) \left(1 - \frac{w}{x}\right)^{x\theta} dw \leq \limsup_{x \to \infty} \int_0^x (e^w - 1)\, e^{-\theta w}\, dw = \int_0^\infty (e^w - 1)\, e^{-\theta w}\, dw < \infty.$$

Furthermore, for any $x_0 \in \mathbb{R}$, we find

$$\liminf_{x \to \infty} \int_0^x (e^w - 1) \left(1 - \frac{w}{x}\right)^{x\theta} dw \geq \liminf_{x \to \infty} \int_0^{x_0} (e^w - 1) \left(1 - \frac{w}{x}\right)^{x\theta} dw = \int_0^{x_0} (e^w - 1)\, e^{-\theta w}\, dw.$$

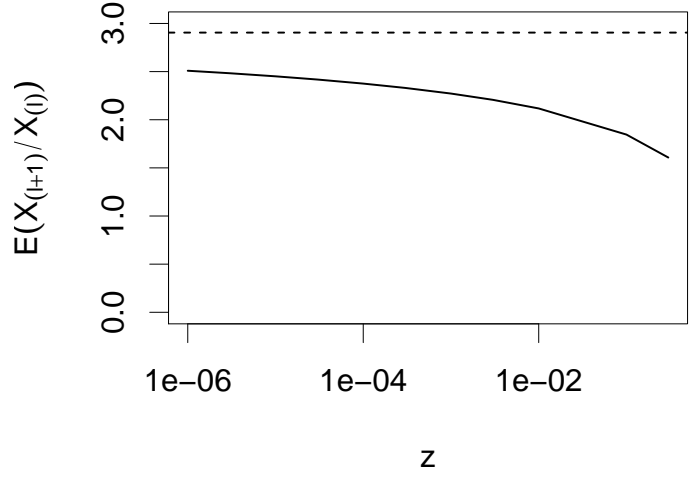

Figure 2: Convergence of $G(\theta, z)$ (solid line) to $G(\theta, 0)$ (dashed line) for $\theta = 1.5$. Note the logarithmic scale of the $x$-axis.

As $\int_0^{x_0} (e^w - 1)\, e^{-\theta w}\, dw \to \int_0^{\infty} (e^w - 1)\, e^{-\theta w}\, dw$ for $x_0 \to \infty$, we conclude that $\int_0^x (e^w - 1)\left(1 - \frac{w}{x}\right)^{x\theta} dw$ converges to $\int_0^{\infty} (e^w - 1)\, e^{-\theta w}\, dw$.

Last, we consider the limiting behaviour of

$$\int_0^x (e^w - 1) \left(1 - \frac{w}{x}\right)^{x\theta} \left(1 - \frac{w}{x}\right)^{-1} dw = \int_0^x (e^w - 1) \left(1 - \frac{w}{x}\right)^{x(\theta - \varepsilon)} \left(1 - \frac{w}{x}\right)^{\varepsilon x - 1} dw.$$

Let $\varepsilon > 0$, s.t. $\theta - \varepsilon > 1$. For $x$ large enough, $\varepsilon x - 1 > 0$ and $\left(1 - \frac{w}{x}\right)^{\varepsilon x - 1} < 1$. With the argument from above,

$$\limsup_{x \to \infty} \int_0^x (e^w - 1) \left(1 - \frac{w}{x}\right)^{x\theta} \left(1 - \frac{w}{x}\right)^{-1} dw \leq \int_0^{\infty} (e^w - 1)\, e^{-(\theta - \varepsilon)w}\, dw.$$

Since this inequality holds true for any $\varepsilon > 0$, we can take $\varepsilon = 0$. The estimate for $\liminf$ from below relies on the same argument as before: for $x_0 \in \mathbb{R}$ fixed we have

$$\liminf_{x \to \infty} \int_0^x (e^w - 1) \left(1 - \frac{w}{x}\right)^{x\theta} \left(1 - \frac{w}{x}\right)^{-1} dw$$
$$\geq \liminf_{x \to \infty} \int_0^{x_0} (e^w - 1) \left(1 - \frac{w}{x}\right)^{x\theta} \left(1 - \frac{w}{x}\right)^{-1} dw = \int_0^{x_0} (e^w - 1)\, e^{-\theta w}\, dw.$$

Hence,

$$\lim_{x \to \infty} \int_0^x (e^w - 1) \left(1 - \frac{w}{x}\right)^{x\theta - 1} dw = \int_0^{\infty} (e^w - 1)\, e^{-\theta w}\, dw = \frac{1}{\theta - 1} - \frac{1}{\theta}.$$

□

For $n \gg 1$ and $z \ll 1$ and $\theta = K/\ln(1/z)$, we have

$$E(X_{(\ell+1)}/X_{(\ell)}) \approx \frac{\theta}{\theta - 1} \qquad (15)$$

From this result, we conclude that $E(X_{(\ell+1)}/X_{(\ell)})$ mainly depends on $\theta = K/\ln(1/z)$; however, the convergence of $G(\theta, z)$ to $G(\theta, 0)$ is rather slow (see figure 2). If we parametrize the model with the rescaled "party creation rate" $\theta$ and the relative minimal party size $z$, the model is sensitive in $\theta$ and insensitive in $z$. It is only necessary to know the rough magnitude of $z$ as long as we know $\theta$ precisely.

### 2.3.4 Logarithm of the group sizes

**Proof:** [of theorem 2.3] Using proposition 2.4, we find

$$\lim_{n \to \infty} E(\ln(X_{(1)}) - \ln(n)) = \lim_{n \to \infty} \sum_{i=n_0}^{n} \ln(i/n) P(X_{(1)} = i)$$

$$= \lim_{n \to \infty} \sum_{i=n_0}^{n} \ln(i/n) \left\{ \left( \sum_{j=i}^{n} \frac{c}{j} \right)^K - \left( \sum_{j=i+1}^{n} \frac{c}{j} \right)^K \right\} = \lim_{n \to \infty} \sum_{i=n_0}^{n} \ln(i/n) \left\{ \left( \sum_{j=i}^{n} \frac{c}{j} \right)^K - \left( \sum_{j=i}^{n} \frac{c}{j} - \frac{c}{i} \right)^K \right\}.$$

Note that $\sum_{j=i}^{n} \frac{c}{j} = \mathcal{O}(n^0)$. Taylor expansion yields

$$\lim_{n \to \infty} E(\ln(X_{(1)}) - \ln(n)) = \lim_{n \to \infty} \sum_{i=n_0}^{n} \ln(i/n) \left\{ \frac{Kc}{i} \left( \sum_{j=i}^{n} \frac{c}{j} \right)^{K-1} + \mathcal{O}(i^{-2}) \right\}$$

$$= \lim_{n \to \infty} \sum_{i=n_0}^{n} \ln(i/n) \left\{ \frac{K c^K}{i/n} \left( \sum_{j=i}^{n} \frac{1}{j/n} \frac{1}{n} \right)^{K-1} + \mathcal{O}((i/n)^{-2}) \frac{1}{n} \right\} \frac{1}{n}$$

We recognize two nested Riemann sums, that converge to the corresponding integrals. If we use that $c$ converges to $1/\ln(1/z)$ for $n \to \infty$, we obtain

$$\lim_{n \to \infty} E(\ln(X_{(1)})) - \ln(n)) = K \ln(1/z)^{-K} \int_z^1 \frac{\ln(x)}{x} \left( \int_x^1 \frac{1}{y} dy \right)^{K-1} dx = -K \int_z^1 \frac{1}{x} \left( \frac{\ln(1/x)}{\ln(1/z)} \right)^K dy.$$

For $\ell > 1$, $E(\ln(X_\ell))$ is handled in a similar way:

$$\lim_{n \to \infty} E(\ln(X_{(\ell)}) - \ln(n)) = \sum_{i=n_0}^{n} \ln(i/n) P(X_{(\ell)} = i)$$

$$= \lim_{n \to \infty} \sum_{i=n_0}^{n} \left\{ \ln(i/n) \frac{c^K}{i_1} \sum_{j=0}^{\ell-1} \binom{K}{j} \left\{ \left( \sum_{w=n_0}^{i_1-1} \frac{1}{w} \right)^{j-1} \left( \sum_{m=i_1+1}^{n} \frac{1}{m} \right)^{K-j-1} \left[ K \left( \sum_{w=n_0}^{i_1-1} \frac{1}{w} \right) - jc^{-1} \right] \right\} + \mathcal{O}(i_1^{-2}) \right\}$$

$$= -\sum_{j=0}^{\ell-1} \binom{K}{j} \int_z^1 \frac{1}{x} \left( 1 - \frac{\ln(1/x)}{\ln(1/z)} \right)^{j-1} \left( \frac{\ln(1/x)}{\ln(1/z)} \right)^{K-j} \left( K - j - K \frac{\ln(1/x)}{\ln(1/z)} \right) dx.$$

$\square$

Note that $E(\log(X_{(\ell+1)})) - E(\log(X_{(\ell)}))$ is not independent on $\ell$, even for $n$ large; the expectation of $\log(X_{(\ell)})$ does not depend exactly in a linear way on the rank $\ell$, but only approximately. The growth law of the group sizes can be better seen in the rations of subsequent groups than in the logarithm of group sizes. However, for practical purpose, the difference of the linear growth law for the logarithmic group sizes (Zipf's law) is negligible.

As as heuristic estimator for $z$ we will use

$$E(\ln(X_{(1)})) - \ln(n) = -K \int_z^1 \frac{1}{x} \left(\frac{\ln(1/x)}{\ln(1/z)}\right)^K dy.$$

We replace $E(X_{(1)})$ by the minimal observed group size, and infer from the relation above the parameter $z$. Though we consider data that are conditioned on the total population (number of voters known), for practical purposes this estimator works fine (see figures sin section 5).

## 2.4 Rank statistics – conditioned case

Now we investigate the corresponding order statistics in the conditioned case: We consider $K$ independent realizations $X_1, \ldots, X_K$ of i.i.d. RV that assume values in $\{n_0, \ldots, n\}$, where $0 < n_0 < n$, $P(X_i = j) = c/j$ for $j \in \{n_0, \ldots, n\}$ and 0 else, where $c^{-1} = \sum_{j=n_0}^n j^{-1}$. We condition on $\sum_{i=1}^K X_i = n$ and order these realizations according to size $X_{(1)} \leq X_{(2)} \ldots \leq X_{(K)}$. In order to distinguish the conditioned and the non-conditioned random variables, let us denote the realizations with condition by

$$X_{(1),n} \leq X_{(2),n} \ldots \leq X_{(K),n}.$$

The objects to investigate are $E(X_{(\ell+1),n}/X_{(\ell),n})$ and $E(\ln(X_{(\ell),n}))$.

### 2.4.1 Joint distribution

**Proposition 2.9** Let

$$M_K = \left\{ (i_1, \ldots i_K) \,|\, n_0 \leq i_1 \leq n/K, \quad i_K = n - \sum_{\ell=1}^K i_\ell, \right.$$

$$\left. i_{j-1} \leq i_j \leq \frac{1}{K-j+1}\left(n - \sum_{\ell=1}^{j-1} i_\ell\right), \quad j = 2, \ldots K-1 \right\}$$

and

$$c_K \approx \left( \int_z^{1/K} \int_{x_1}^{(1-x_1)/(K-1)} \int_{x_2}^{(1-x_1-x_2)/(K-2)} \cdots \int_{x_{K-1}}^{(1-\sum_{j=1}^{K-2} x_j)/2} \frac{1}{1 - \sum_{\ell=1}^{K-1} x_\ell} \prod_{\ell=1}^{K-1} \frac{1}{x_\ell} dx_{K-1} \cdots dx_1 \right)^{-1}.$$

Then, for $(i_1, \ldots, i_K) \in M_K$, we have

$$P(X_{(1),n} = i_1, \ldots, X_{(K),n} = i_K) = c_K \prod_{\ell=1}^K \frac{1}{i_\ell} + \mathcal{O}(n^{-1}).$$

**Proof:** The values that $(X_{(1),n} \leq X_{(2),n} \ldots \leq X_{(K),n})$ can assume is given by $(i_1, \ldots i_K) \in M_K$ with

$$M_k = \{(i_1, \ldots i_K) \,|\, n_0 \leq i_1 \leq i_2 \ldots \leq i_K, \sum_{\ell=1}^K i_\ell = n\}.$$

In order to obtain a conditioned realization, we may draw unconditioned realizations until the condition is hit, and only accept those. Hence, the probability for an admissible value is proportional to the unconditioned probability distribution:

$$P(X_{(1),n} = i_1, \ldots X_{(K),n} = i_K) = C \prod_{\ell=1}^{K} \frac{1}{i_\ell};$$

The constant $C$ can be determined by

$$C^{-1} = \sum_{(i_1,\ldots i_K) \in M_K} \prod_{\ell=1}^{K} \frac{1}{i_\ell}.$$

We characterize $M_K$ better. For $(i_1, \ldots i_K) \in M_K$, we mey write

$$i_K = n - \sum_{\ell=1}^{K-1} i_\ell.$$

Thus,

$$i_{K-1} \leq i_K = n - \sum_{\ell=1}^{K-1} i_\ell \quad \Rightarrow \quad i_{K-1} \leq \frac{1}{2}\left(n - \sum_{\ell=1}^{K-2} i_\ell\right).$$

We can proceed recursively,

$$i_{K-2} \leq i_{K-1} \leq \frac{1}{2}\left(n - \sum_{\ell=1}^{K-2} i_\ell\right) \quad \Rightarrow \quad i_{K-2} \leq i_{K-1} \leq \frac{1}{3}\left(n - \sum_{\ell=1}^{K-3} i_\ell\right)$$

$$\cdots \quad i_{K-j-1} \leq i_{K-j} \leq \frac{1}{j+1}\left(n - \sum_{\ell=1}^{K-j-1} i_\ell\right)$$

or

$$i_{j-1} \leq i_j \leq \frac{1}{K-j+1}\left(n - \sum_{\ell=1}^{j-1} i_\ell\right)$$

For $i_1$, we obtain the maximal value given if all indices are equal, $i_1 \leq n/K$. Hence,

$$M_K = \Bigg\{(i_1, \ldots i_K) \,\Big|\, n_0 \leq i_1 \leq n/K, \quad i_K = n - \sum_{\ell=1}^{K} i_\ell,$$

$$i_{j-1} \leq i_j \leq \frac{1}{K-j+1}\left(n - \sum_{\ell=1}^{j-1} i_\ell\right), \quad j = 2, \ldots K-1\Bigg\}.$$

Therewith,

$$C^{-1} = \sum_{(i_1,\ldots i_K) \in M_K} \frac{1}{1 - \sum_{\ell=1}^{K-1} i_\ell/n} \prod_{\ell=1}^{K-1} \frac{1}{i_\ell/n} n^{-K}$$

$$\approx n^{-1} \int_z^{1/K} \int_{x_1}^{(1-x_1)/(K-1)} \int_{x_2}^{(1-x_1-x_2)/(K-2)} \cdots \int_{x_{K-1}}^{(1-\sum_{j=1}^{K-2} x_j)/2} \frac{1}{1 - \sum_{\ell=1}^{K-1} x_\ell} \prod_{\ell=1}^{K-1} \frac{1}{x_\ell} dx_{K-1} \cdots dx_1$$

$$=: (c_K\, n)^{-1}.$$

For symmetry reasons, we may write as well

$$c_K^{-1} = \frac{1}{K!} \int_{z_0}^{1-(K-1)z_0} \cdots \int_{z_0}^{1-(K-1)z_0} \frac{1}{1-\sum_{\ell=1}^{K-1} x_\ell} \prod_{\ell=1}^{K-1} \frac{1}{x_\ell} dx_{x_{K-1}} \cdots dx_1.$$

Note that $(X_1,..,X_K)/n$ follows for $n \to \infty$ a truncated Dirichlet distribution with parameters $\alpha_i = 0$; while for the original Dirichlet distribution necessarily $\alpha_i > 0$ due to integrability conditions, the truncated Dirichet distribution is also well defined for $\alpha_i = 0$.

### 2.4.2 Size ratio

We do not compute the expectation of the quotient for general $K$ but only for $K = 3$. As discussed above, the joint distribution of $(X_{(1)}, X_{(2)})$ is given by

$$P(X_{(1)} = i_1, X_{(2)} = i_2) = \frac{c_3}{i_1 \, i_2 (n - i_1 - i_2)}$$

respectively

$$P(X_{(2)} = i_2, X_{(3)} = i_3) = c_3 \frac{1}{(n - i_2 - i_3) \, i_2 \, i_3}$$

where

$$c_3^{-1} = \frac{2}{3!} \int_z^{1-2z} \frac{1}{x(1-x)} \ln\left(\frac{1-z-x}{z}\right) dx.$$

Therewith,

$$c_3 \lim_{n \to \infty} E(X_{(2)}/X_{(1)}) = \int_{z_0}^{1/3} \int_x^{(1-x)/2} \frac{1}{x^2 (1-x-y)} dy \, dx = \int_{z_0}^{1/3} x^{-2} \ln\left(\frac{2(1-2x)}{(1-x)}\right) dx.$$

Furthermore,

$$c_3 \lim_{n \to \infty} E(X_{(3)}/X_{(2)}) = \int_{z_0}^{1/3} \int_x^{(1-x)/2} \frac{1}{x \, y^2} dy \, dx = 2\ln(2) - 1 + \frac{1}{z_0} - 2\ln(1/z_0 - 1).$$

Obviously,

$$\lim_{n \to \infty} E(X_{(2)}/X_{(1)}) \neq \lim_{n \to \infty} E(X_{(3)}/X_{(2)}).$$

The magnitude of both expectation are in the same range, though (see figure 3).

### 2.4.3 Logarithm of group sizes

**Theorem 2.10** *For $k < K$,*

$$E(\ln(X_{(k),n})) - \ln(n)$$
$$= c_K \int_z^{1/K} \int_{x_1}^{(1-x_1)/(K-1)} \int_{x_2}^{(1-x_1-x_2)/(K-2)} \cdots \int_{x_{K-1}}^{(1-\sum_{j=1}^{K-2} x_j)/2} \frac{\ln(x_k)}{1 - \sum_{\ell=1}^{K-1} x_\ell} \prod_{\ell=1}^{K-1} \frac{1}{x_\ell} dx_{K-1} \cdots dx_1$$

and

$$E(\ln(X_{(K)})) - \ln(n)$$
$$= c_K \int_z^{1/K} \int_{x_1}^{(1-x_1)/(K-1)} \int_{x_2}^{(1-x_1-x_2)/(K-2)} \cdots \int_{x_{K-1}}^{(1-\sum_{j=1}^{K-2} x_j)/2} \frac{\ln\left(1 - \sum_{\ell=1}^{K-1} x_\ell\right)}{1 - \sum_{\ell=1}^{K-1} x_\ell} \prod_{\ell=1}^{K-1} \frac{1}{x_\ell} dx_{K-1} \cdots dx_1.$$

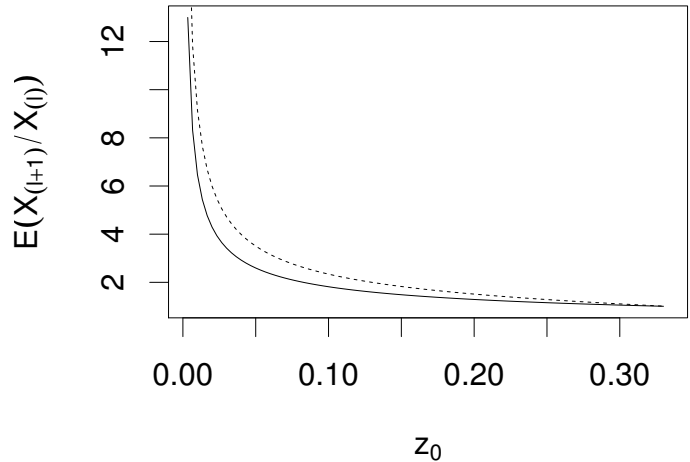

Figure 3: $E(X_{(2)}/X_{(1)})$ (solid line) and $E(X_{(3)}/X_{(2)})$ (dashed line) for $K = 3$.

The proof consists of an obvious calculation. Particularly, for $n$ large, $E(\ln(X_{(\ell),n})) = \ln(n)$ plus a term only depending on $z$, $\ell$, $K$, but not on $n$.

Though we see that, strictly spoken, there is no linear relation between the logarithmic size of groups and their order, simulations indicate that the dependency is almost linear, even if $K$ is small (see figure 4).

### 2.4.4 Simulations

Direct simulations of $X_{(k),n}/n$ in a naive way is costly for large $n$. The convergence of $X_{(k),n}/n$ to the truncated Dirichlet distribution opens the way for a simple method to construct realizations. We fix a population size of $\hat{n} = 10^{5+m}$, where $m \in \mathbb{N}_0$ is the minimal non-negative integer to ensure that $z\,\hat{n} > 5$. Then, we draw $K$ independent realizations of $X_k$ as introduced in section 2.3. We accept a realization if $\sum_{i=1}^{K} X_k \in [0.95\,\hat{n}, 1.05\,\hat{n}]$. In order to obtain (approximate) realizations for population size $n$, we rescale $n\,X_k/\hat{n}$. This algorithm is able to handle the populations sizes at hand for the data we consider.

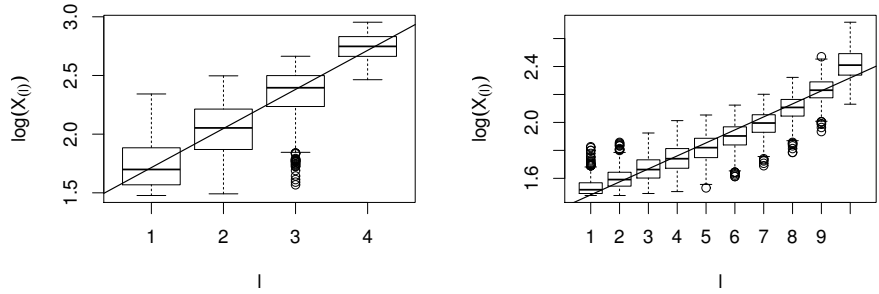

Figure 4: $E(log(X_{(\ell)}))$ over $\ell$. Boxplot from 1000 realizations, line is a linear fit to the expectations of $E(log(X_{(\ell)}))$; $n = 1000$, $z = 0.03$. Left: $K = 4$. Right: $K = 10$.

# 3 Visualization of Data

## 3.1 Data sources

The data for the election of the FRG ("Bundestagswahl") can be downloaded from
https://www.bundeswahlleiter.de/en/bundeswahlleiter.html
We only take parties into account that can be vote for by second votes ("Zweitstimmen").

The data for the elections in France can be found in
https://www.interieur.gouv.fr/Elections/Les-resultats

The data for the Republican primaries 2016 in the US can be found in:
Iowa caucuses: https://edition.cnn.com/election/2016/primaries/states/ia
retrieved from
https://www.iowagop.org/
New Hampshire : http://www.thegreenpapers.com/P16/NH-R
retrieved from
http://sos.nh.gov/2016RepPresPrim.aspx?id=8589957185
Nevada: http://www.thegreenpapers.com/P16/NV-R retrieved
from
http://nevadagop.org/nevada-republican-presidential-caucus-results/
Massachusetts: http://www.thegreenpapers.com/P16/MA-R
retrieved from
http://electionstats.state.ma.us/elections/search/year_from:2016/year_to:2016/office_id:1/stage:Republican
Tennessee: http://www.thegreenpapers.com/P16/TN-R
retrieved from
https://sos.tn.gov/products/elections/election-results
Texas: http://www.thegreenpapers.com/P16/TX-R
retrieved from
http://elections.sos.state.tx.us/elchist273_state.htm
Michigan: http://www.thegreenpapers.com/P16/MI-R
retrieved from
http://miboecfr.nictusa.com/election/results/2016PPR_CENR.html
Wisconsin: http://www.thegreenpapers.com/P16/WI-R
retrieved from
http://elections.wi.gov/elections-voting/results/2016/spring-election-presidential-preference

## 3.2 Overall results

We show below data from the US elections (Republicans, Primaries, 2016, $n = 8$), from France (Presidential elections, first round, 2007, 2012, and 2017, $n = 8$), and the Federal Republic of Germany (Federal elections 1949-2017, $n = 95$) from different organizational units (city, federal state, country). We present a semi-logarithmic representation, together with a linear fit of the data. In order to obtain a first, overall impression about the quality of the fits, we consider $R^2_{adj}$ in figure 5. We find that the linear model mostly explains far more than 90% of the variability in the data.

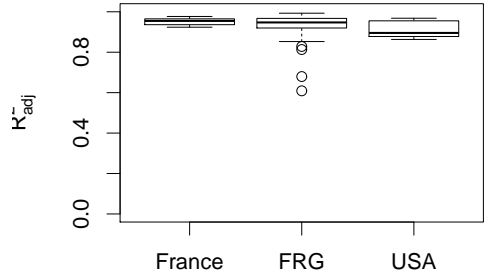

Figure 5: Boxplot for the $R^2_{adj}$ values in France, Germany and the US.

In section 5, we fit according to the heuristic estimator (proposed at page 17) the parameter $z$; the histogram of the logarithm of this parameter obtained from the elections in the FRG (in each election, for two cities Munich and Stuttgart, two states Bavaria and Baden Wüttemberg, and for the complete FRG) is shown in figure 6. We find an unimodal distribution. A closer analysis of the dependency of the number of voters and number of parties for the unit under consideration reveals a significant, but weak dependency ($R^2_{adj} = 0.4$; see table in figure 6).

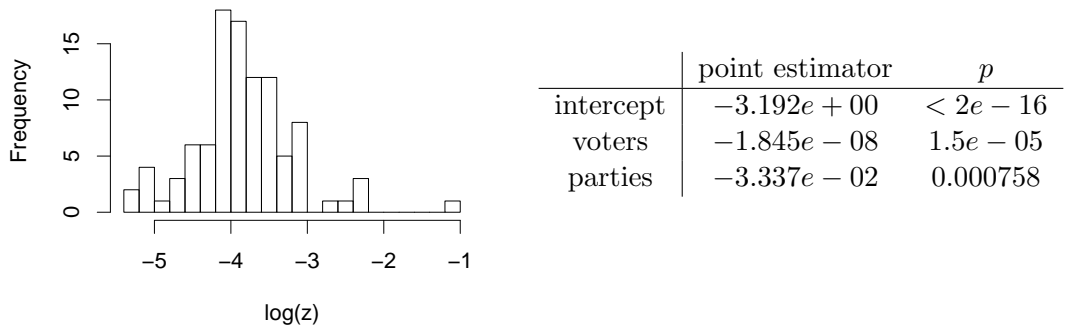

Figure 6: Left: Logarithmic histogram of the parameter $z$. Right: Result of a linear fit of $\log(z)$ by the number of voters $n$.

# 4 Elections in semilogarithmic representation with linear fit

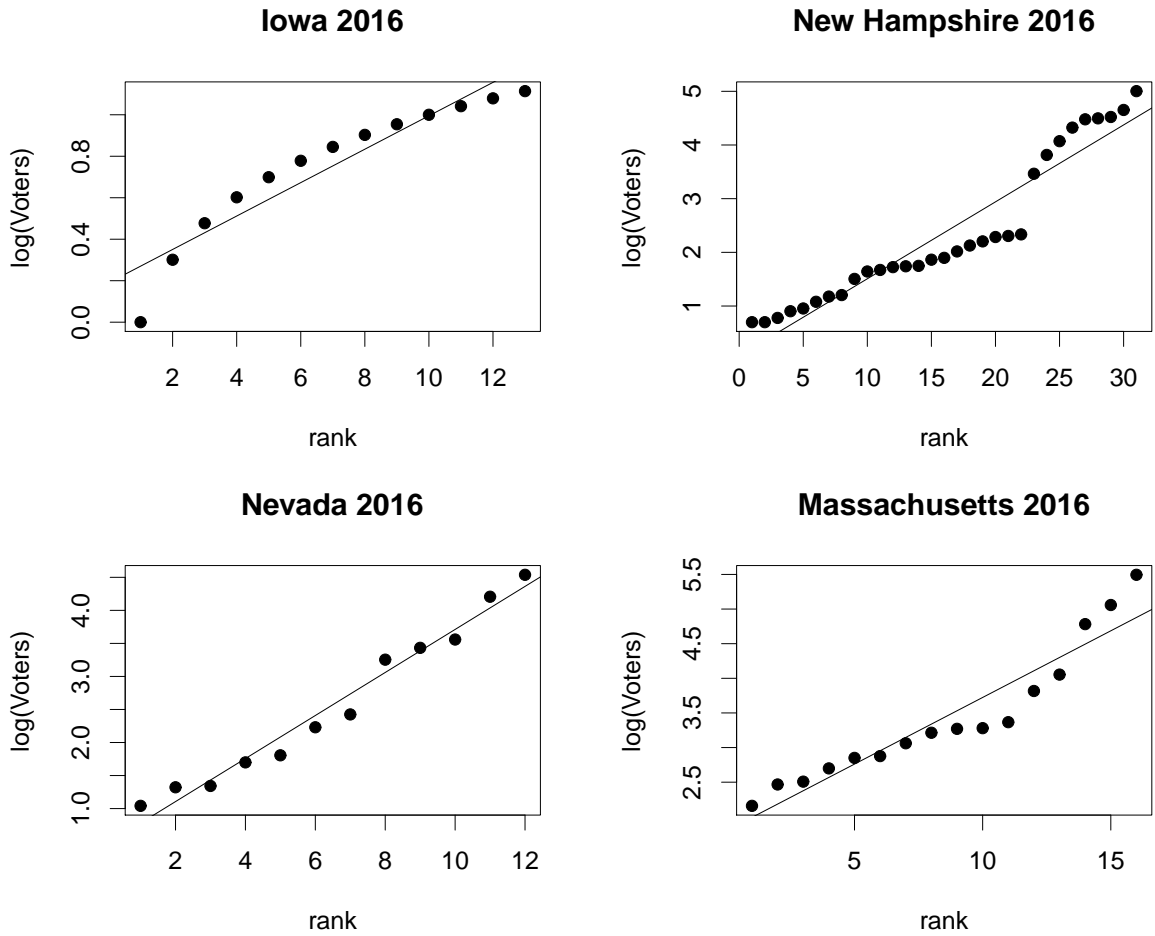

Figure 7: Election US (republican primaries), 2016 (bullets: data, line: linear fit).

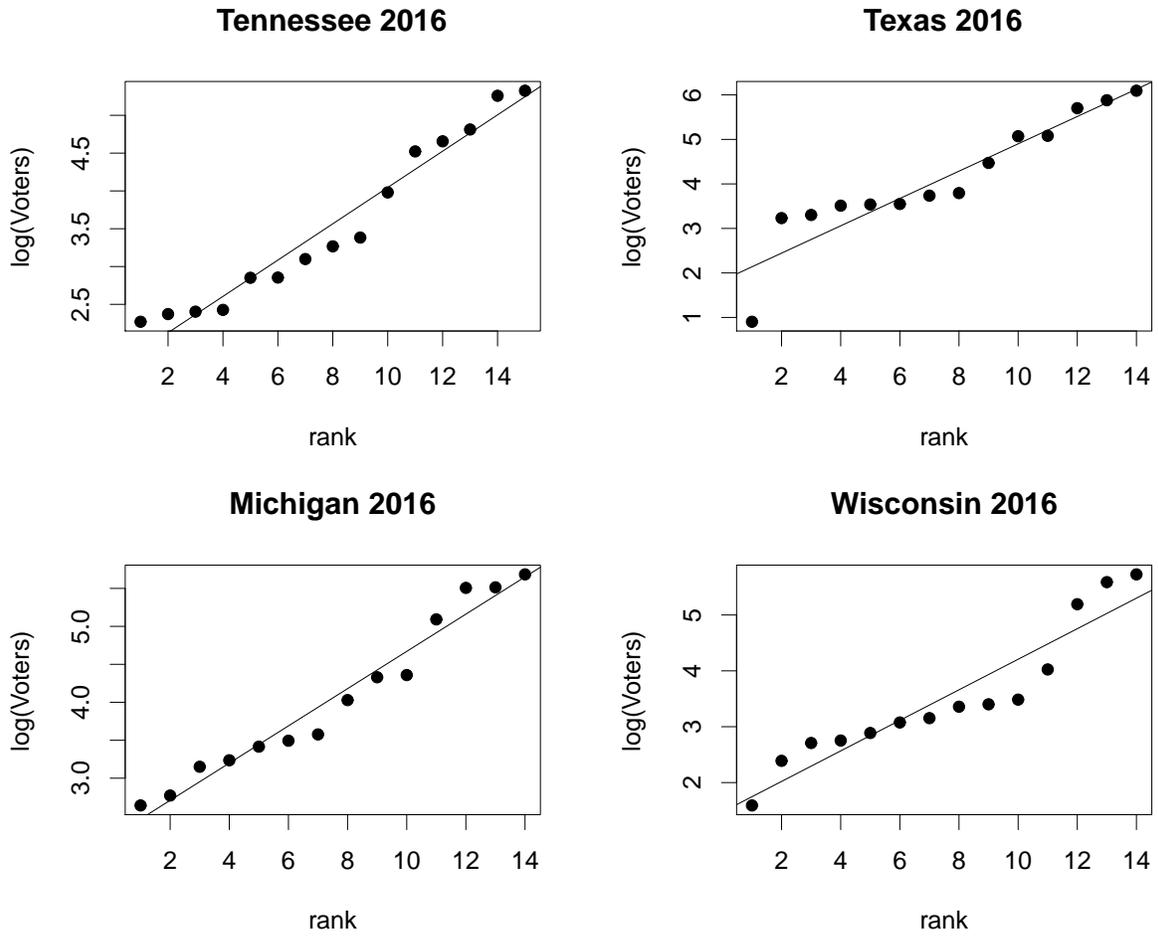

Figure 8: Election US (republican primaries), 2016 (bullets: data, line: linear fit).

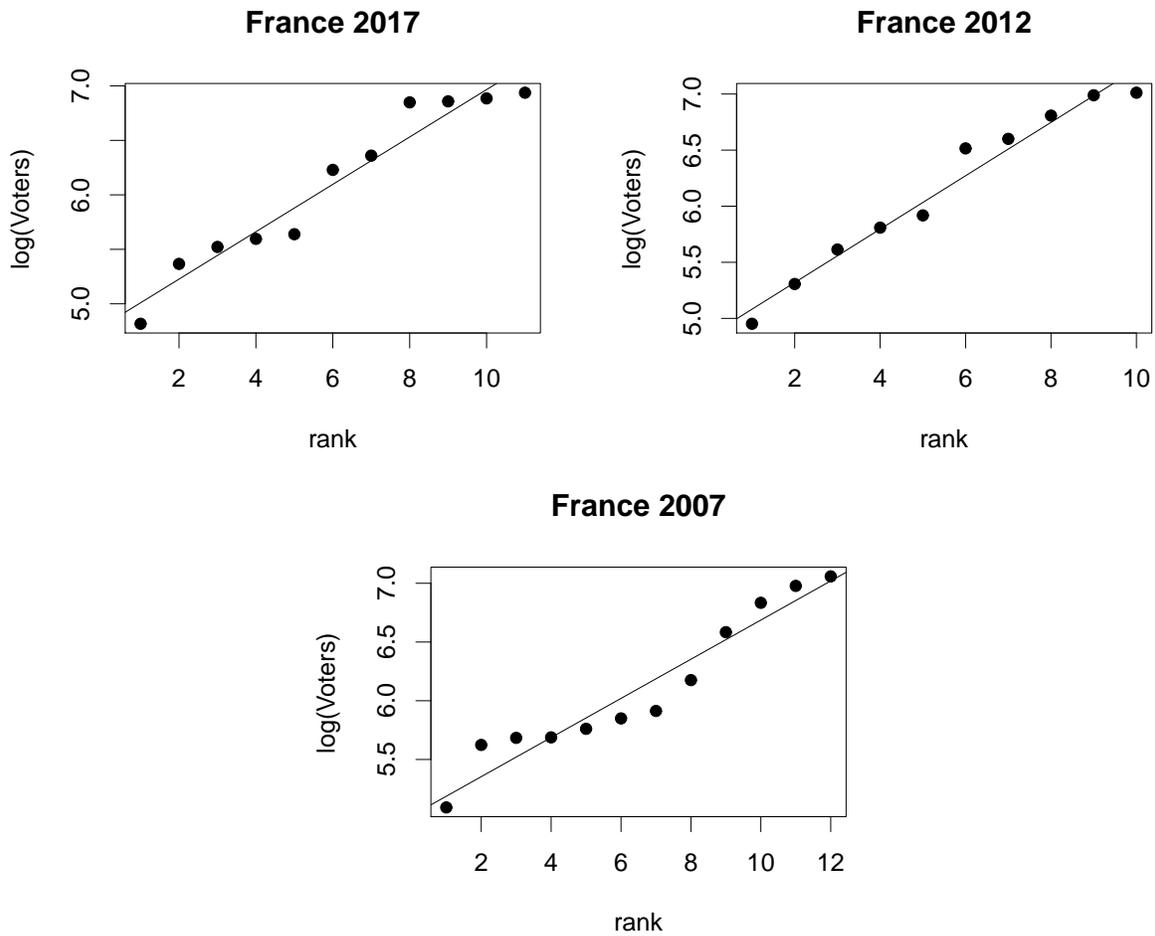

Figure 9: Election France, 2017, 1012, 2007 (bullets: data, line: linear fit).

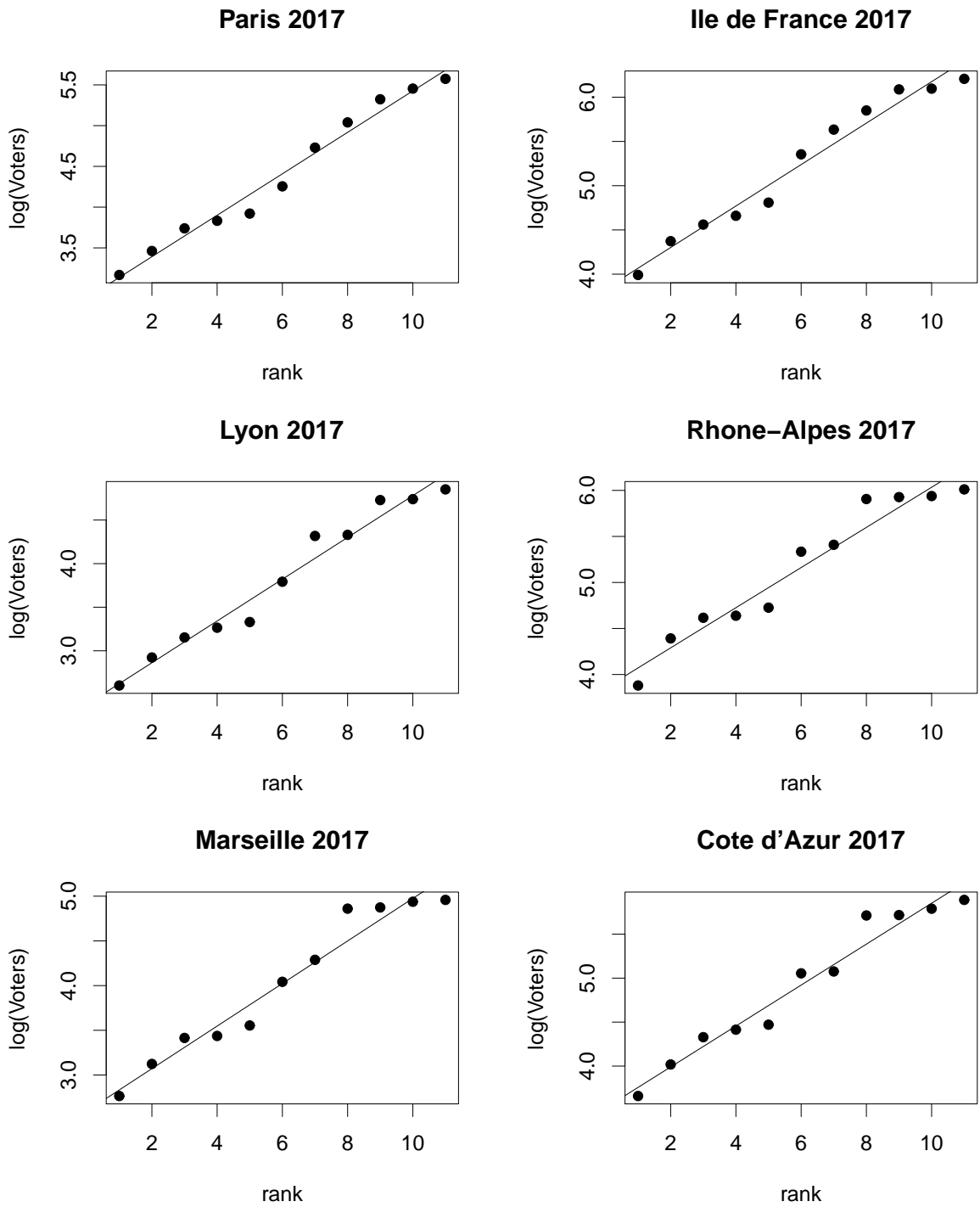

Figure 10: Election France, 2017 (bullets: data, line: linear fit).

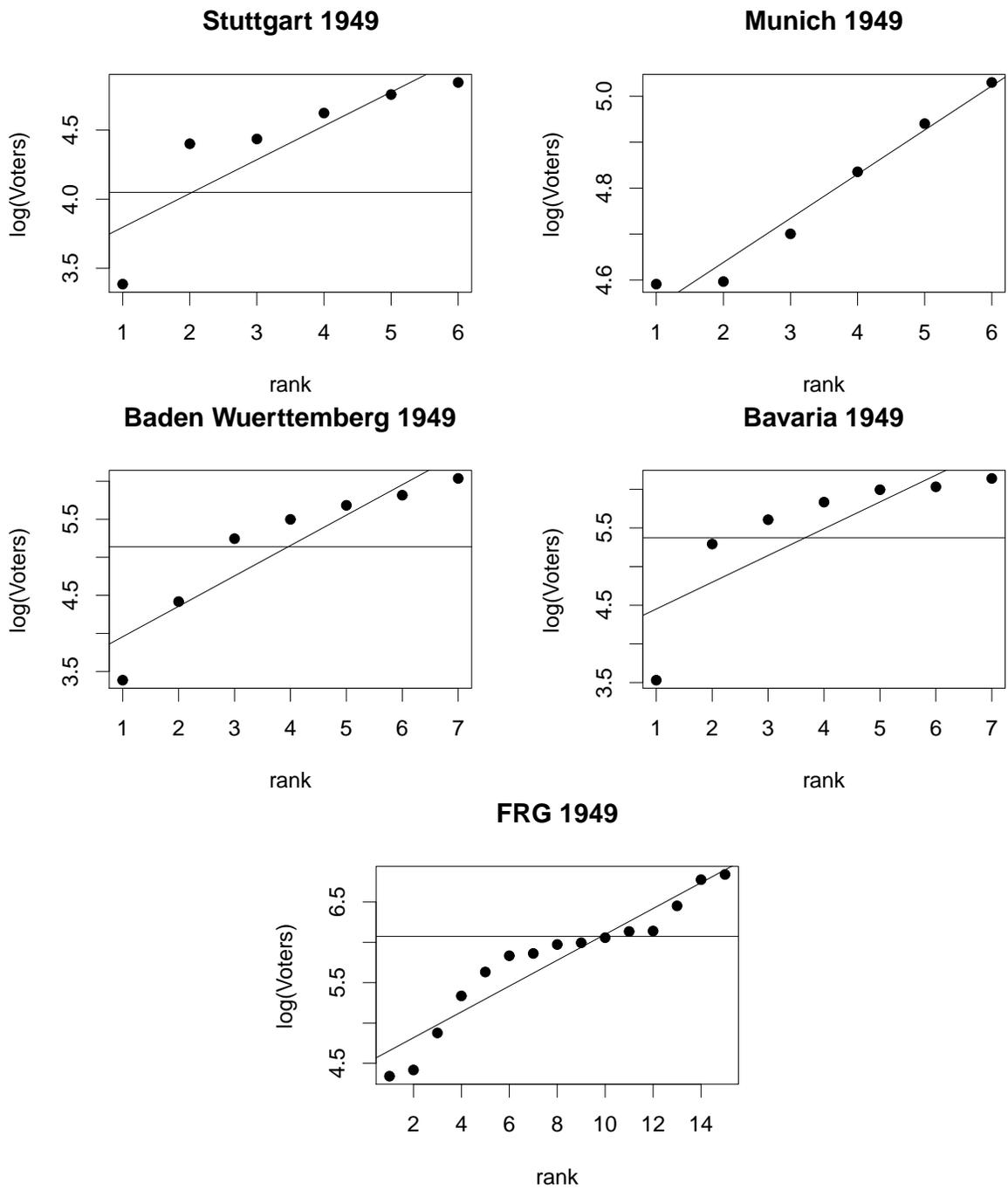

Figure 11: Election FRG, 1949 (bullets: data, line: linear fit).

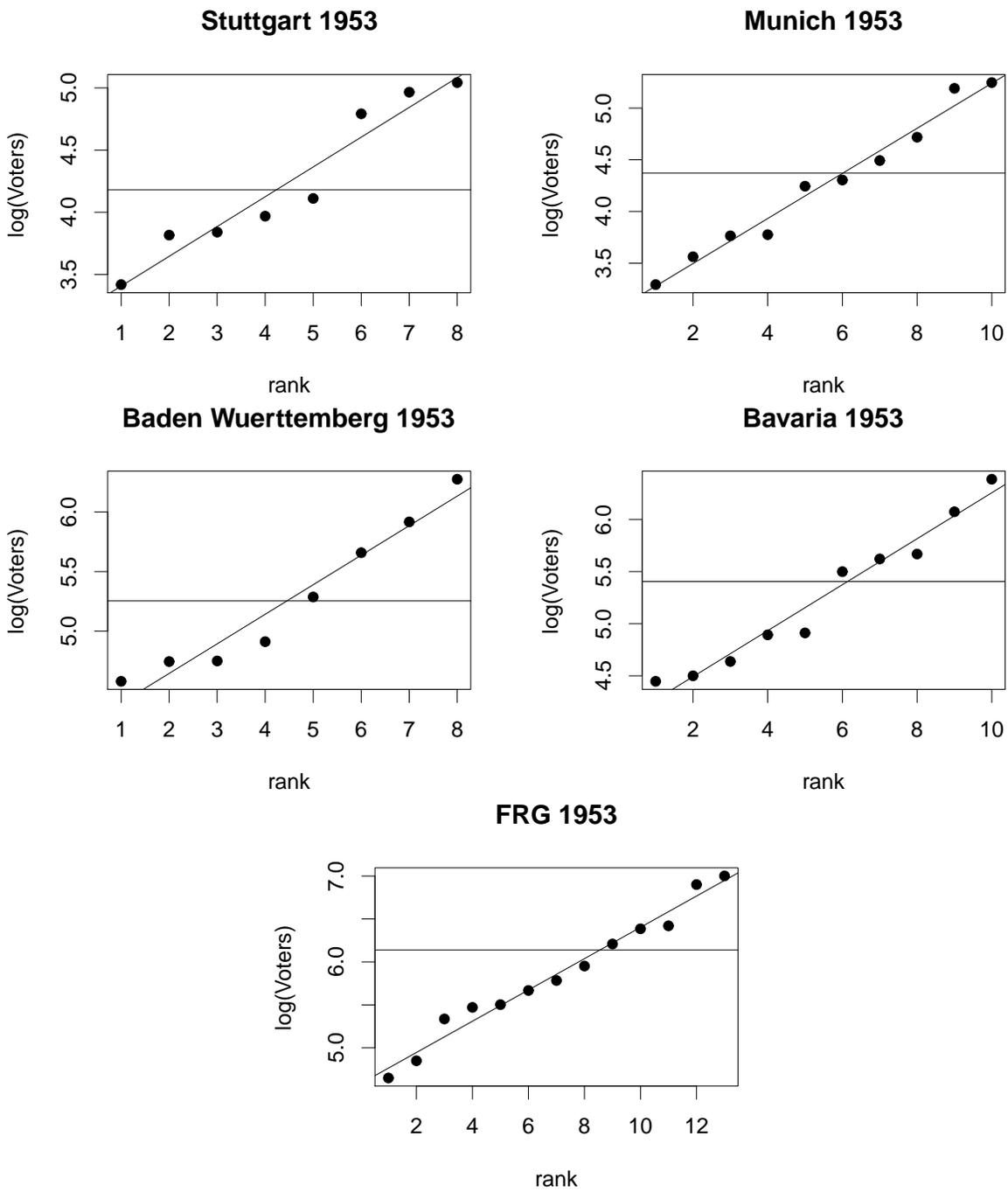

Figure 12: Election FRG, 1953 (bullets: data, line: linear fit).

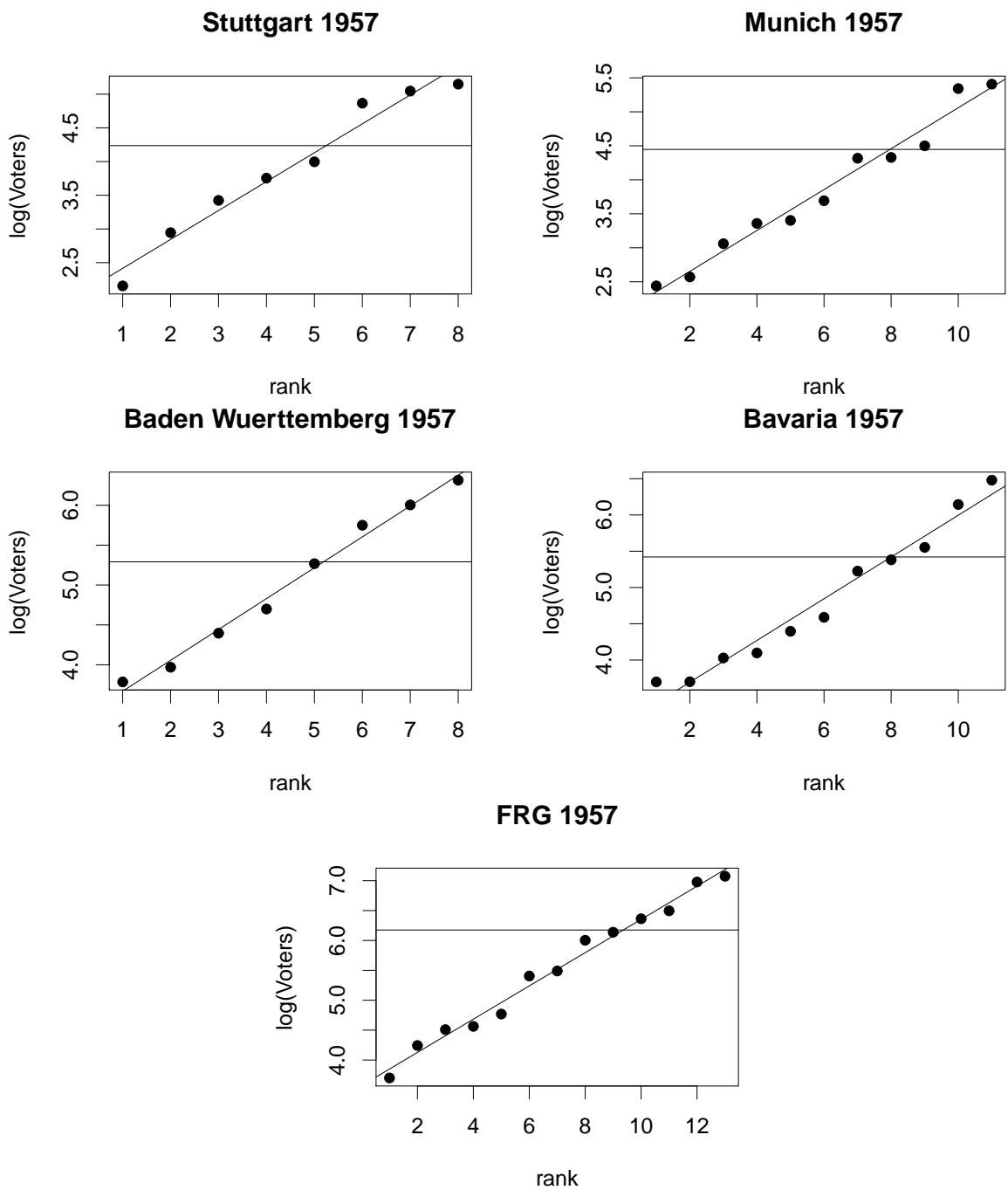

Figure 13: Election FRG, 1957 (bullets: data, line: linear fit).

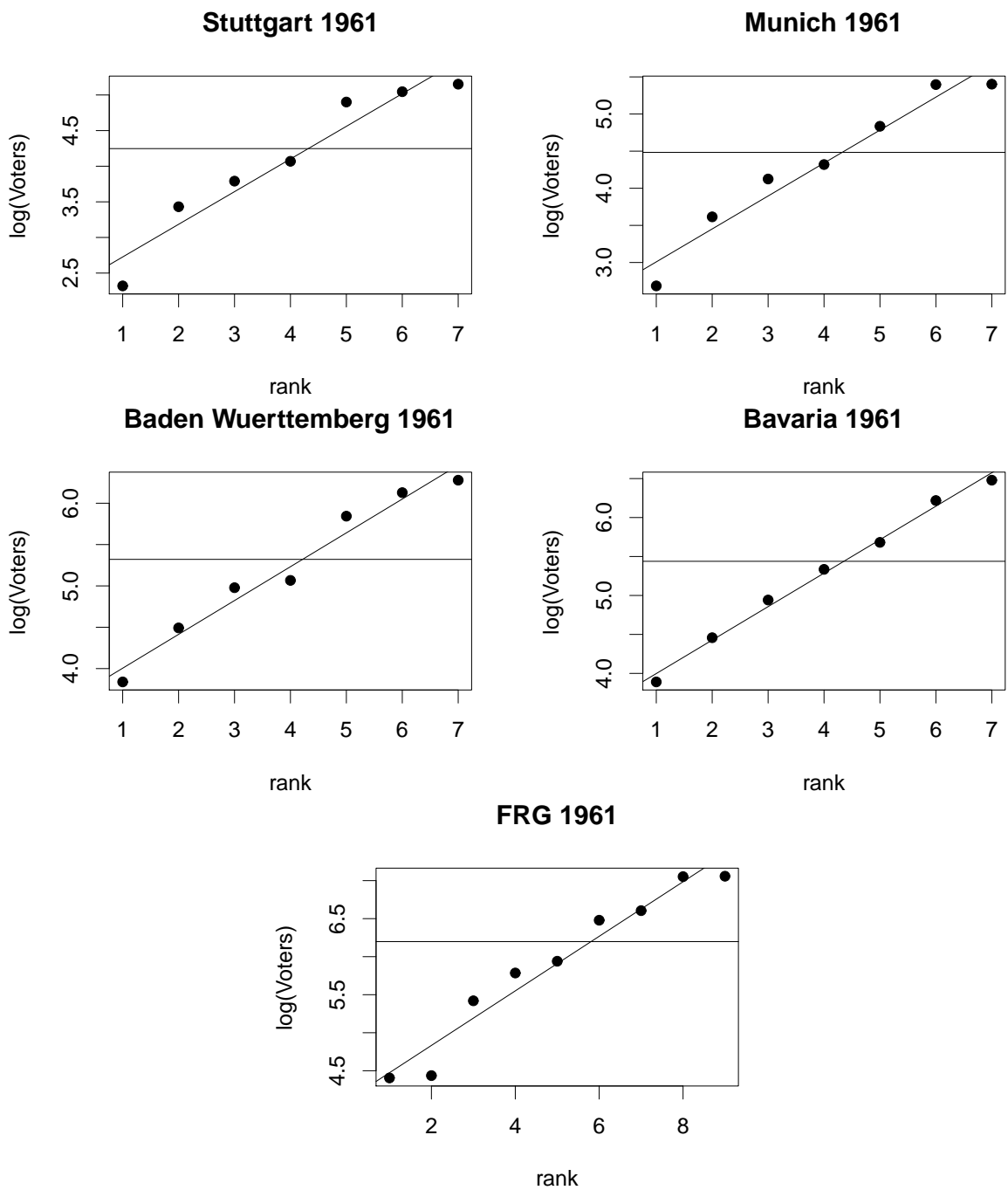

Figure 14: Election FRG, 1961 (bullets: data, line: linear fit).

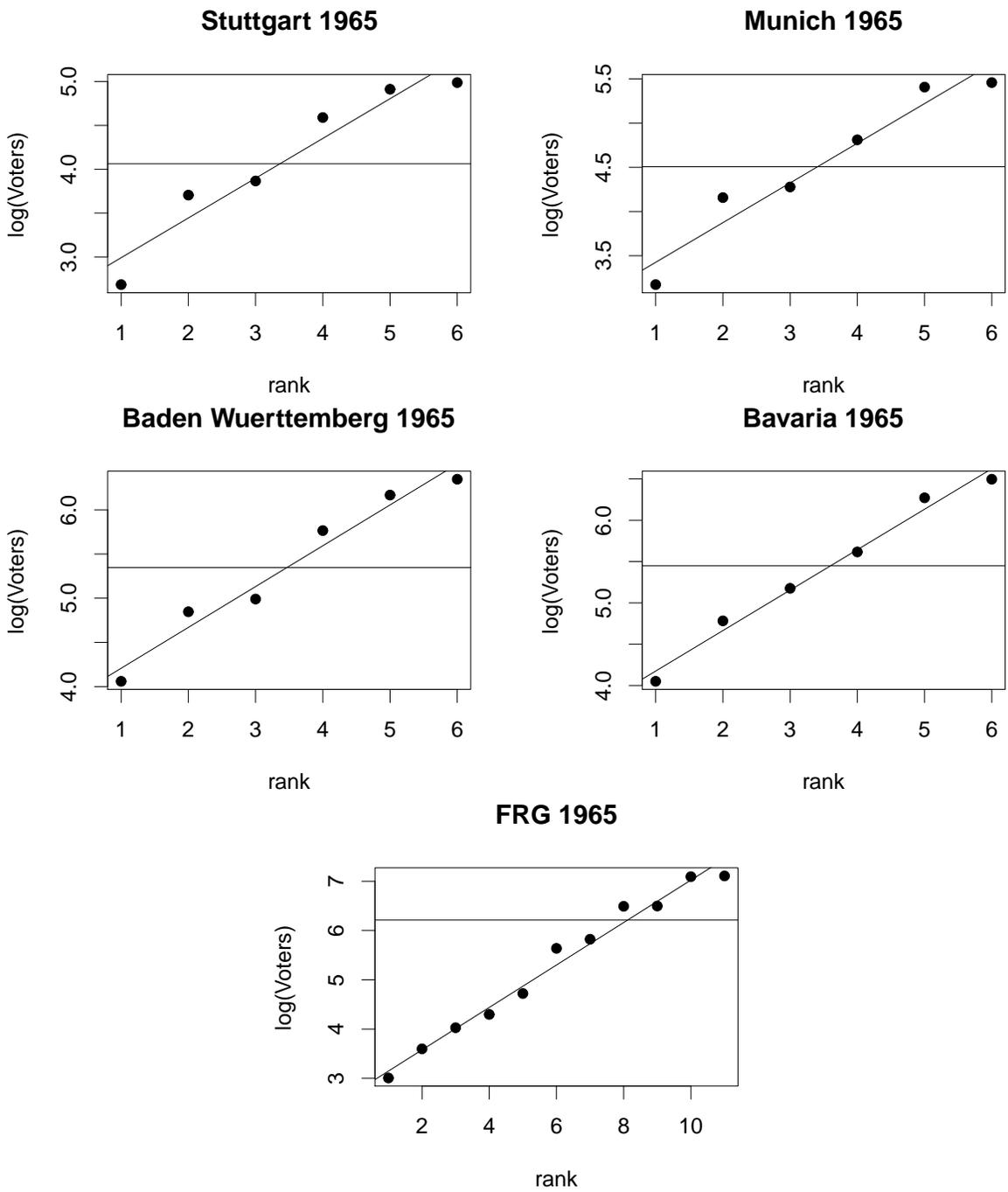

Figure 15: Election FRG, 1965 (bullets: data, line: linear fit).

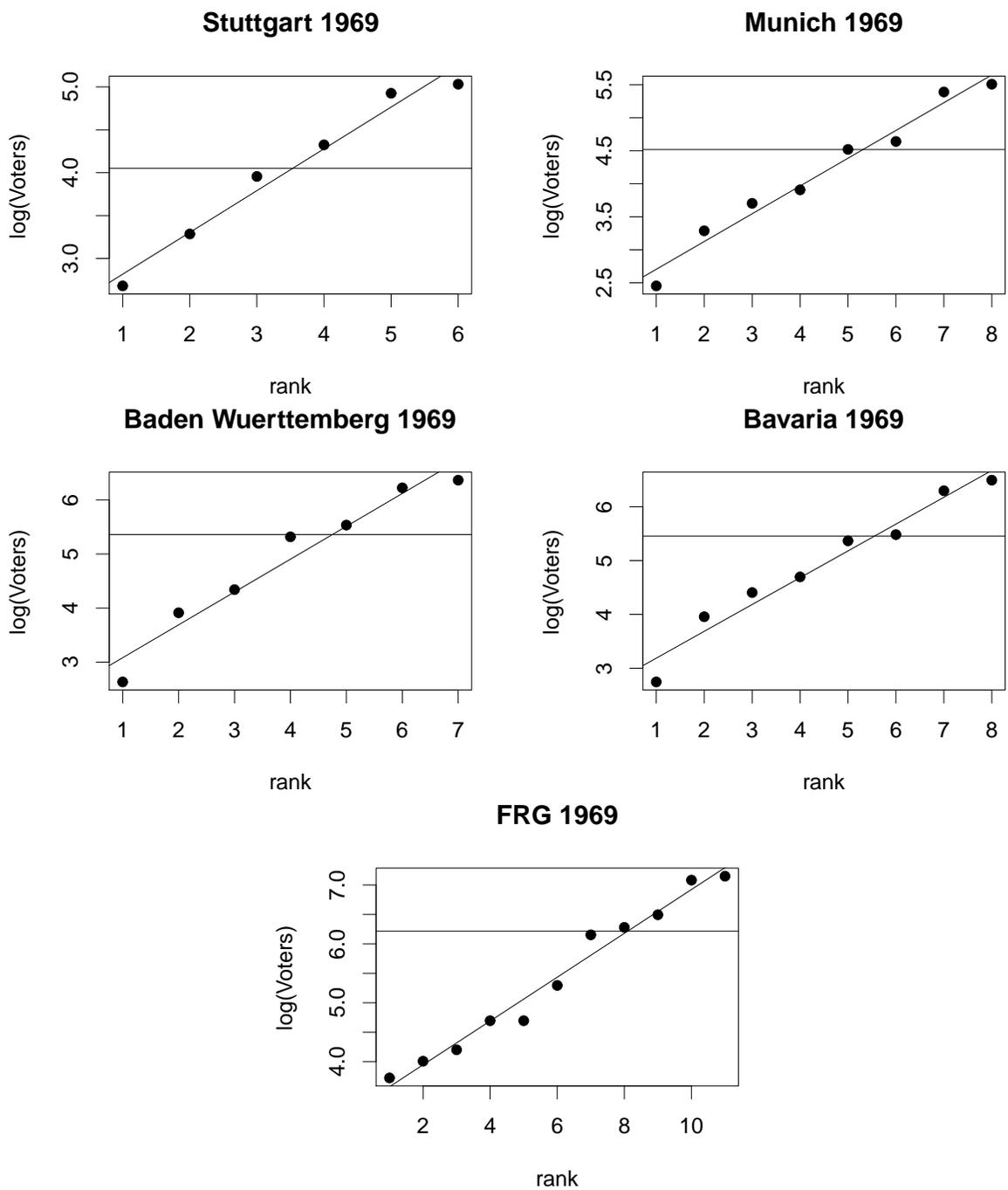

Figure 16: Election FRG, 1969 (bullets: data, line: linear fit).

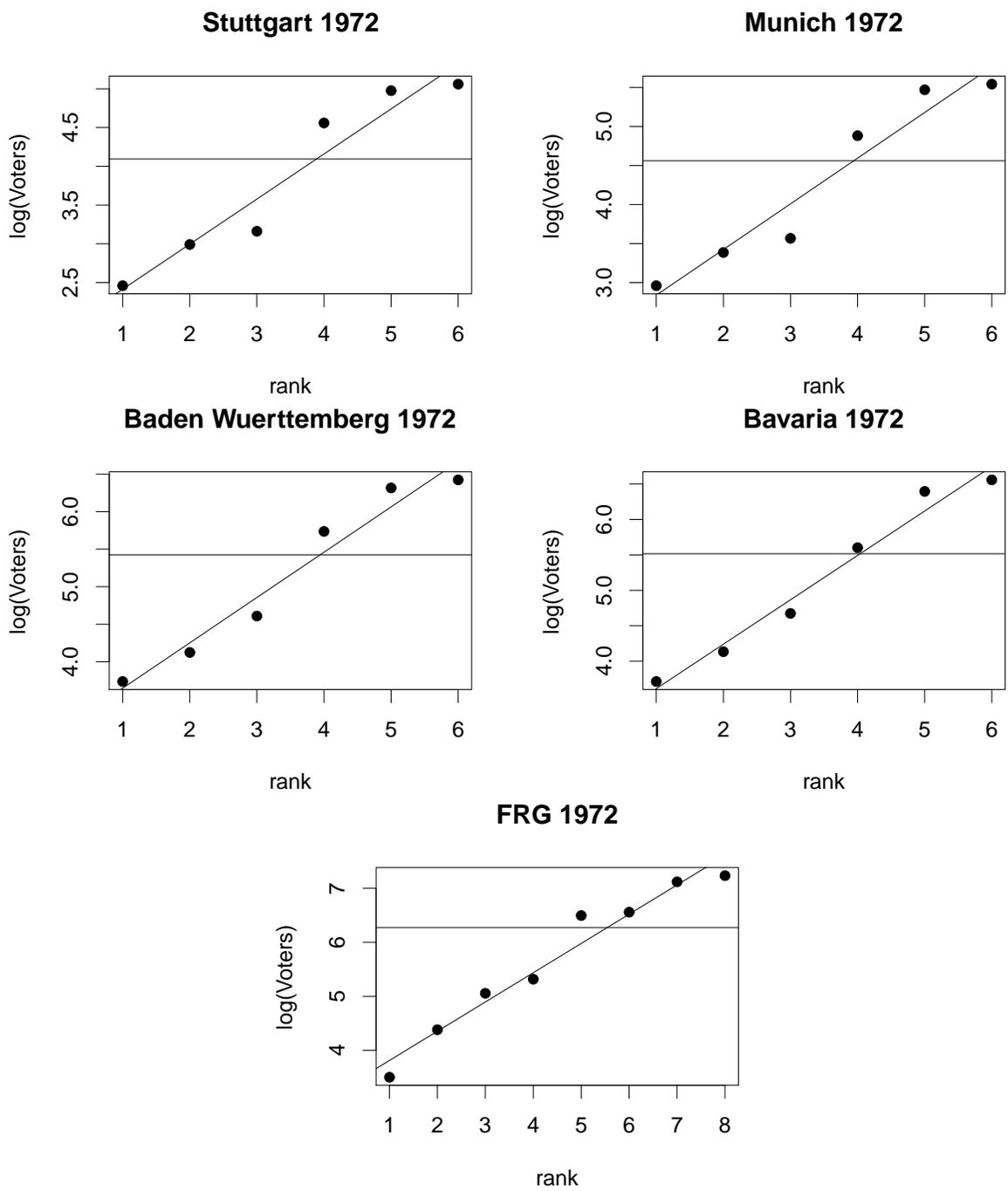

Figure 17: Election FRG, 1972 (bullets: data, line: linear fit).

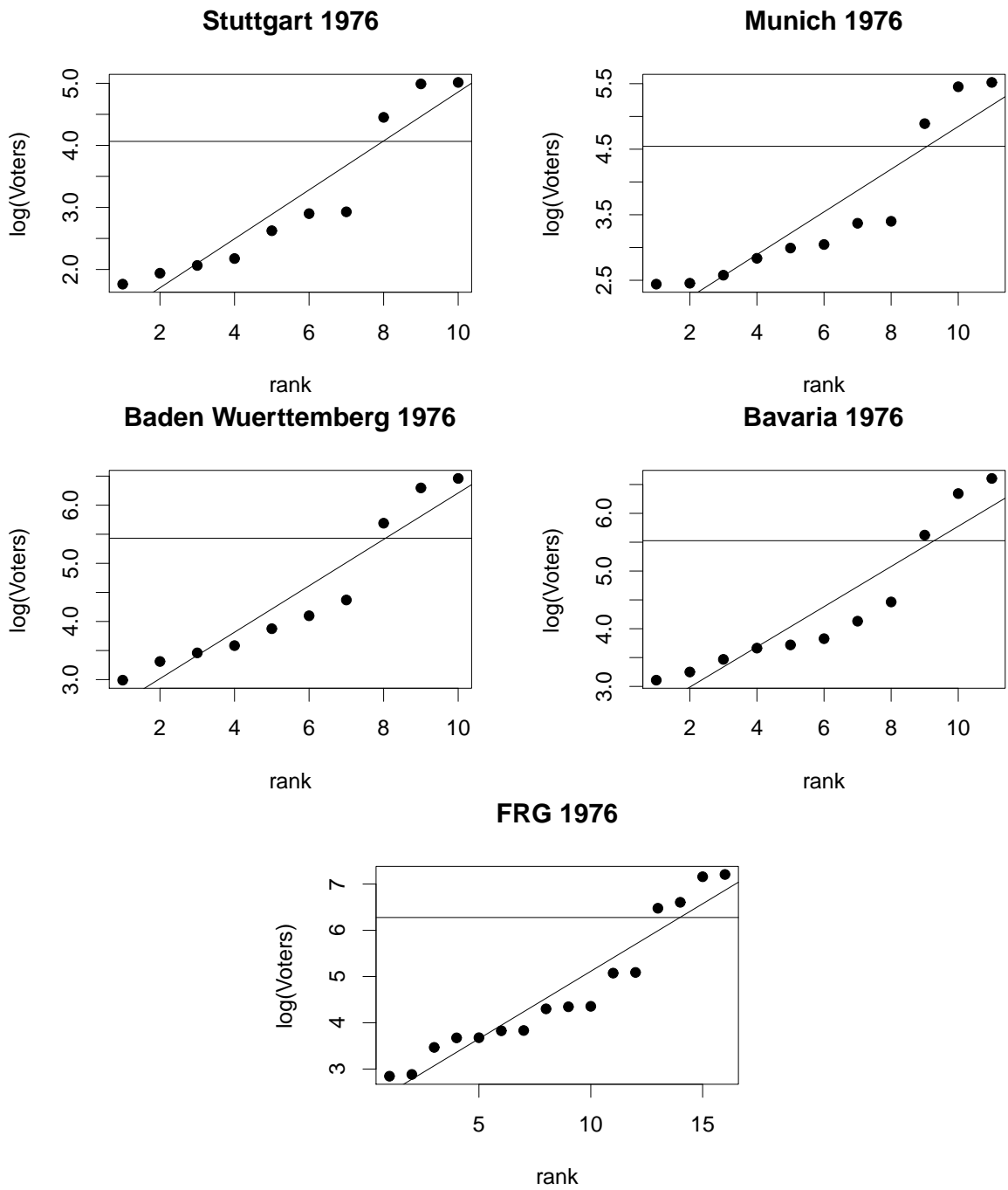

Figure 18: Election FRG, 1976 (bullets: data, line: linear fit).

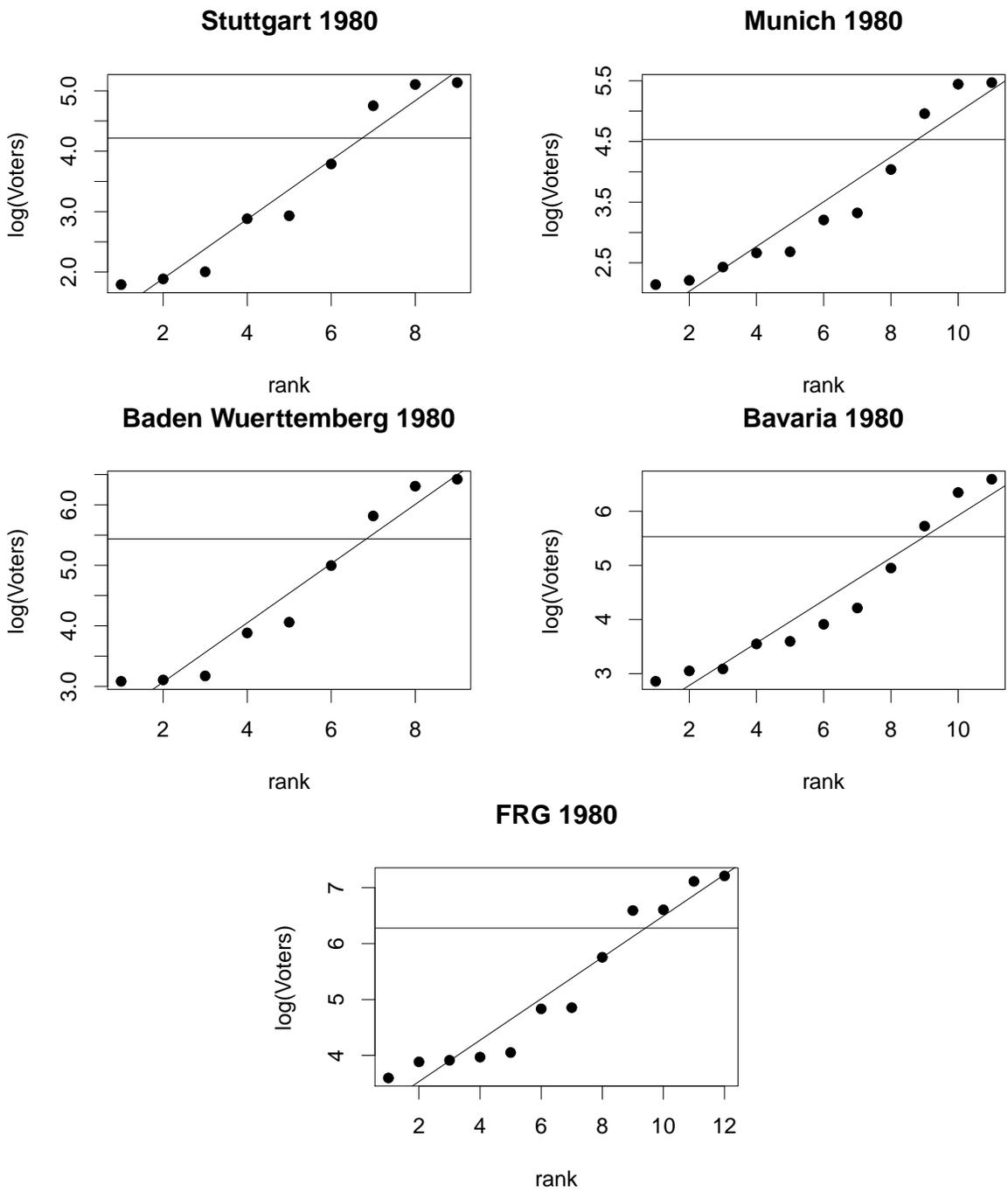

Figure 19: Election FRG, 1980 (bullets: data, line: linear fit).

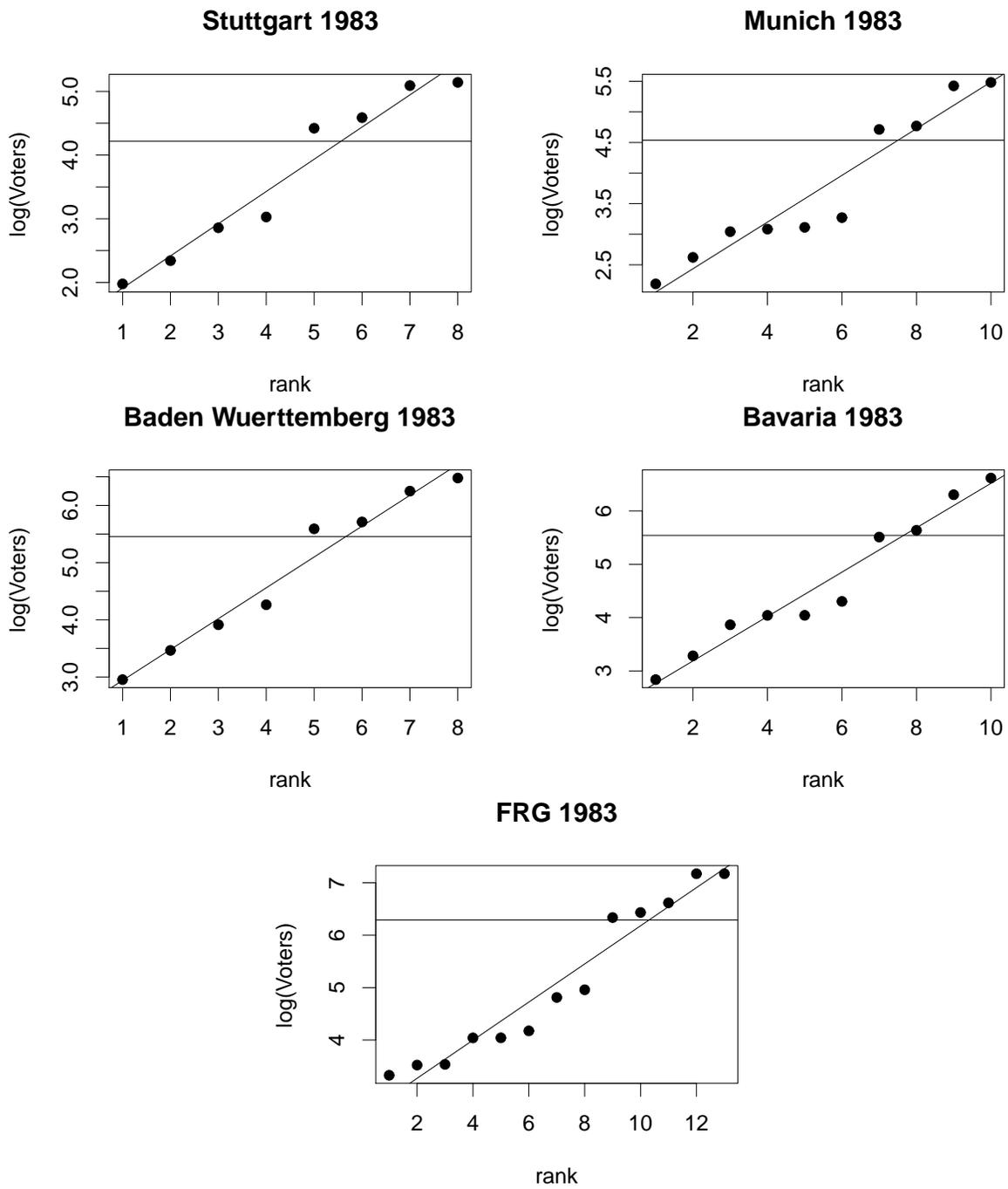

Figure 20: Election FRG, 1983 (bullets: data, line: linear fit).

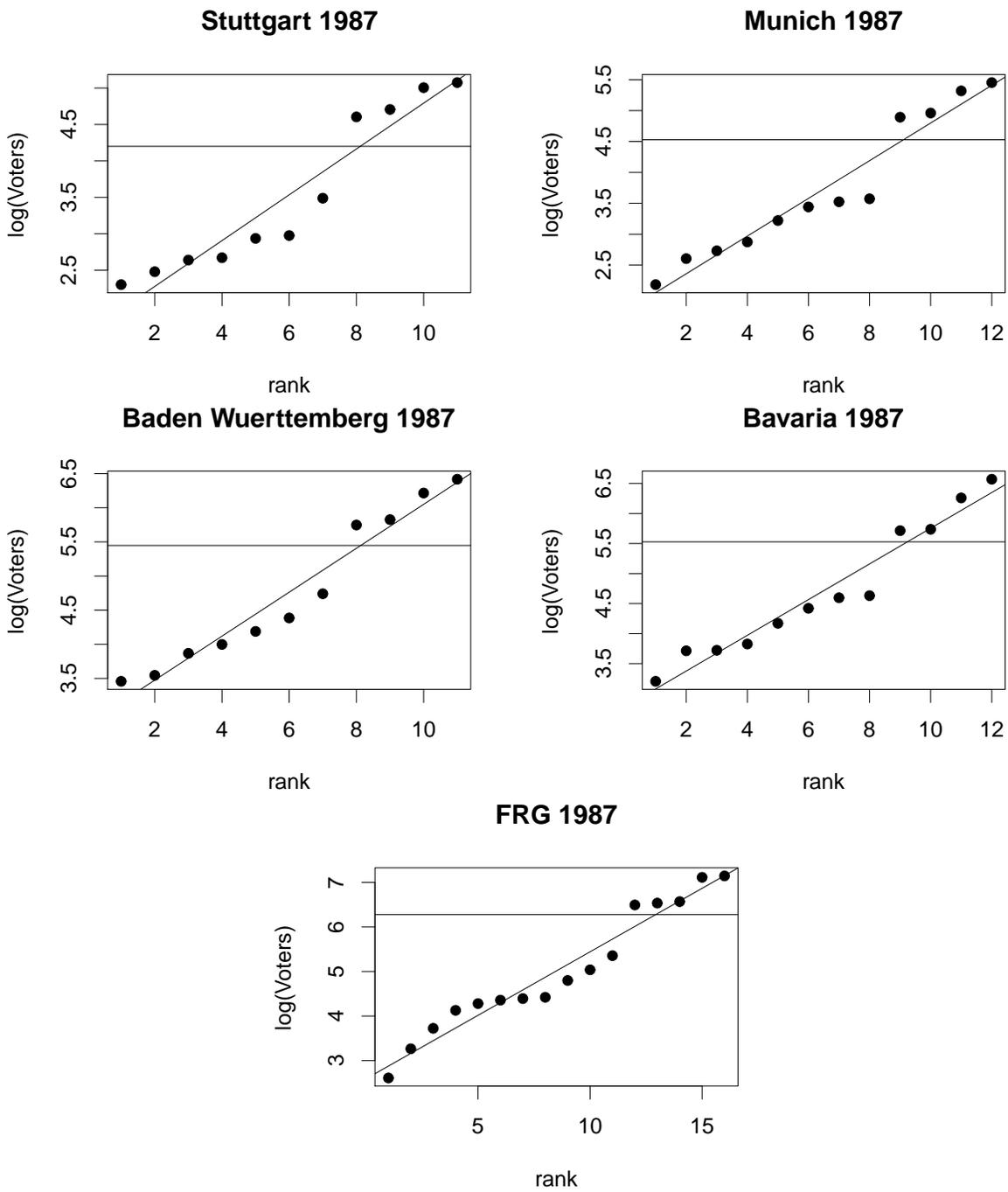

Figure 21: Election FRG, 1987 (bullets: data, line: linear fit).

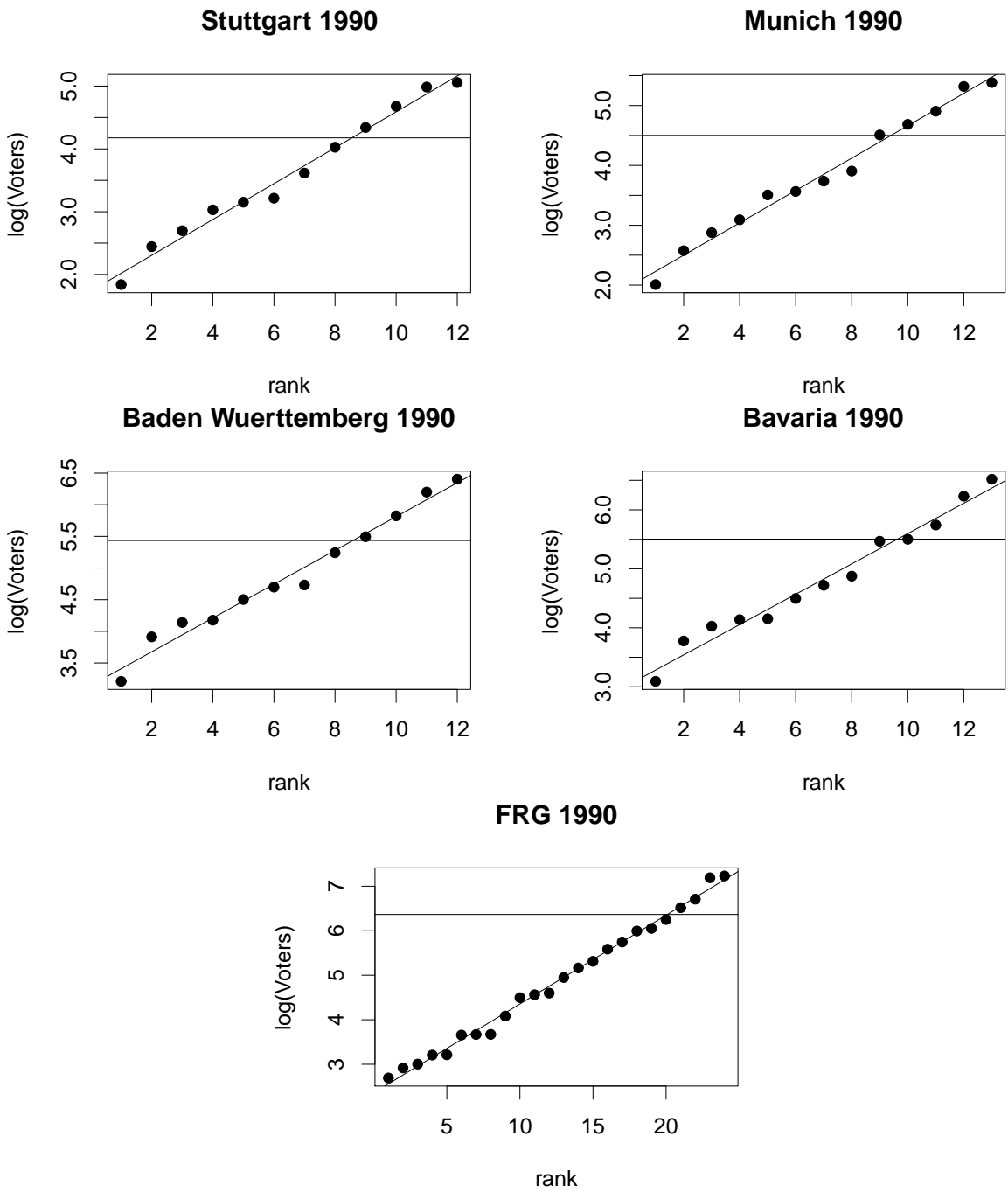

Figure 22: Election FRG, 1990 (bullets: data, line: linear fit).

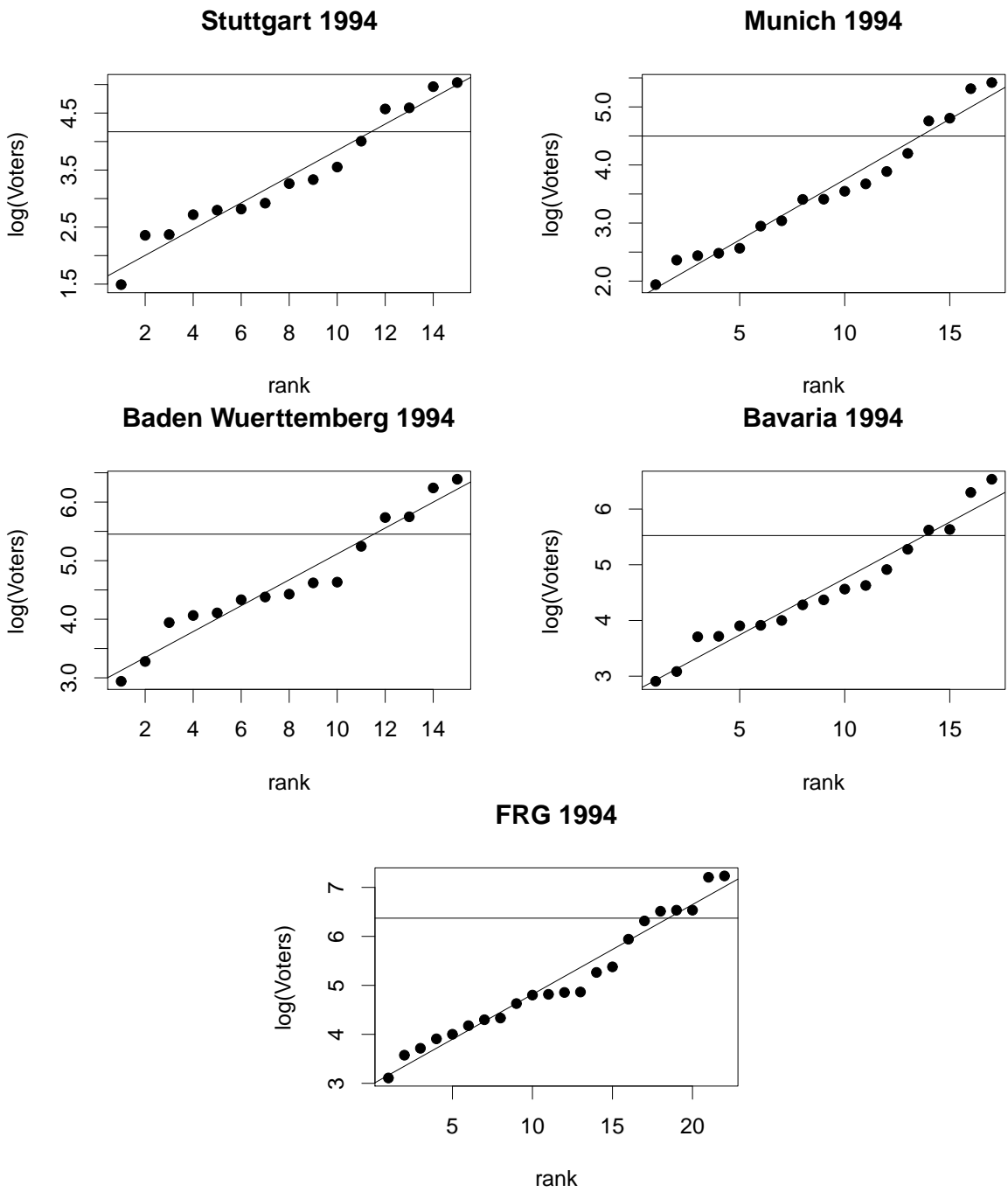

Figure 23: Election FRG, 1994 (bullets: data, line: linear fit).

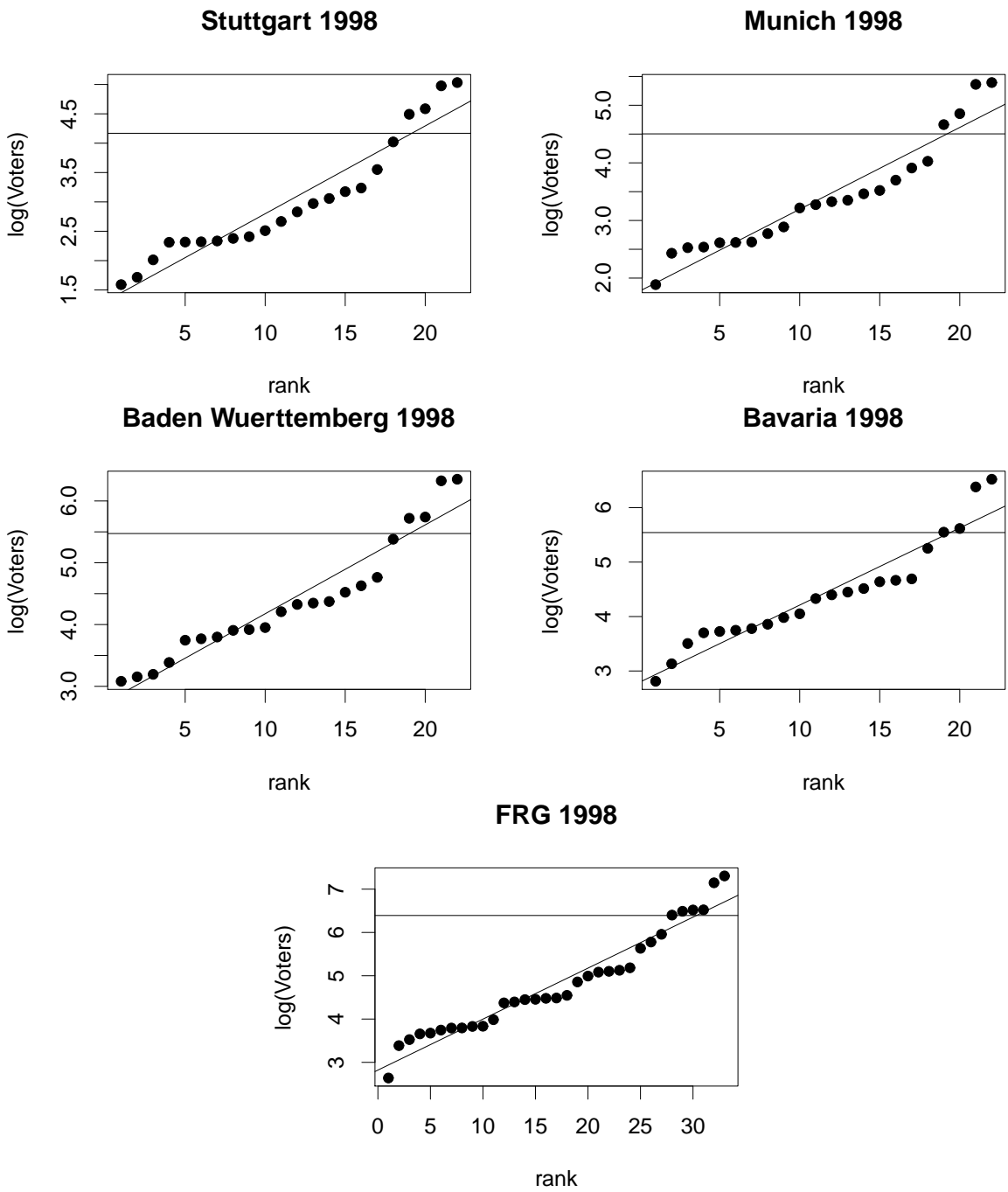

Figure 24: Election FRG, 1998 (bullets: data, line: linear fit).

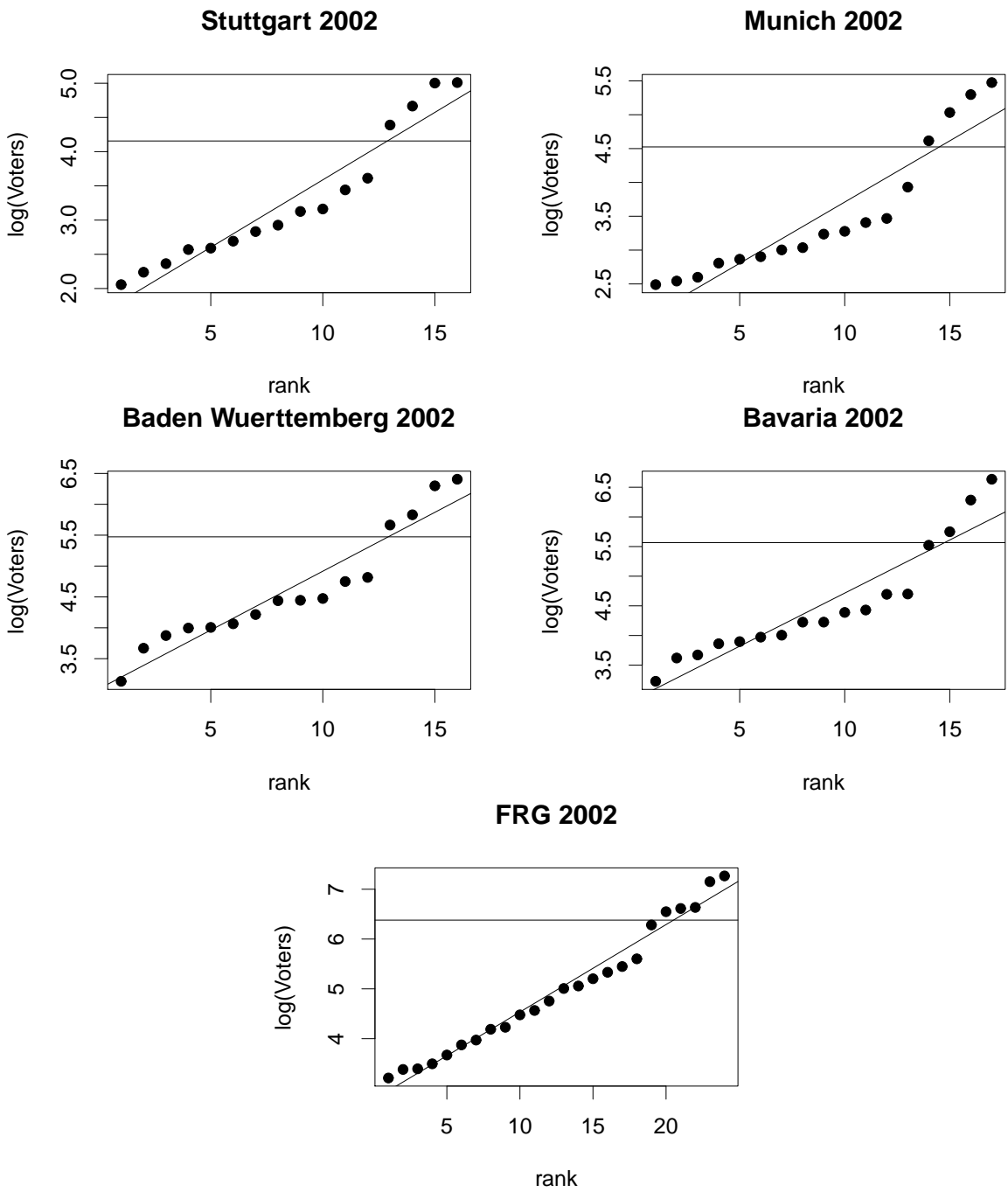

Figure 25: Election FRG, 2002 (bullets: data, line: linear fit).

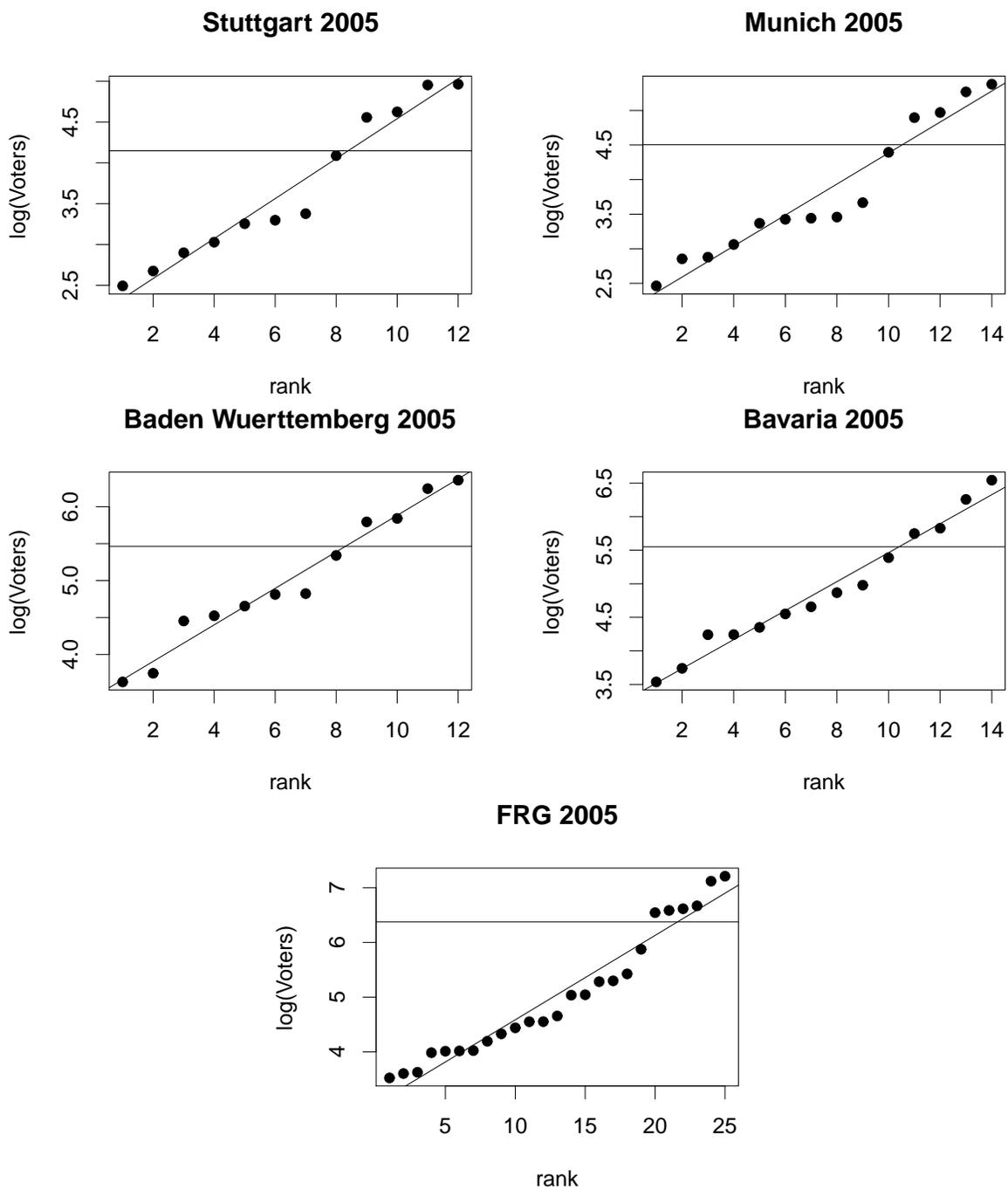

Figure 26: Election FRG, 2005 (bullets: data, line: linear fit).

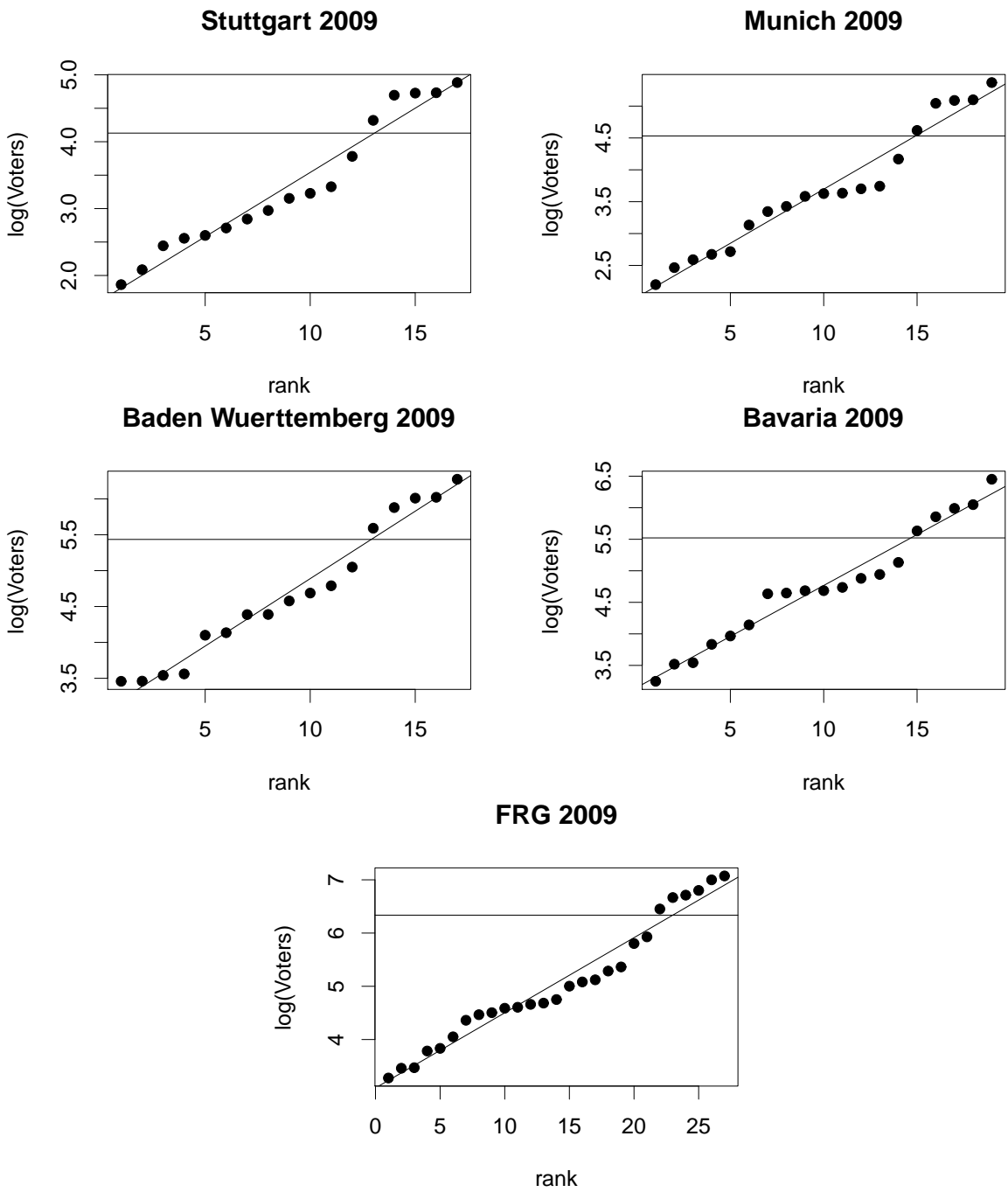

Figure 27: Election FRG, 2009 (bullets: data, line: linear fit).

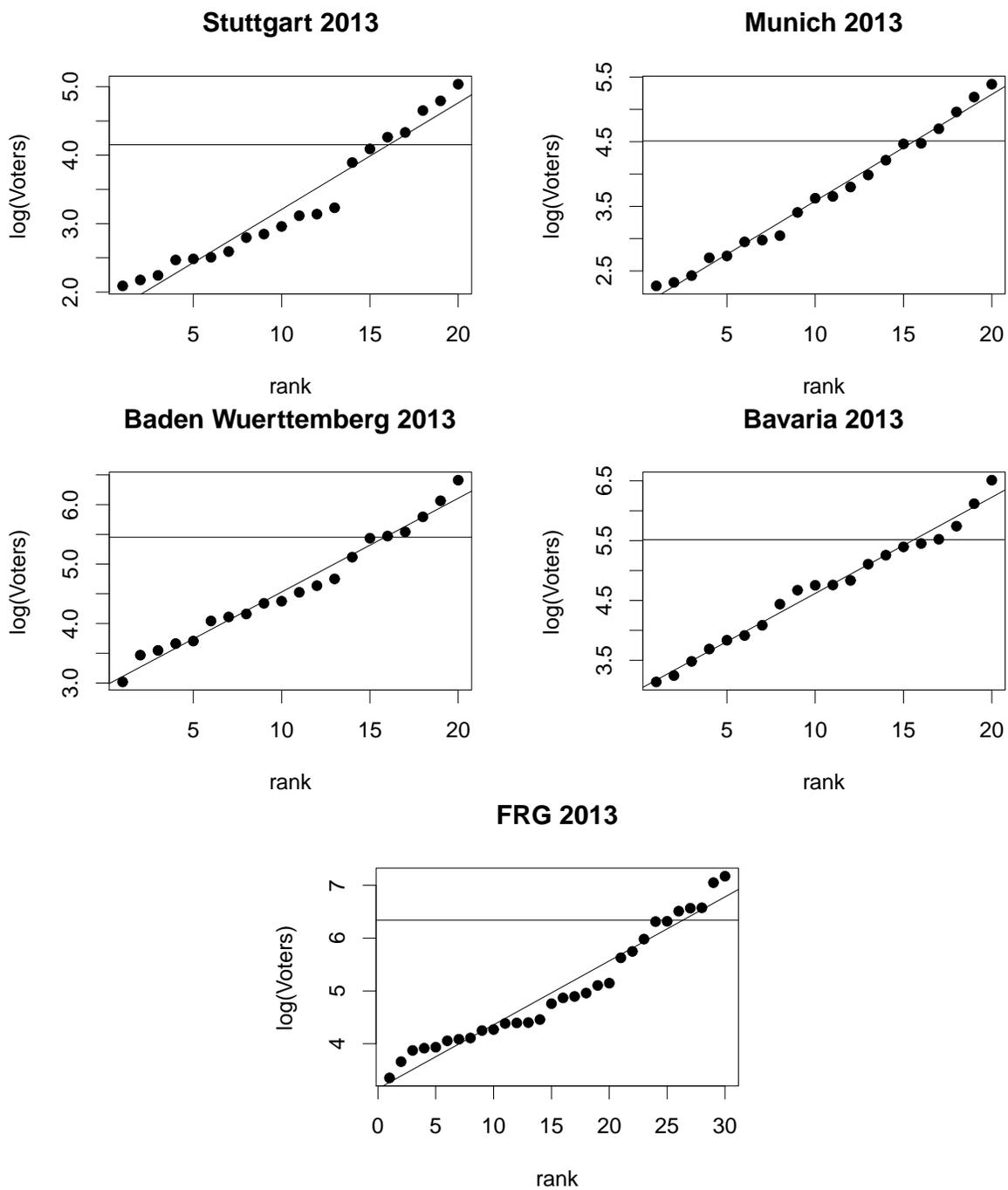

Figure 28: Election FRG, 2013 (bullets: data, line: linear fit).

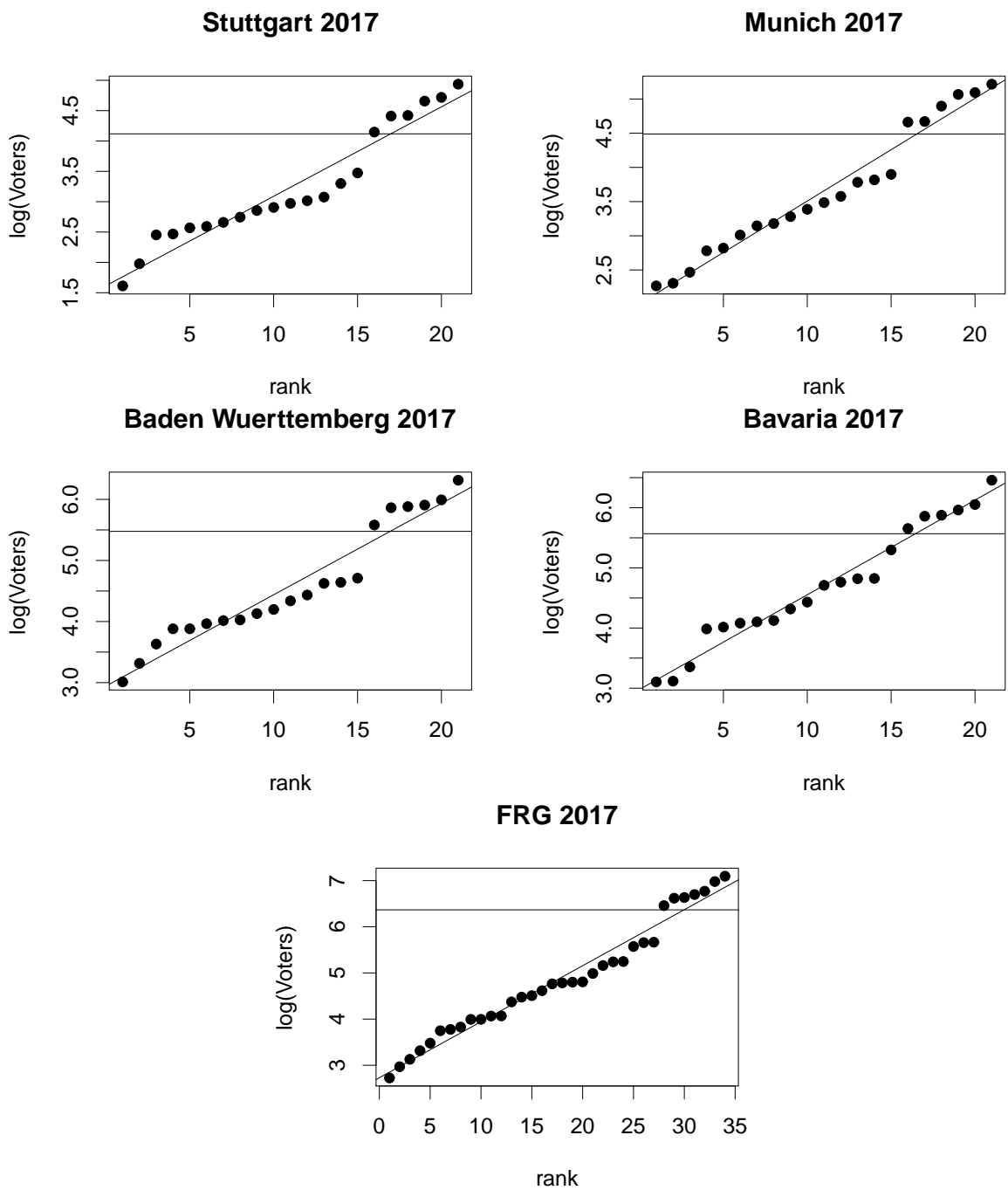

Figure 29: Election FRG, 2017 (bullets: data, line: linear fit).

# 5 Elections in semilogarithmic representation with model simulations

In this section, we show a boxplot of 100 realization of the model ($n$, $K$ and $z$ adapted) together with the data from the corresponding election according to the algorithm described in section 2.4.4. The number of voters $n$ and the number of candidates/parties $K$ are directly taken from the data, the relative minimal group size $z$ is estimated according to the estimator described at page 17 (this supplement).

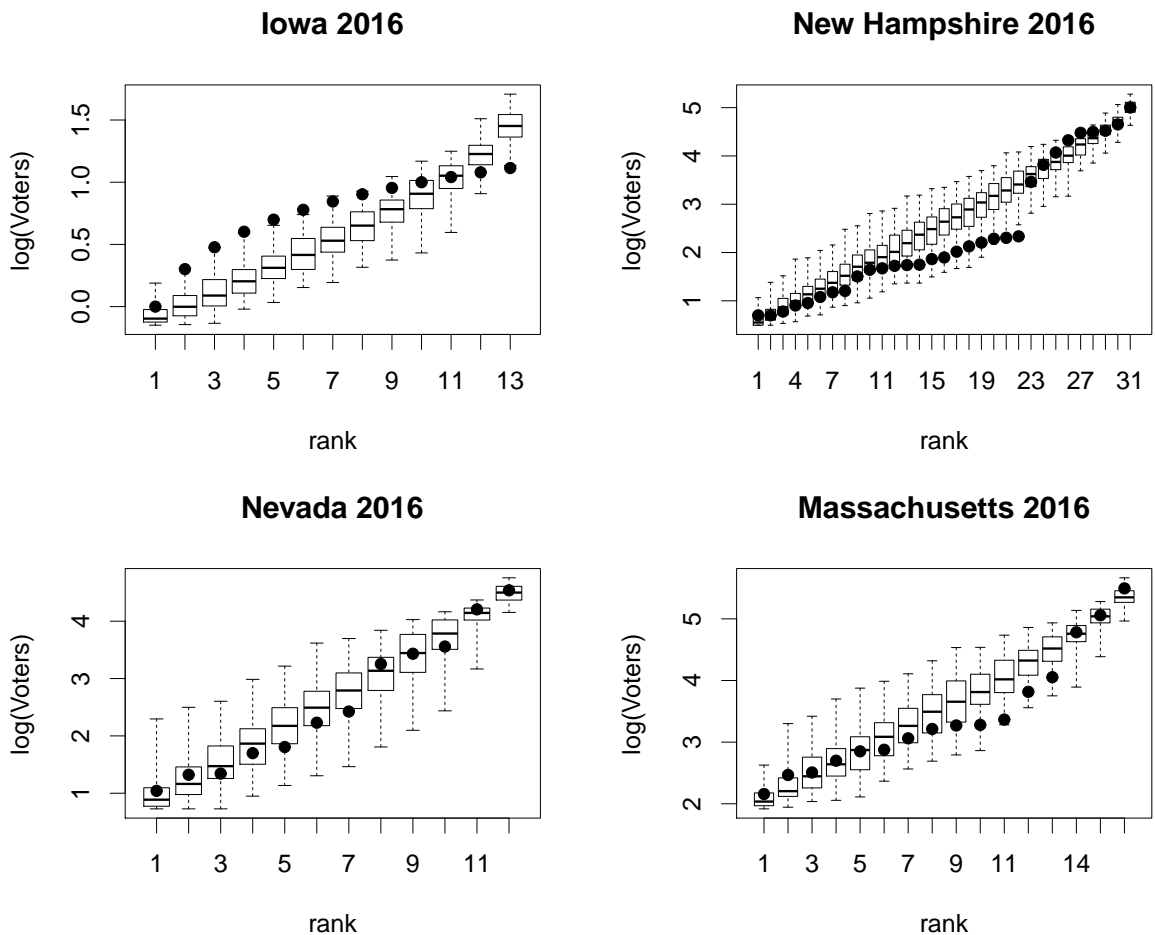

Figure 30: Election US (republicans), 2016 (boxplot of 100 realizations og the model, bullets: data).

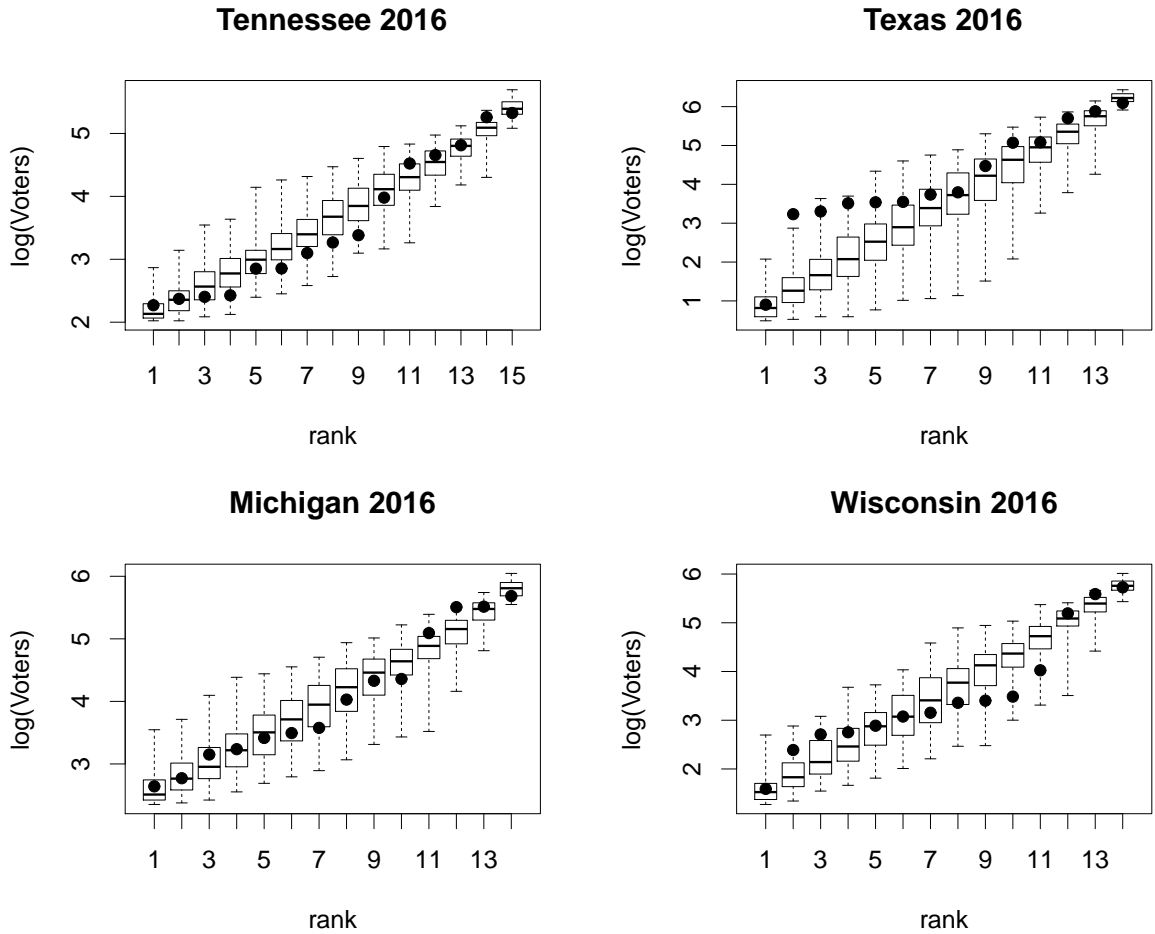

Figure 31: Election US (republicans), 2016. Note that, in case of Texas, $z$ was estimated according to an outlier (boxplot of 100 realizations og the model, bullets: data).

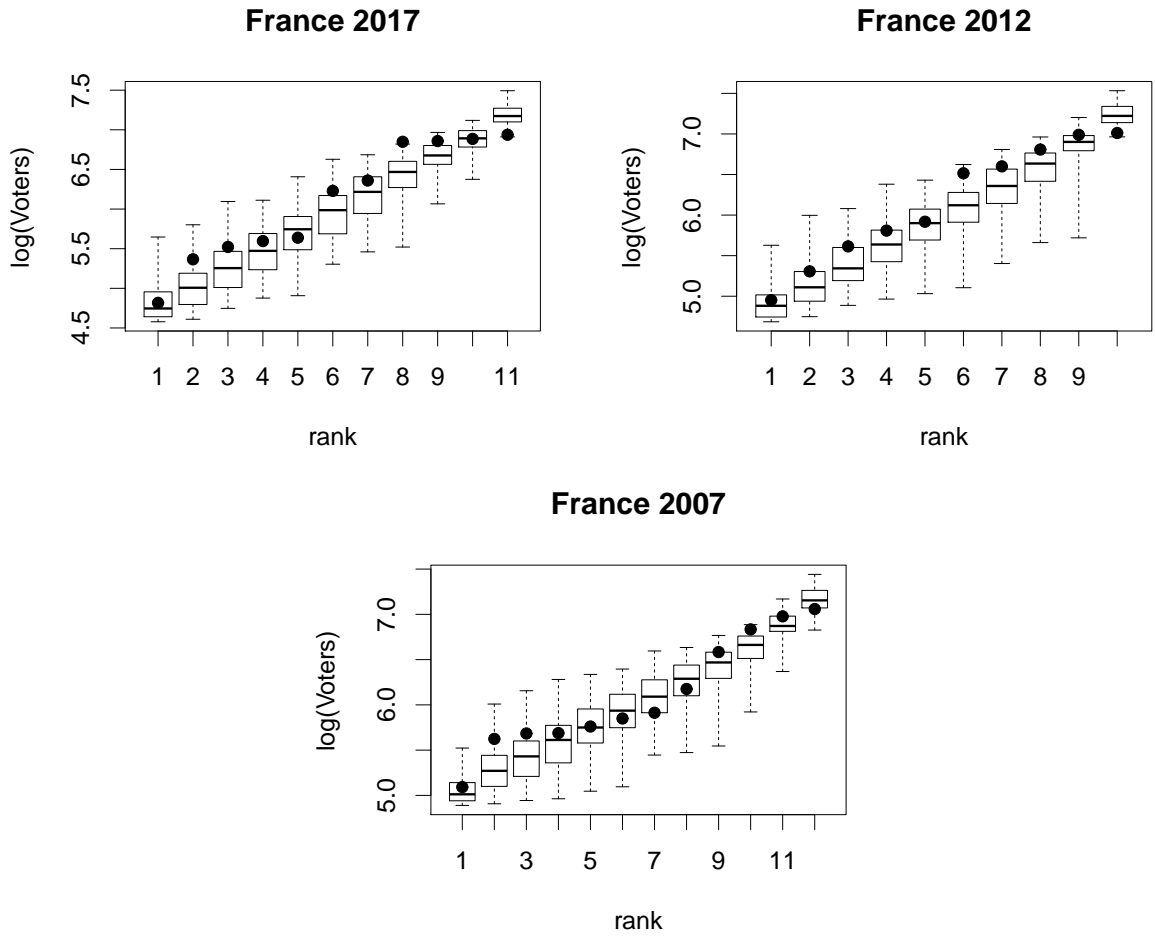

Figure 32: Election France, 2017, 1012, 2007 (boxplot of 100 realizations og the model, bullets: data).

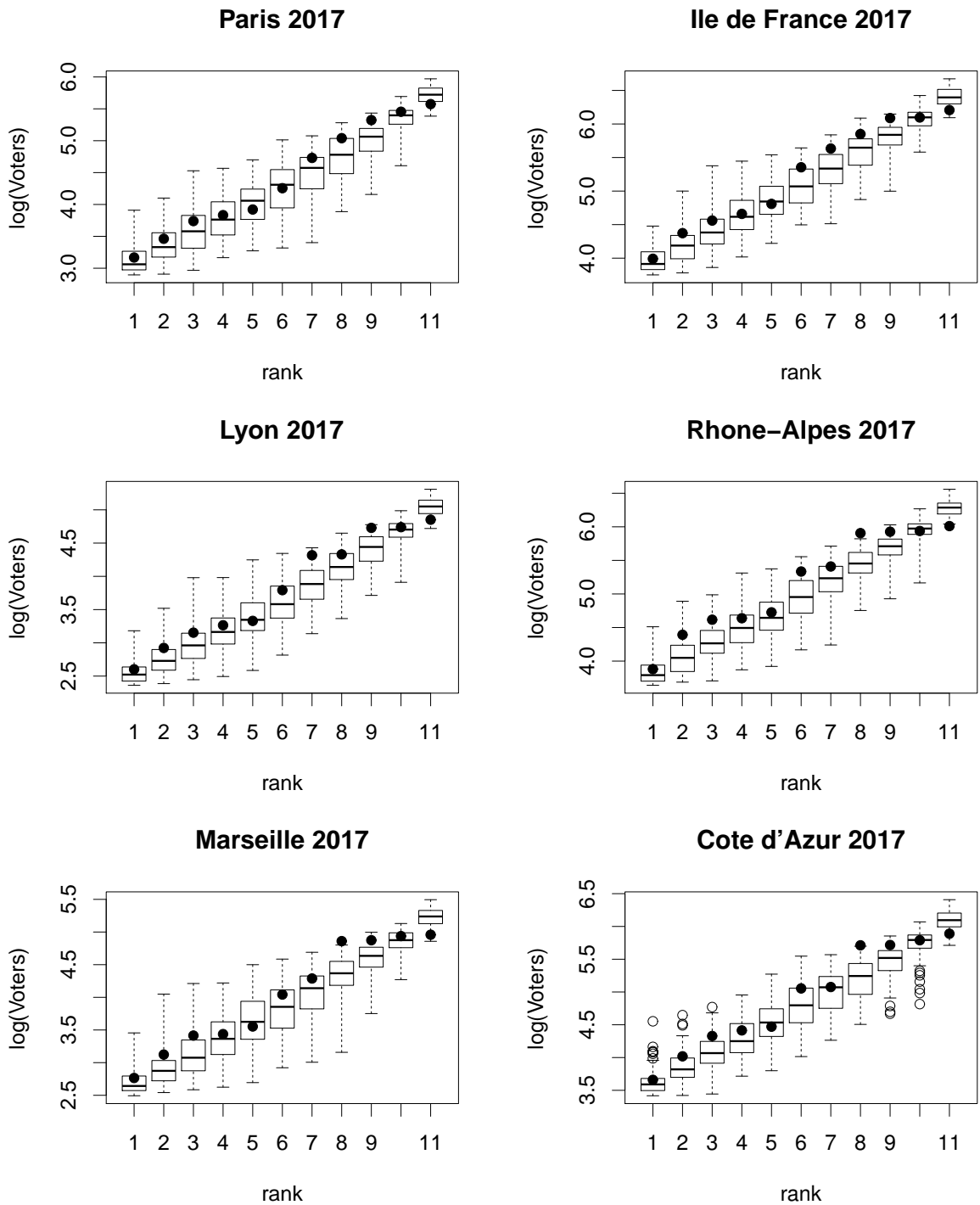

Figure 33: Election France, 2017 (boxplot of 100 realizations og the model, bullets: data).

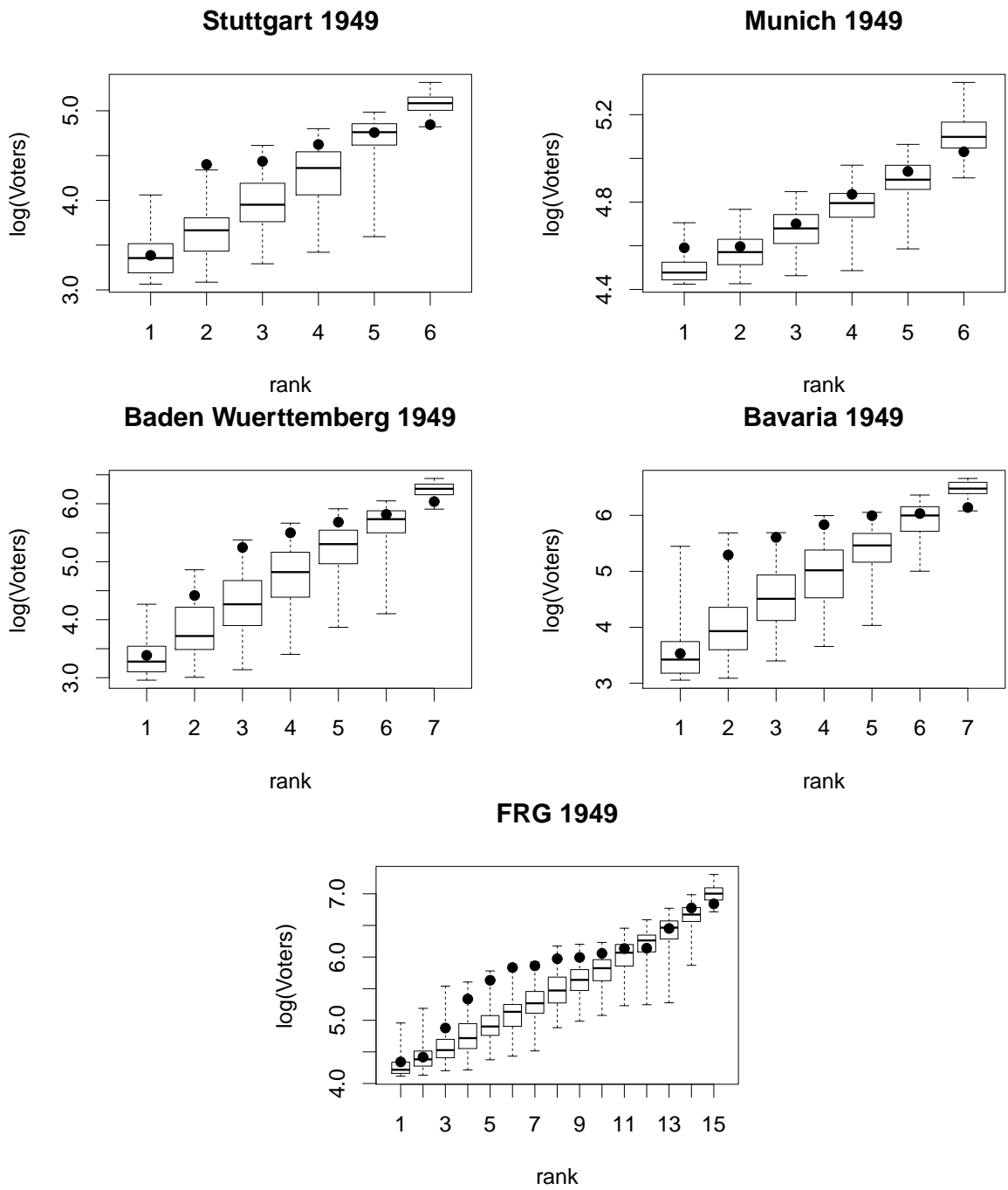

Figure 34: Election FRG, 1949 (boxplot of 100 realizations og the model, bullets: data).

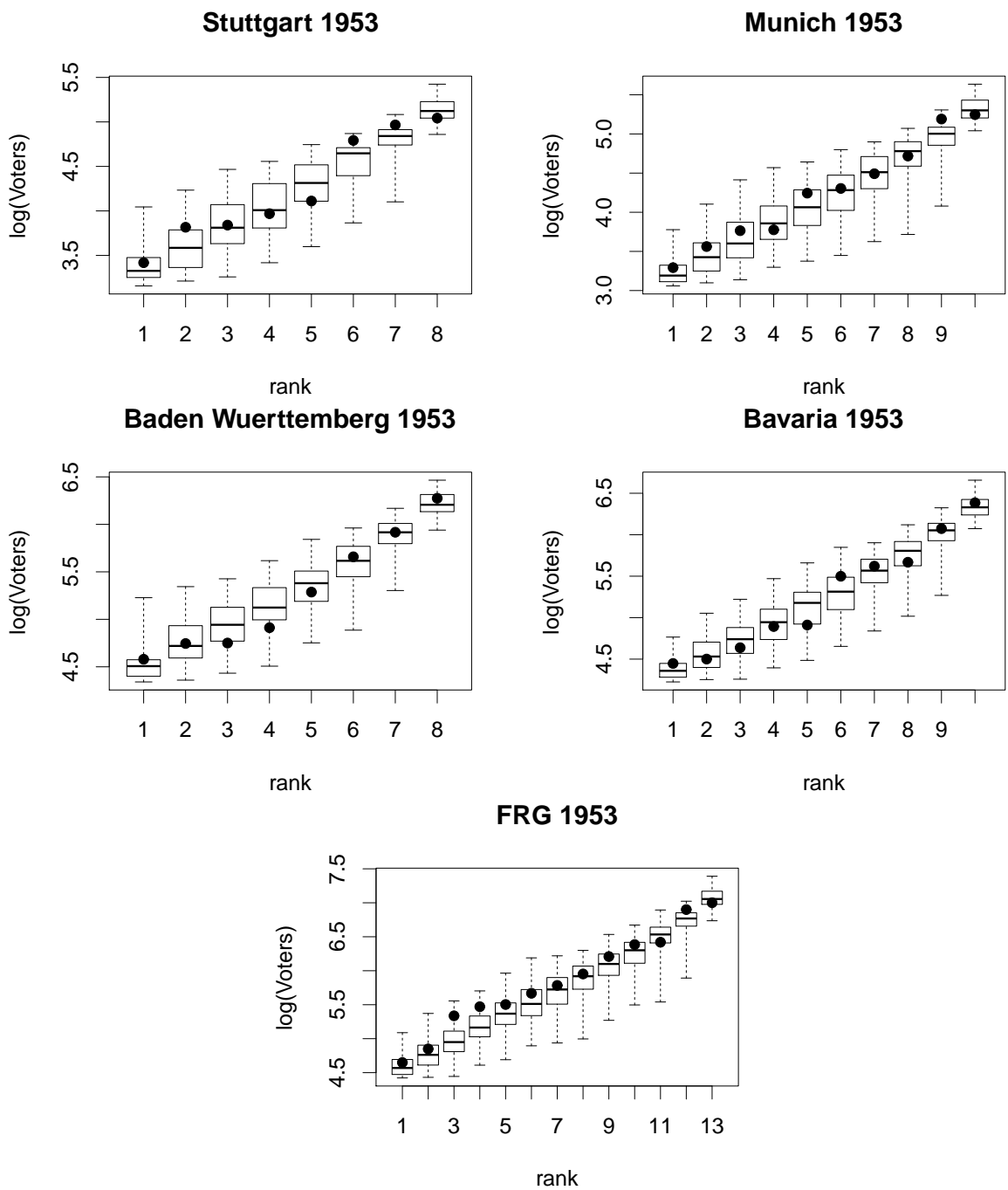

Figure 35: Election FRG, 1953 (boxplot of 100 realizations og the model, bullets: data).

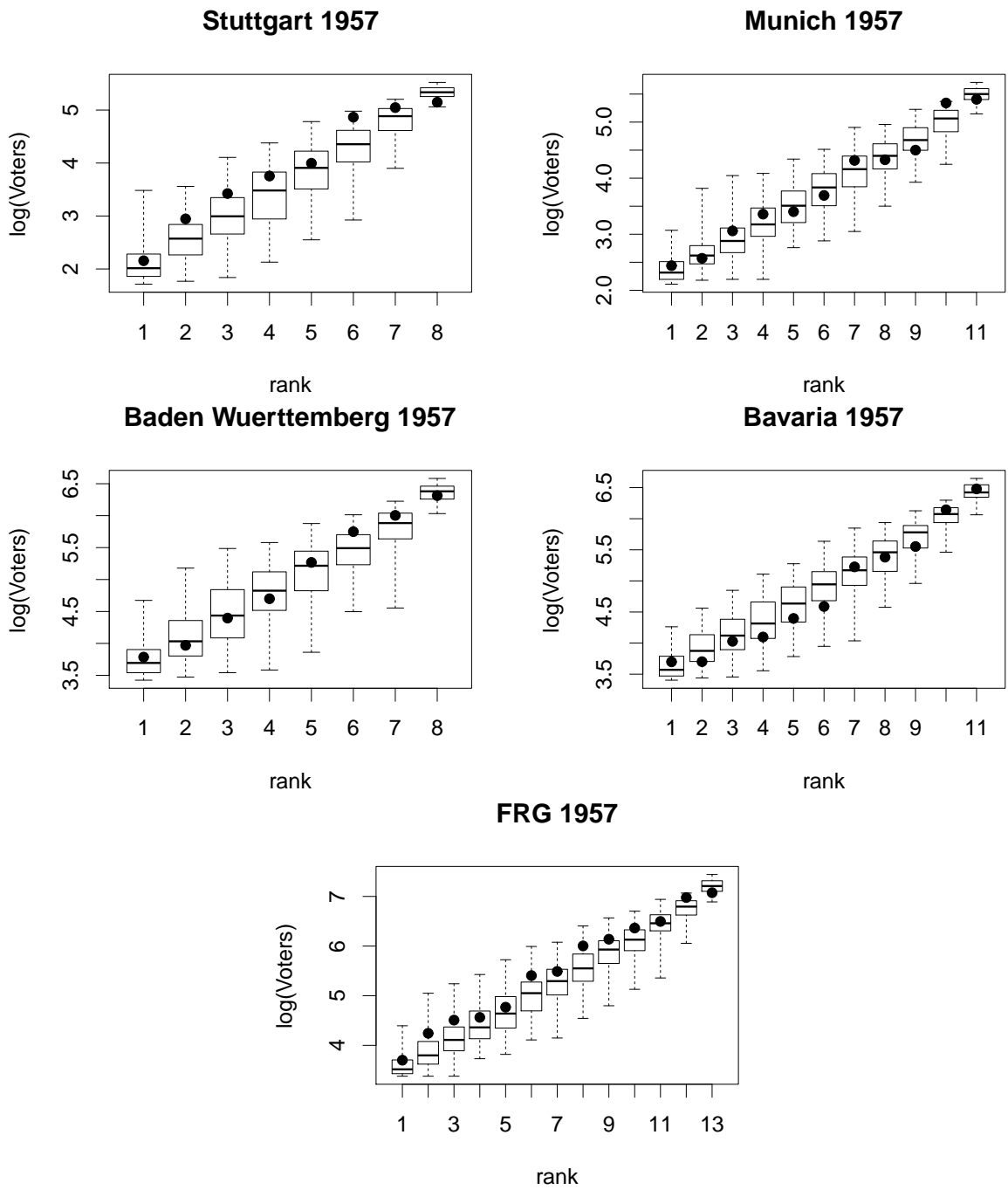

Figure 36: Election FRG, 1957 (boxplot of 100 realizations og the model, bullets: data).

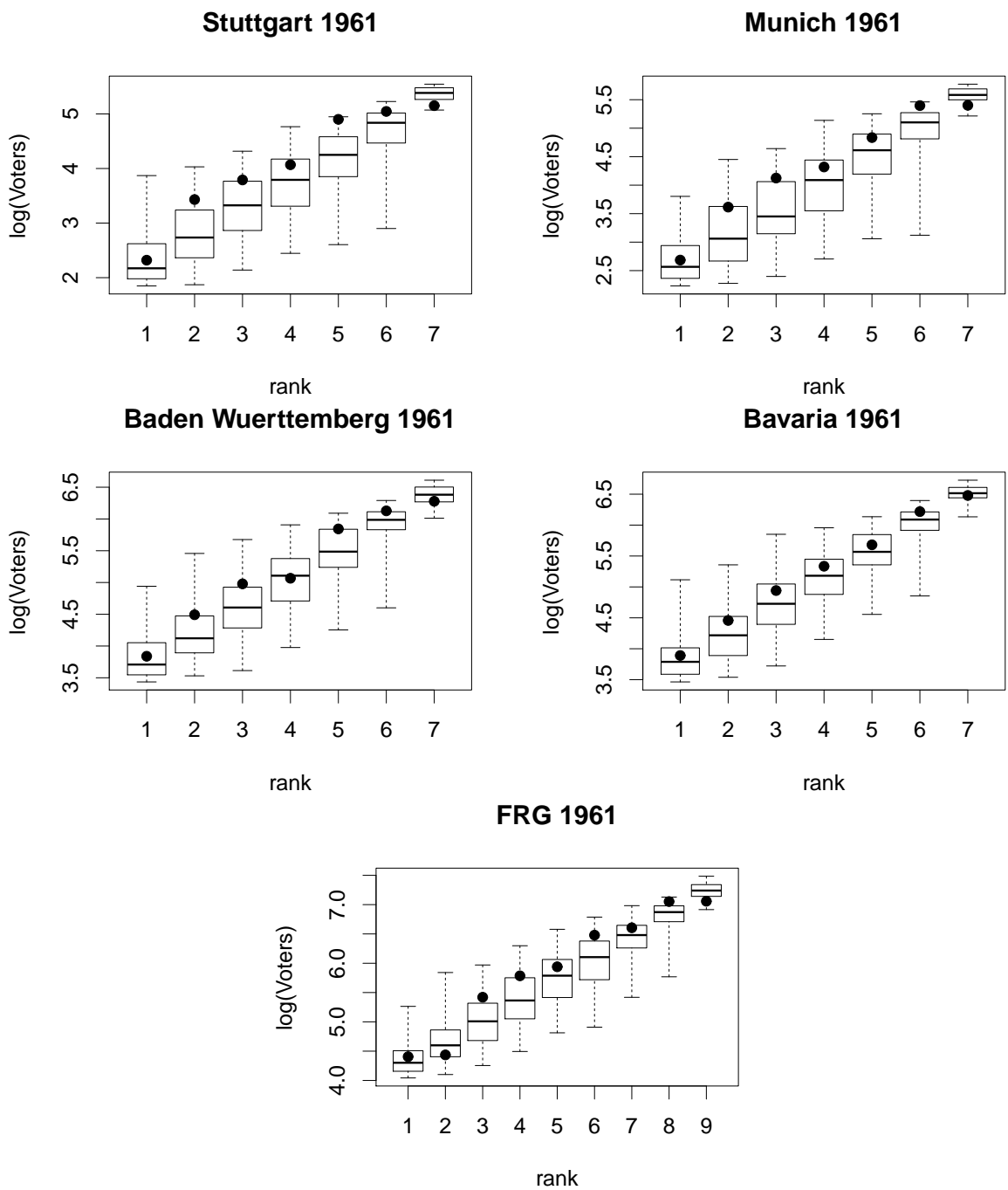

Figure 37: Election FRG, 1961 (boxplot of 100 realizations og the model, bullets: data).

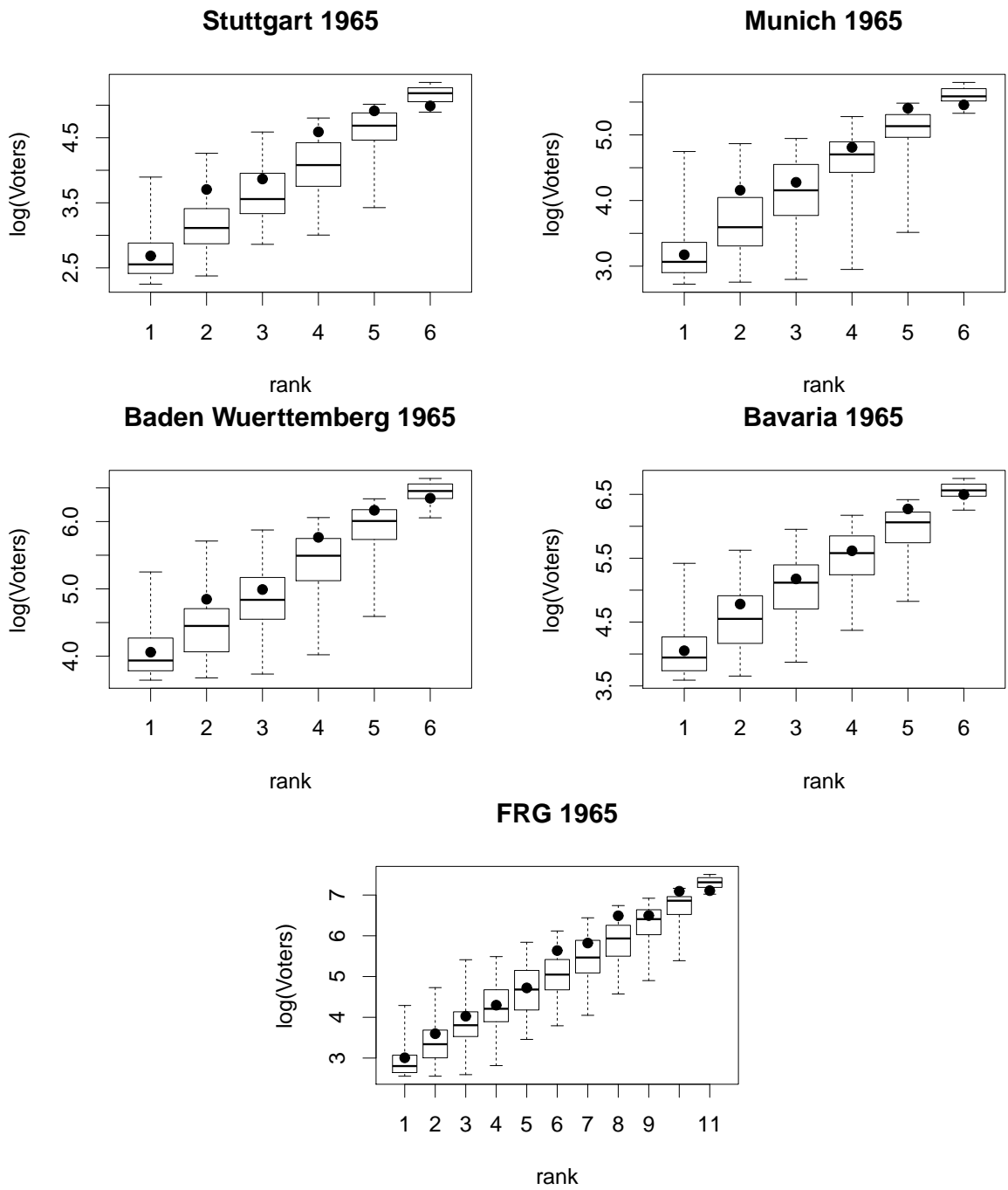

Figure 38: Election FRG, 1965 (boxplot of 100 realizations og the model, bullets: data).

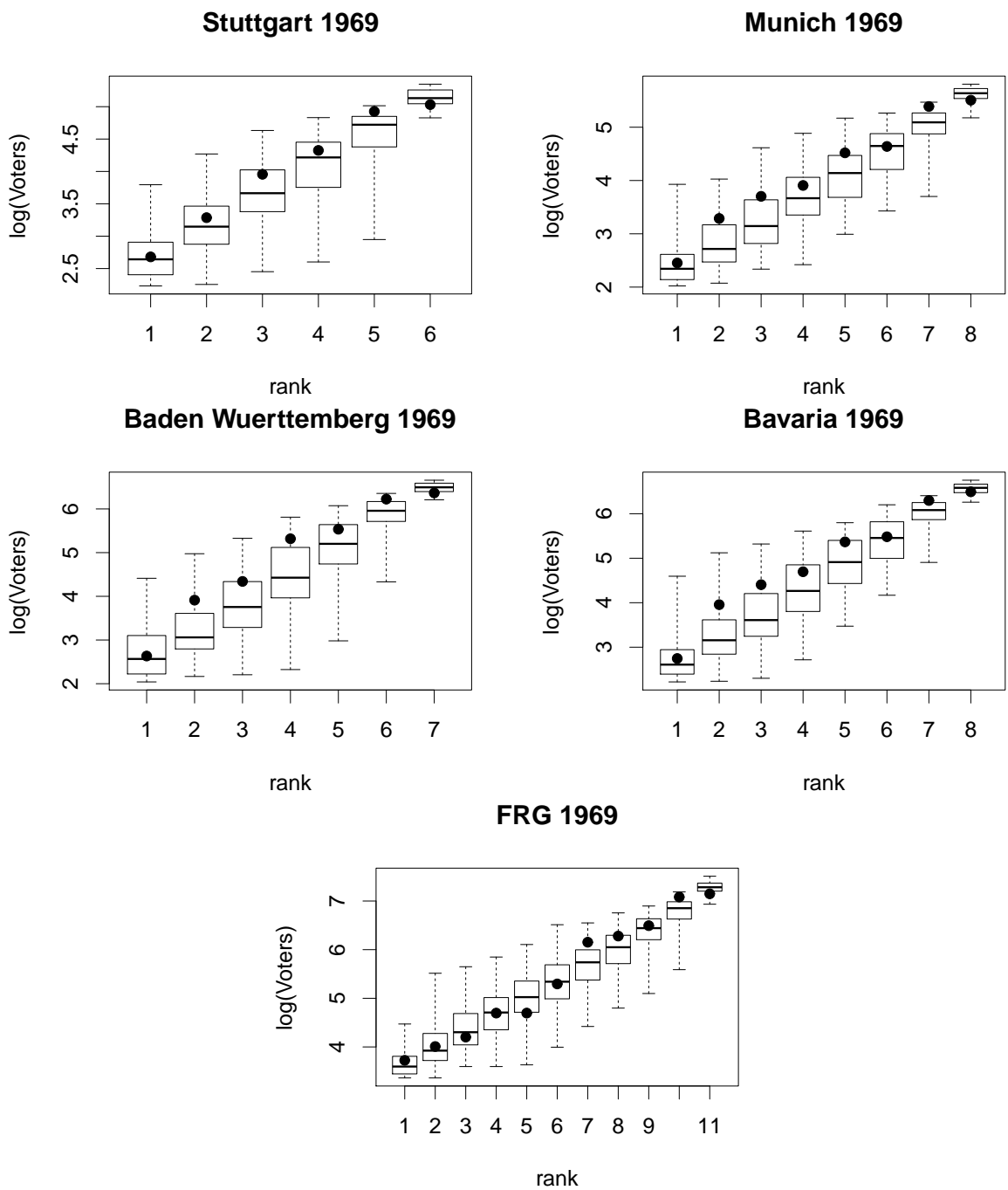

Figure 39: Election FRG, 1969 (boxplot of 100 realizations og the model, bullets: data).

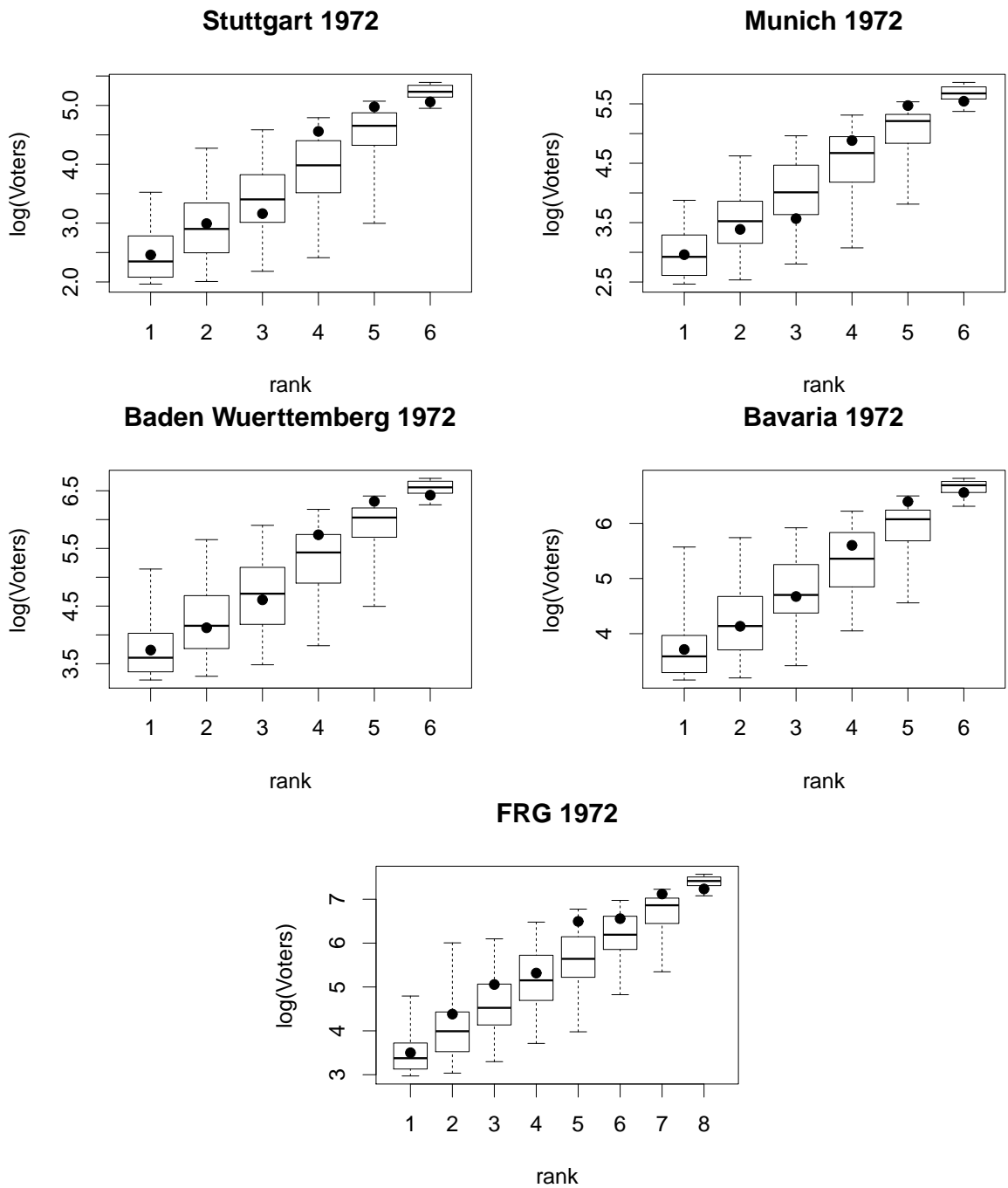

Figure 40: Election FRG, 1972 (boxplot of 100 realizations og the model, bullets: data).

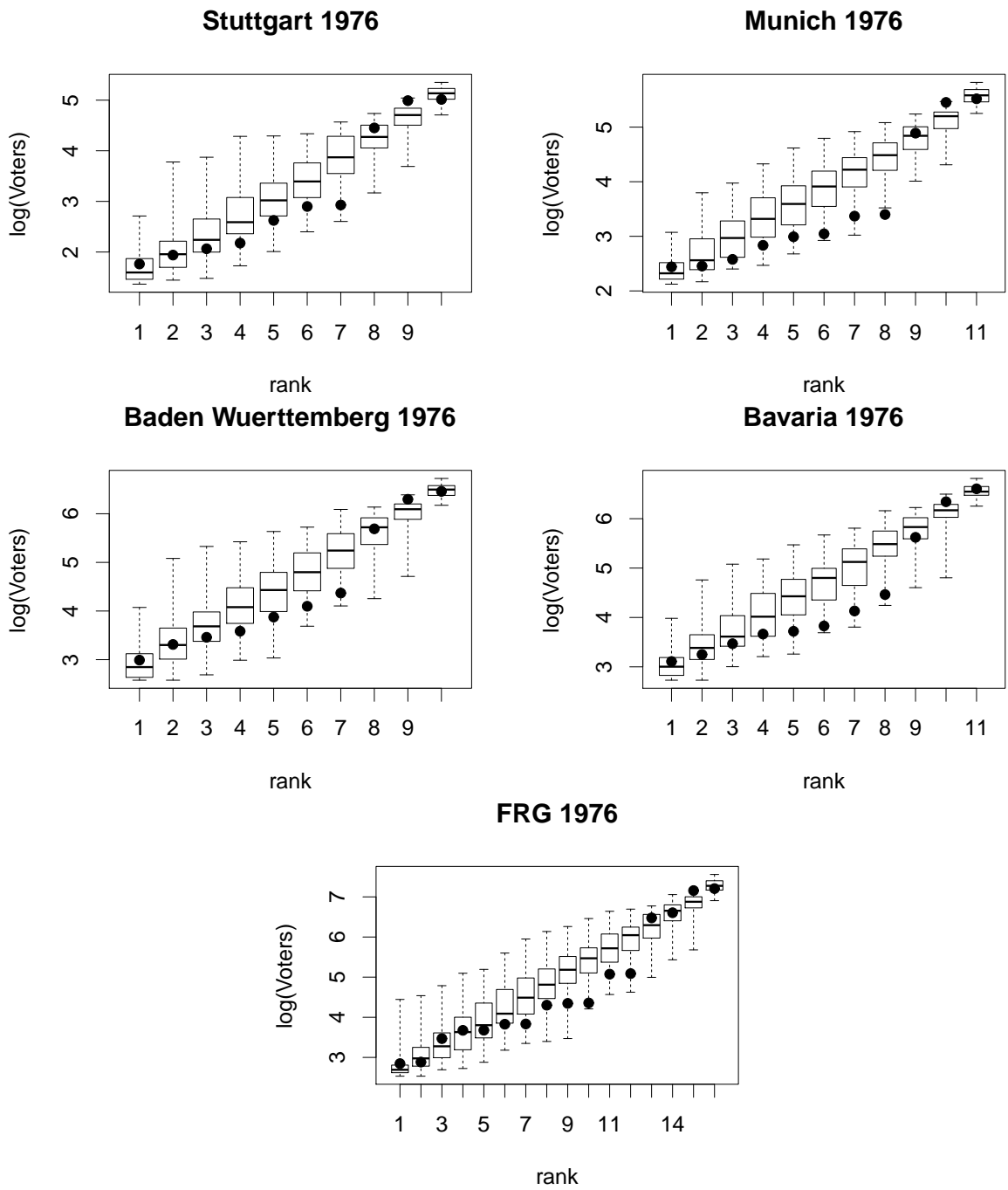

Figure 41: Election FRG, 1976 (boxplot of 100 realizations og the model, bullets: data).

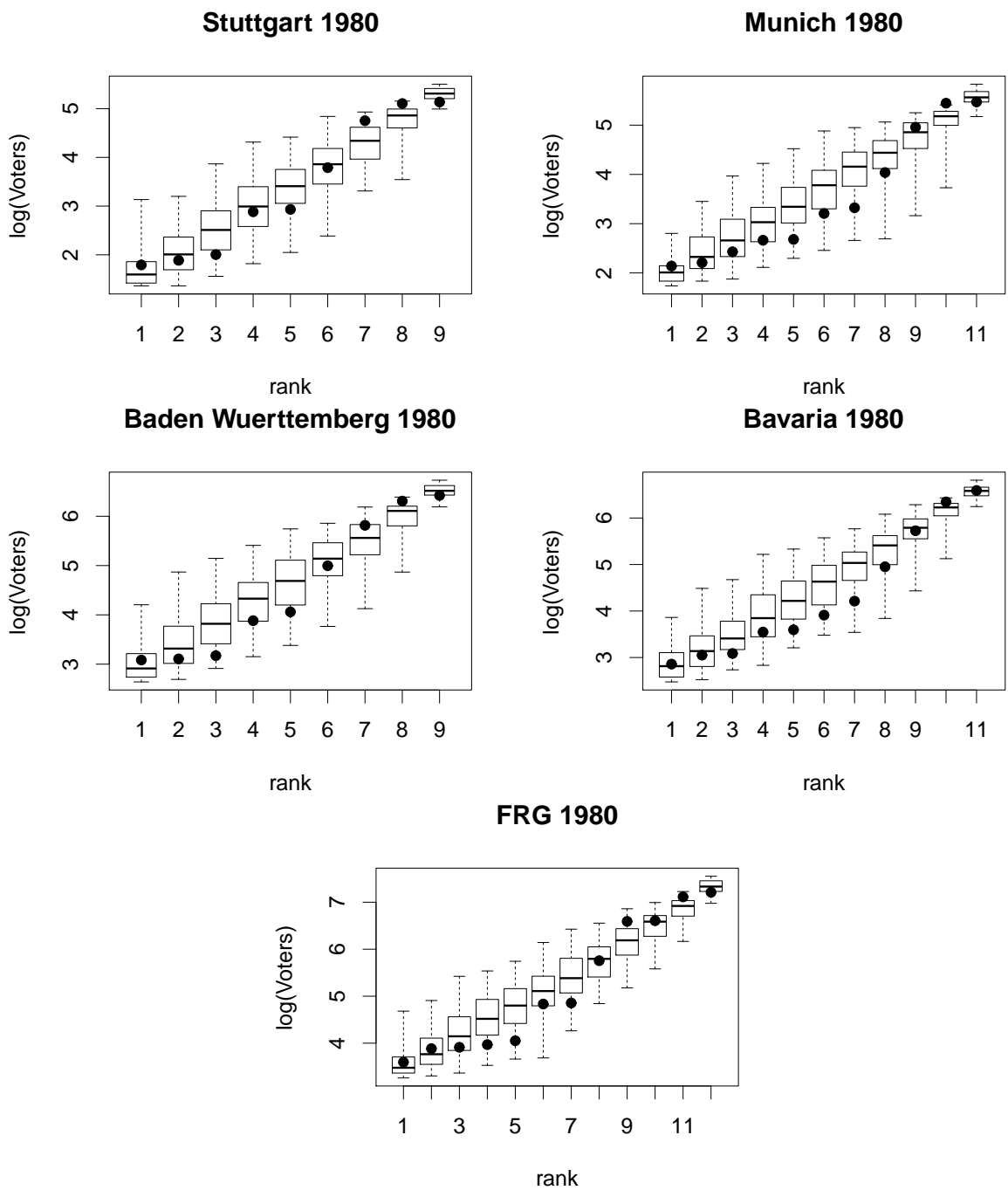

Figure 42: Election FRG, 1980 (boxplot of 100 realizations og the model, bullets: data).

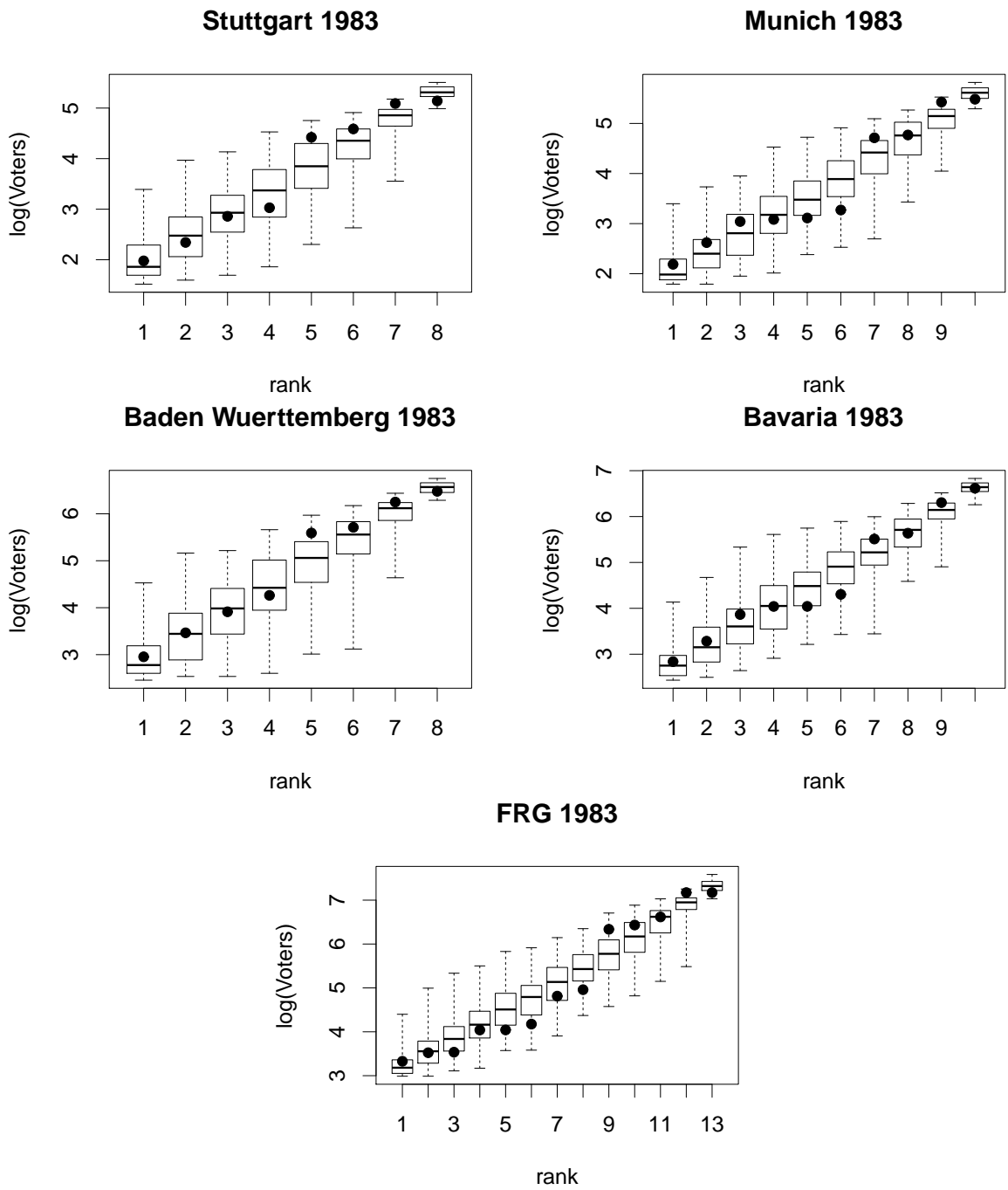

Figure 43: Election FRG, 1983 (boxplot of 100 realizations og the model, bullets: data).

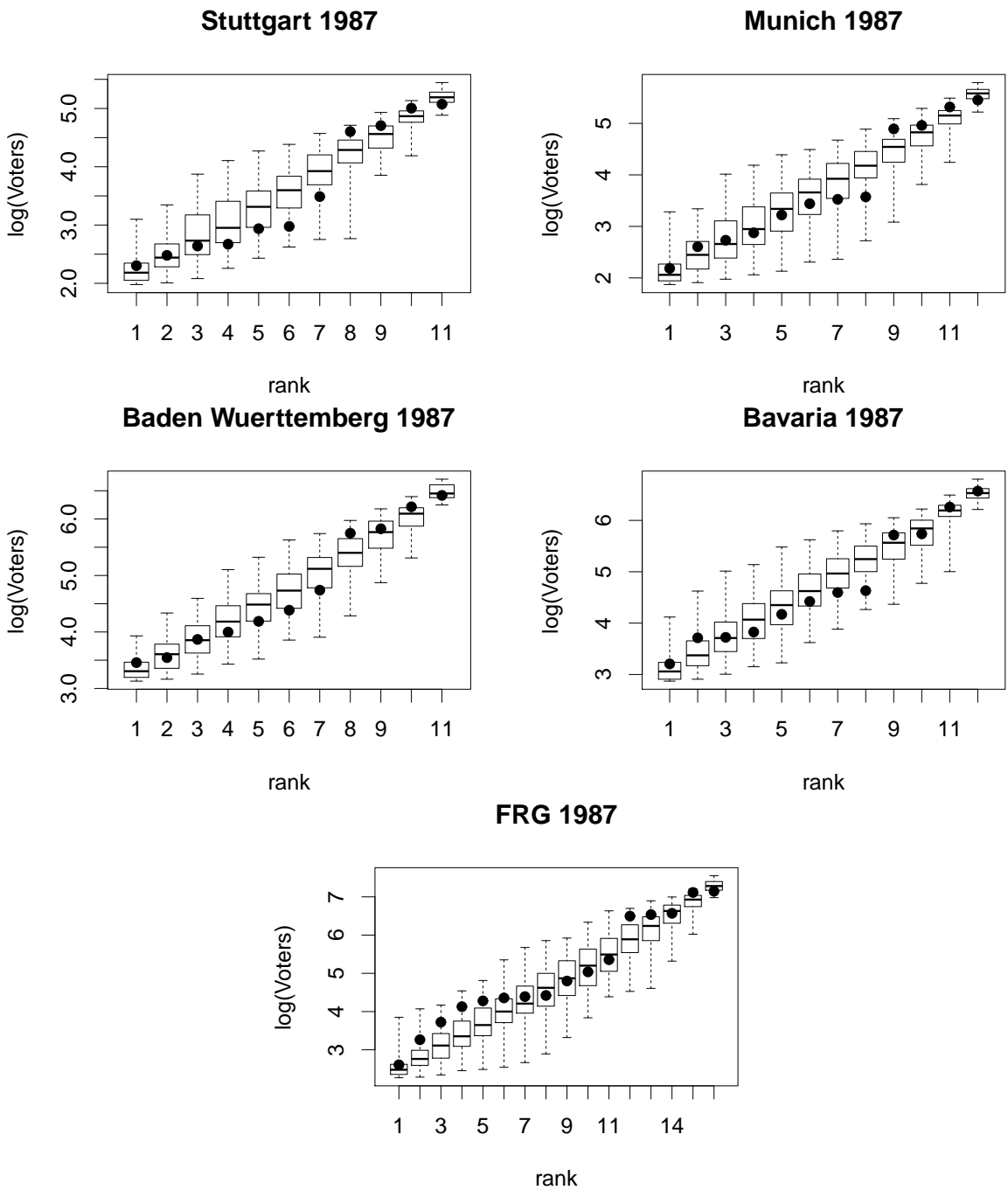

Figure 44: Election FRG, 1987 (boxplot of 100 realizations og the model, bullets: data).

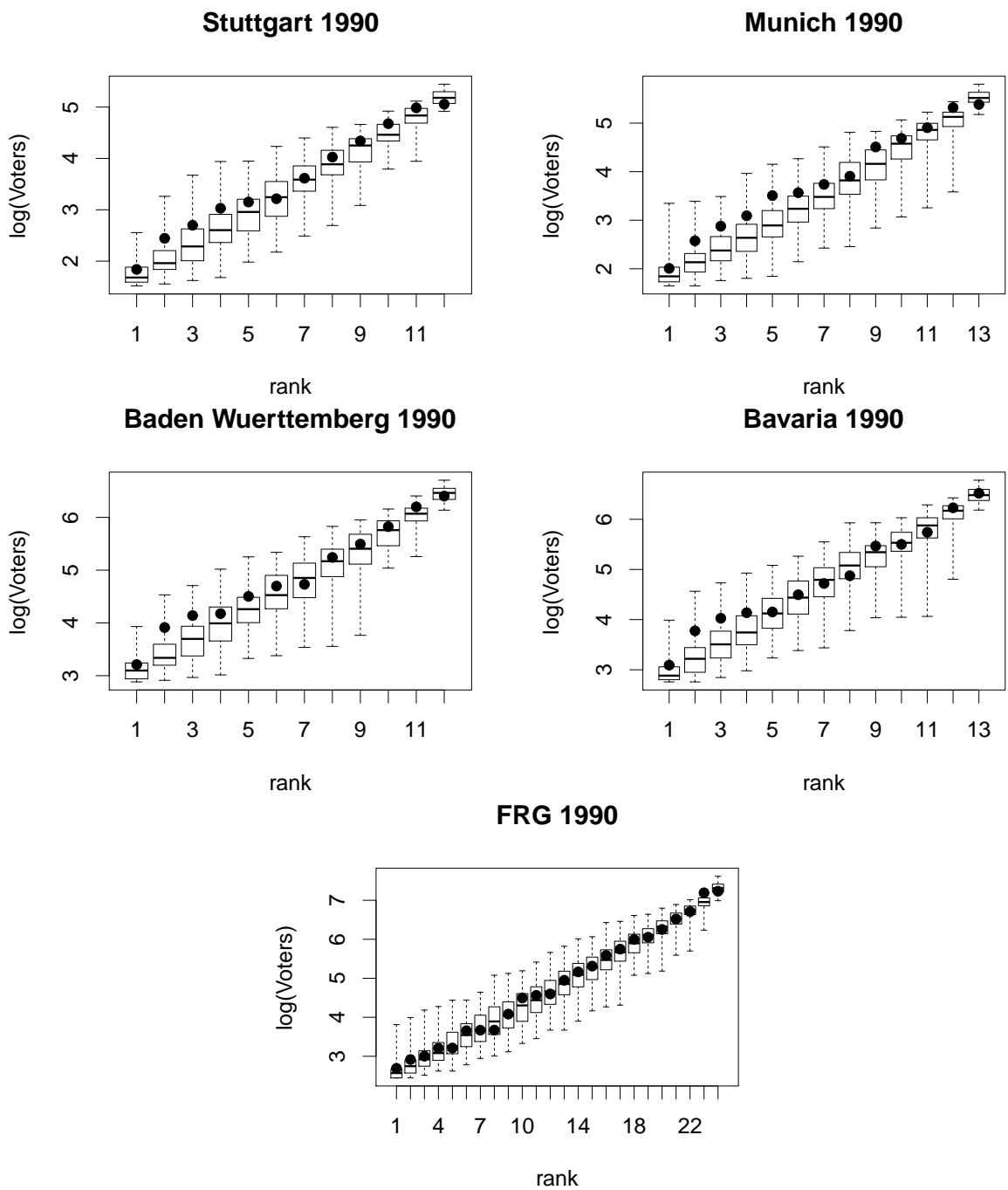

Figure 45: Election FRG, 1990 (boxplot of 100 realizations og the model, bullets: data).

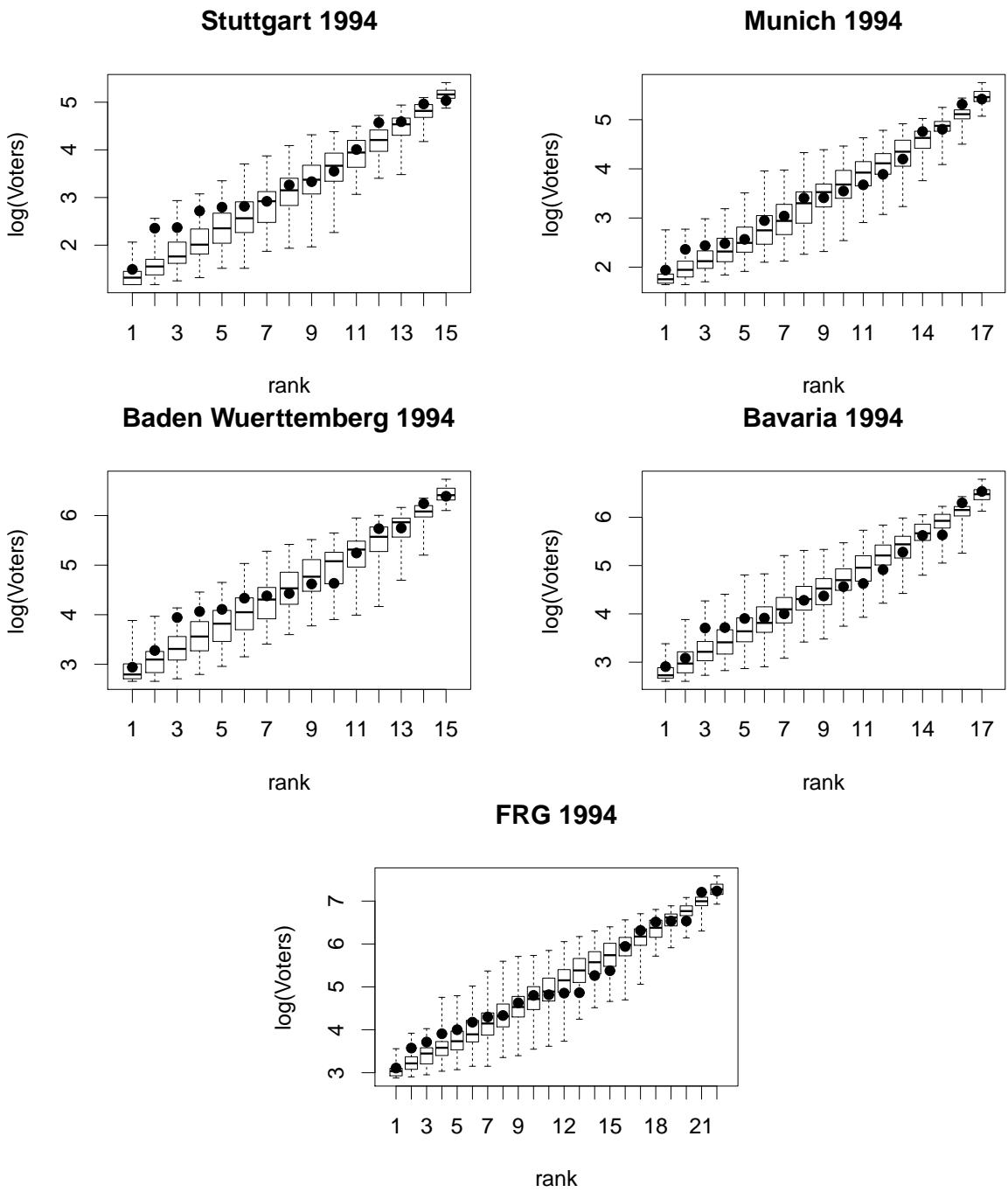

Figure 46: Election FRG, 1994 (boxplot of 100 realizations og the model, bullets: data).

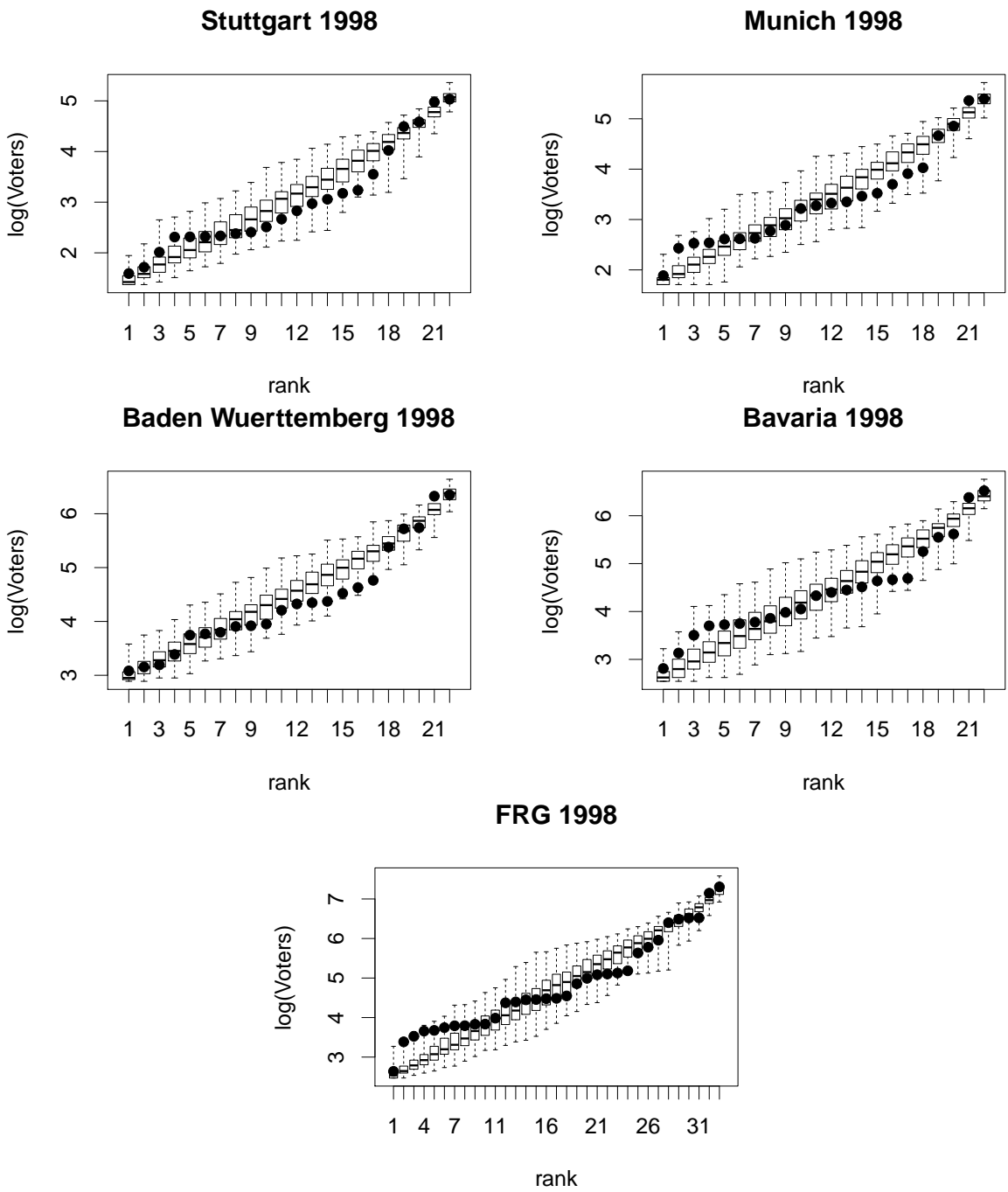

Figure 47: Election FRG, 1998 (boxplot of 100 realizations og the model, bullets: data).

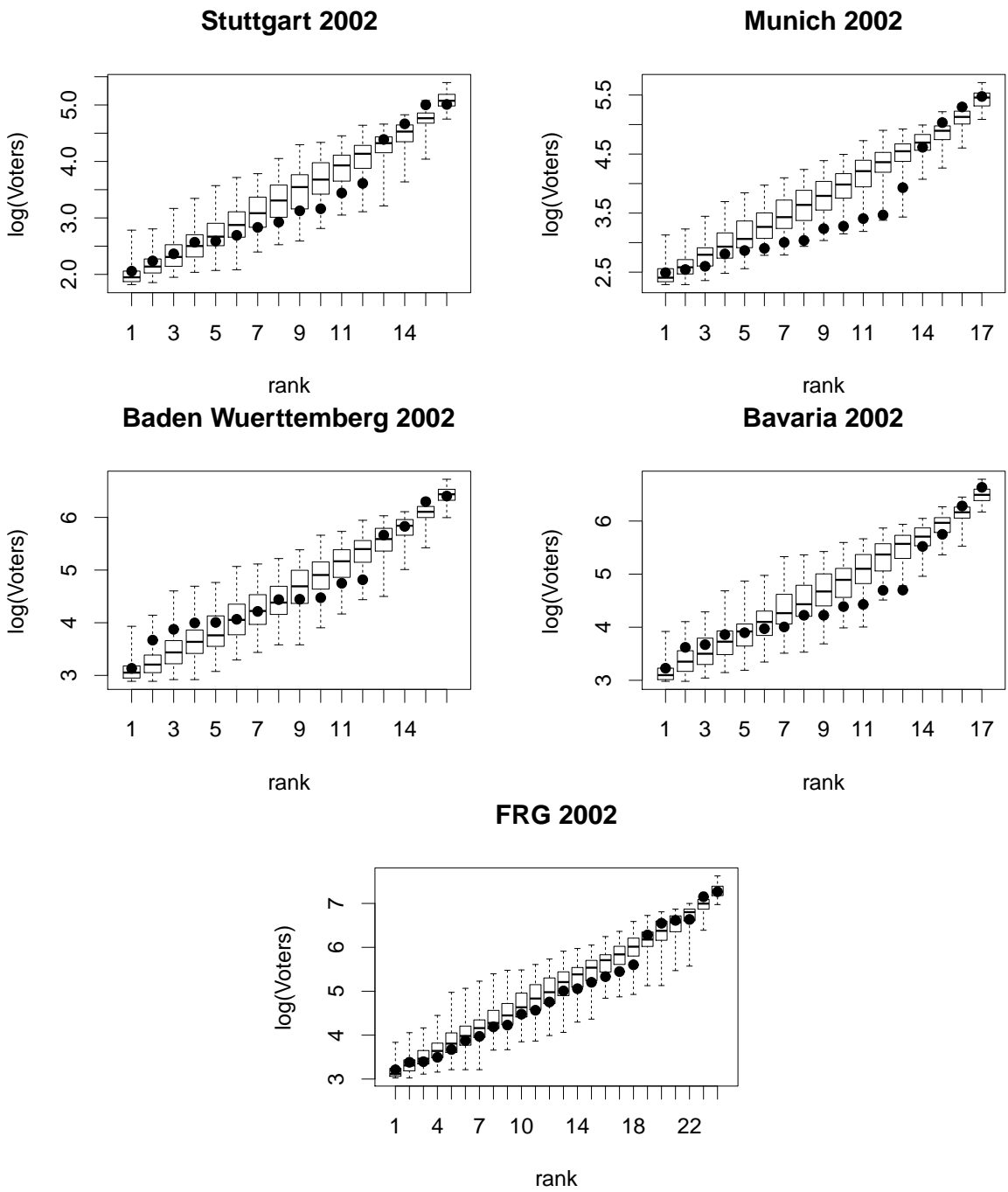

Figure 48: Election FRG, 2002 (boxplot of 100 realizations og the model, bullets: data).

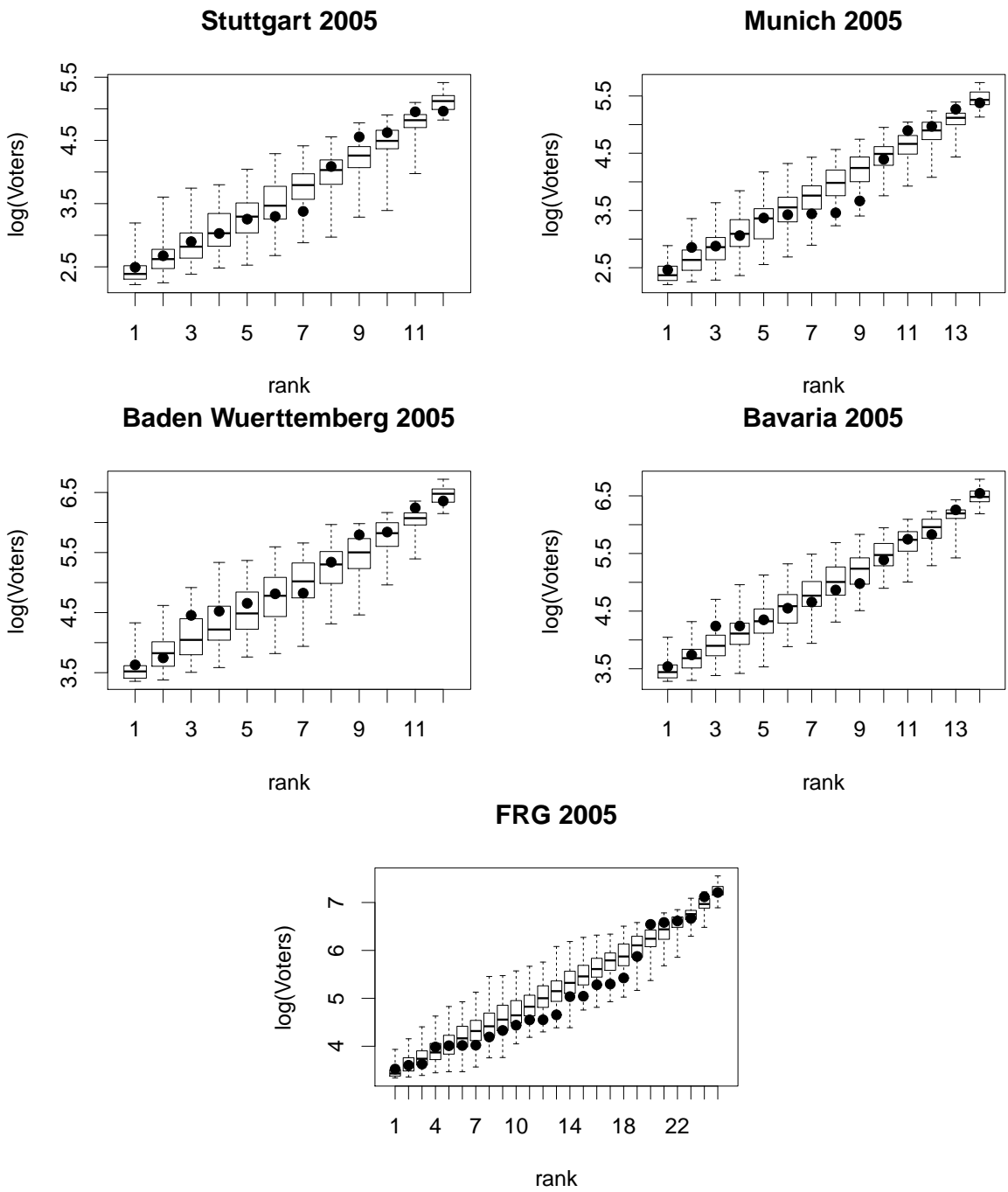

Figure 49: Election FRG, 2005 (boxplot of 100 realizations og the model, bullets: data).

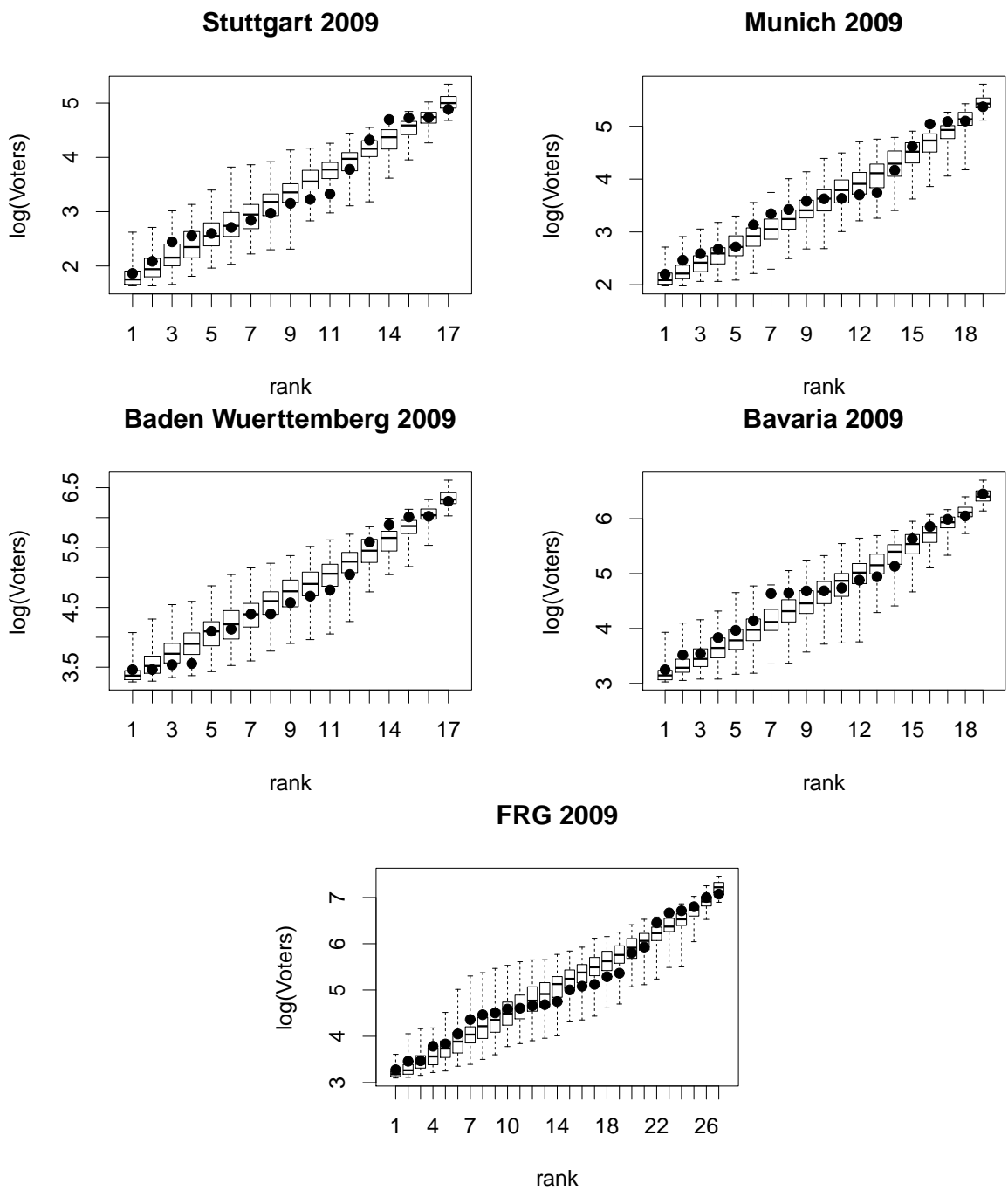

Figure 50: Election FRG, 2009 (boxplot of 100 realizations og the model, bullets: data).

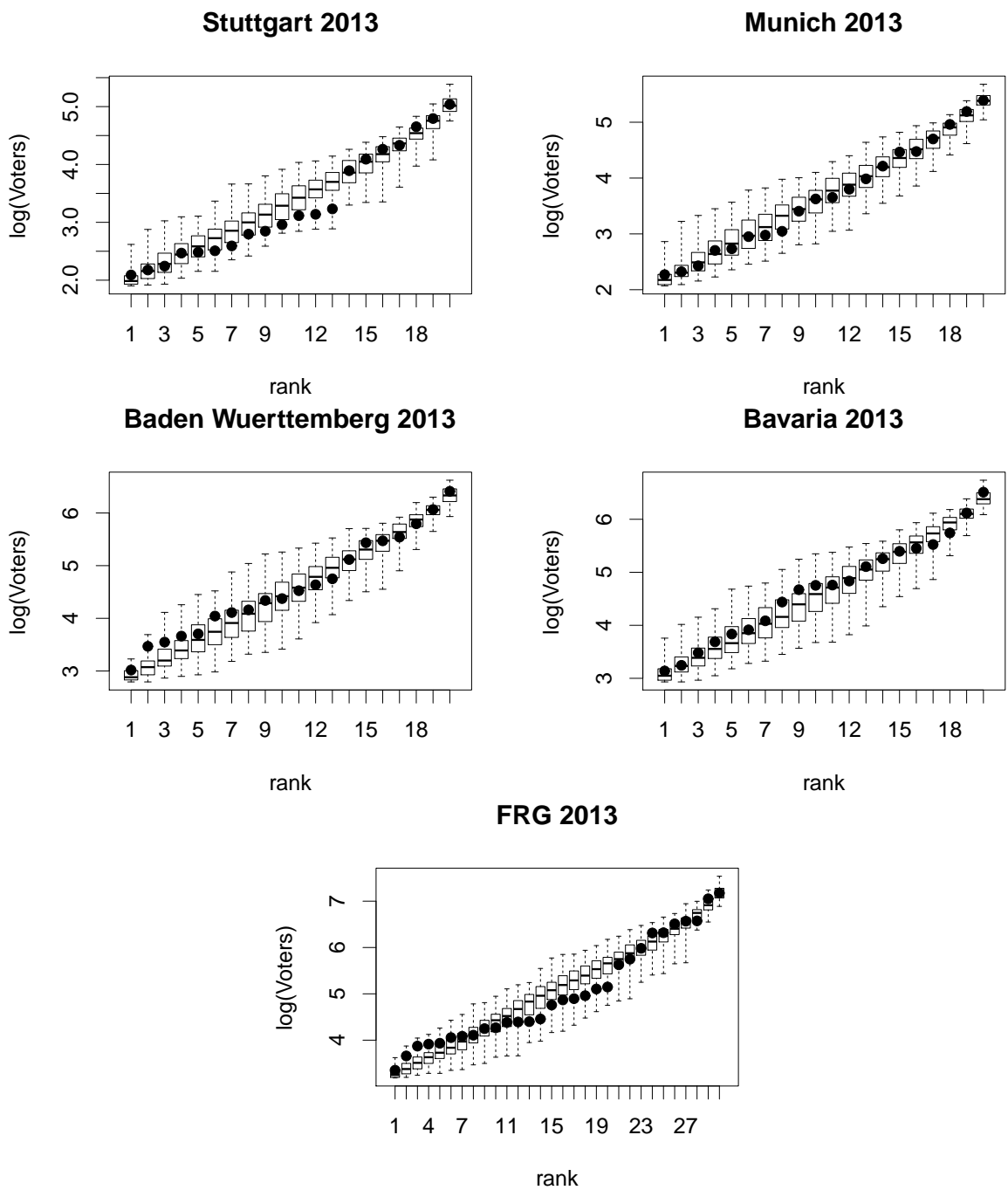

Figure 51: Election FRG, 2013 (boxplot of 100 realizations og the model, bullets: data).

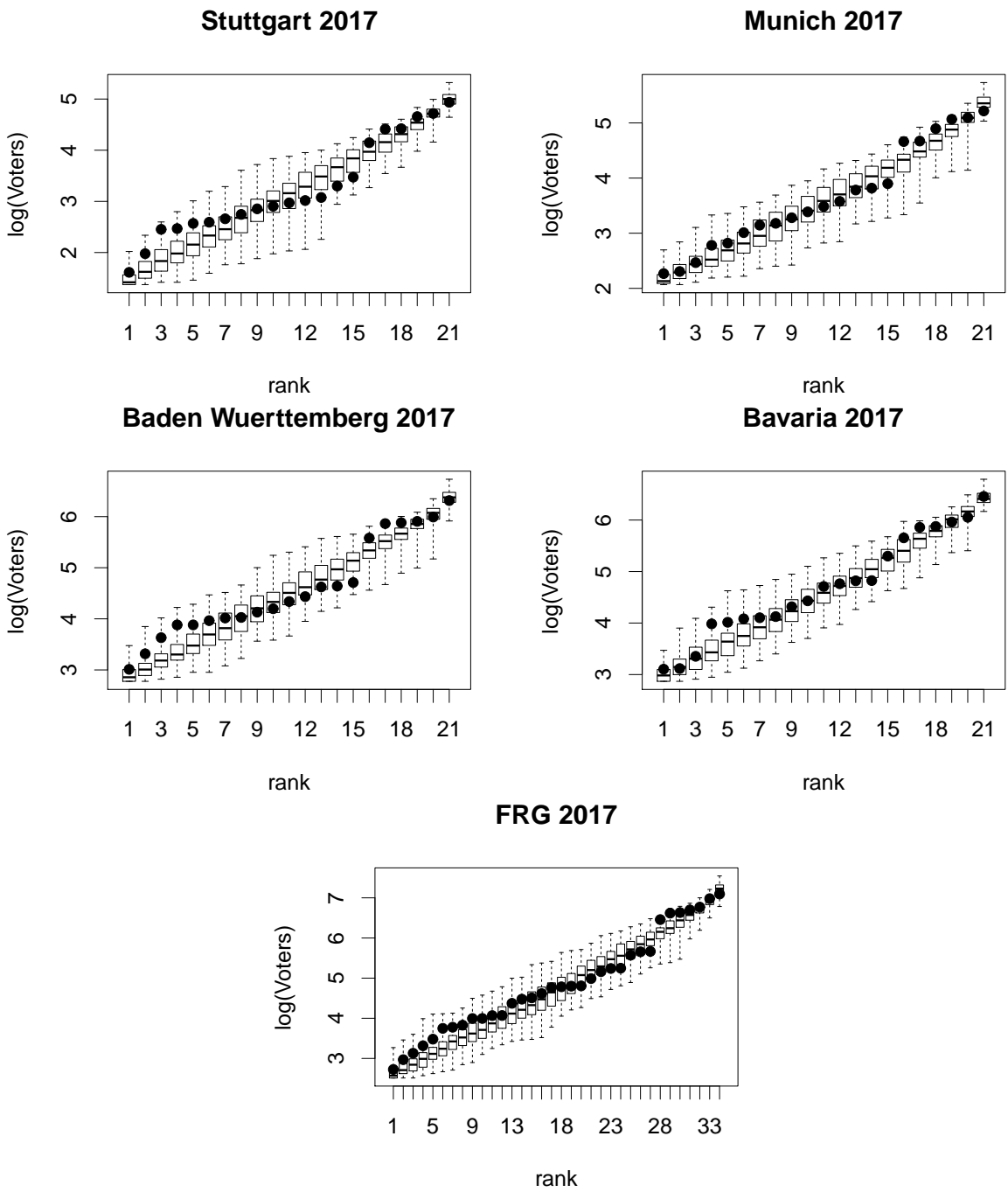

Figure 52: Election FRG, 2017 (boxplot of 100 realizations og the model, bullets: data).